%% file: final.tex
\documentclass[journal,final]{IEEEtran}
\usepackage[utf8]{inputenc}
\usepackage{amsmath,amstext,amsfonts,amssymb,dsfont,bbm}
\usepackage{graphicx, tikz, pgfplots}
\include{macros}

\begin{document}

\title{The Second-Order Coding Rate of the MIMO Quasi-Static Rayleigh Fading Channel}
\author{Jakob~Hoydis,~\IEEEmembership{Member,~IEEE}, Romain~Couillet,~\IEEEmembership{Member,~IEEE}, and Pablo~Piantanida,~\IEEEmembership{Member,~IEEE}
\thanks{This work has been presented in part at the IEEE International Symposium on Information Theory (ISIT'12), Cambridge, MA, USA, Jul. 2012, and the IEEE International Symposium on Information Theory (ISIT'13), Istanbul, Turkey, Jul. 2013.}
\thanks{J. Hoydis is with Bell Laboratories, Alcatel-Lucent, Route de Villejust, 91620 Nozay, France (jakob.hoydis@alcatel-lucent.com).}
\thanks{R. Couillet is with Laboratoire de Signaux et Systèmes (L2S, UMR8506), CNRS-CentraleSupélec-Université Paris-Sud, 3 rue Joliot-Curie, 91192 Gif-sur-Yvette, France (romain.couillet@centralesupelec.fr). Couillet's work is supported by the ERC MORE EC--120133.}
\thanks{P. Piantanida is with Laboratoire de Signaux et Systèmes (L2S, UMR8506), CNRS-CentraleSupélec-Université Paris-Sud, 3 rue Joliot-Curie, 91192 Gif-sur-Yvette, France (pablo.piantanida@centralesupelec.fr). The work of Piantanida was  partially supported by the DIM COREPHY project and  the FP7 Network of Excellence in Wireless COMmunications NEWCOM\#.}
\thanks{Copyright (c) 2014 IEEE. Personal use of this material is permitted.  However, permission to use this material for any other purposes must be obtained from the IEEE by sending a request to pubs-permissions@ieee.org.}
}
\maketitle

\begin{abstract}
The second-order coding rate of the multiple-input multiple-output (MIMO) quasi-static Rayleigh fading channel is studied. We tackle this problem via an information-spectrum approach and statistical bounds based on recent random matrix theory techniques. We derive a central limit theorem (CLT) to analyze the information density in the regime where the block-length $n$ and the number of transmit and receive antennas $K$ and $N$, respectively, grow simultaneously large. This result leads to the characterization of closed-form upper and lower bounds on the optimal average error probability when the coding rate is within $\Oc(1/\sqrt{nK})$ of the asymptotic capacity.
\end{abstract}


\section{Introduction}
In real-world wireless communications, the codeword (or block) length of the transmission is naturally limited due to delay and complexity constraints. It is thus unfortunate that only few tractable performance limits of wireless communication scenarios under the finite block-length regime are available. In general, only bounds on the optimal error probability for a given coding rate and block-length are derivable, e.g.,~\cite{FEI54,SHA59}, which are for most relevant cases difficult to analyze and evaluate. This is in particular the case for non-ergodic channels (e.g., quasi-static or block-fading channels), for which the error probability is fundamentally limited by the outage probability~\cite{BIG98}. The evaluation of these non-asymptotic bounds becomes even more challenging in presence of  multiple-input multiple-output (MIMO) channels.

Feinstein~\cite{FEI54} and Shannon \cite{shannon1957certain} were among the first to explore the tradeoff between
coding rate, error probability, and block-length and developed bounds on the optimal error probability in the finite block-length regime. Bounds on the limit of the scaled logarithm of the error probability---known as the exponential rate of decrease---were derived in \cite{SHA59}. A simpler formula for the latter was then provided by Gallager~\cite{GAL65}, which is still difficult to evaluate for wireless  channel models. In \cite{HYU09}, an explicit expression of Gallager's error exponent was found for the block-fading MIMO channel, but the computation of this result remains quite involved. 

Since the aforementioned bounds are in general not amenable to simple evaluation, asymptotic considerations were made, in particular by Strassen \cite{STR62} who derived a general expression of the error probability for the discrete memoryless channel with unconstrained inputs of code length $n$ in the regime where the coding rate is within $\Oc(1/\sqrt{n})$ of the capacity, which is referred to as the \emph{second-order coding rate}. In his work, the variance of the ``mutual information density" appears to be the fundamental quantity when focusing on Gaussian approximations of the error probability. Nevertheless, Strassen's approach could not be generalized to channels with input constraints, such as the additive white Gaussian noise (AWGN) channel. Hayashi \cite{HAY09} focused on the second-order coding rate and provided an exact characterization of the optimal error probability for different channel models and input constraints. Further considerations were made by Polyanskiy-Poor-Verd\'u in \cite{POL10} where several novel results are provided for memoryless channels, among which new upper and lower bounds on the maximal achievable rate for a fixed error probability and block-length. Along the same lines, the scalar AWGN block-fading channel was addressed in the coherent and non-coherent settings in~\cite{POL11}  and \cite{DBLP:journals/corr/abs-1204-2927}, respectively. 

Additional work on the asymptotic block-length regime via information-spectrum methods comprises the general capacity formula by Verd\'u-Han \cite{VER94} proving the converse via a novel lower bound on the error probability from~\cite{DBLP:journals/iandc/Wolfowitz68,HAN93b}. A very comprehensive literature survey on related aspects can also be found in~\cite{POL10}. During the revision of this article, we became aware of the related works~\cite{YAN13a} and~\cite{YAN13b} which study respectively the quasi-static fading single-input multiple-output (SIMO) and MIMO channel at finite block-length in great detail. 

In this paper, we investigate \emph{closed-form} bounds on the average error probability of the $N\times K$ MIMO quasi-static Rayleigh fading channel where  the transmission takes place over $n$ channel uses during which the channel realization is randomly drawn  but remains constant, and where $N$, $K$, and $n$ are of similar order of magnitude. 

\subsection{Contribution and outline}
We focus on the asymptotic behavior of the error probability when the coding rate is a small perturbation of the ergodic capacity, and hence  follow the line of work of~\cite{HAY09} on the second-order coding rate (see also \cite[Section~IV]{POL10}). We take the approach of inducing ergodicity in the inherently non-ergodic quasi-static fading channel  by growing the channel matrix dimensions. Indeed, assuming an $N\times K$ channel matrix with independent standard Gaussian  entries, letting $K, N\to\infty$ at the same speed, 
the channel becomes ergodic in the limit (even for a single channel use). This ensures that communications at rates arbitrarily close to the asymptotic capacity are possible in this regime and it becomes natural to investigate the optimal average error probability for the second-order coding rate when $K$, $N$, and the block-length $n$ grow simultaneously, i.e., the asymptotically achievable error probability for rates within $\Oc(1/\sqrt{nK})$ of the ergodic capacity ($nK$ being the total number of symbols in each codeword).

Our approach closely follows the information spectrum methodology of \cite{HAY09}. We first start from some basic variations of Feinstein's and Verd{\'u}--Han's lemma that provide, respectively, lower and upper bounds on the optimal error probability. These bounds are exploited to study the second-order statistics of the information density, seen as a real functional of three large-dimensional random matrices, i.e., the $N\times K$ channel, the $K\times n$ input, and the $N\times n$ noise matrices. The analysis of such statistics naturally requires the use of random matrix tools, and in particular here of Gaussian methods such as developed by Pastur \cite{pastur2011eigenvalue}. 

The main contribution of this paper is to derive a central limit theorem (CLT) uniformly over the set of admissible channel inputs. From this result it entails that the optimal average error probability $\mathbb{P}_e(r|\beta,c)$ for the second order coding rate $r<0$ (defined in \eqref{errorprob_def3} below) can be bounded as
\begin{align}
	\label{eq:main_result}
	\Phi \left( \frac{r}{\theta_-} \right) \leq \mathbb{P}_e(r|\beta,c) \leq \Phi \left( \frac{r}{\theta_+} \right)
\end{align}
where $\beta=n/K$, $c=N/K$, $\Phi(\cdot)$ is the \emph{Gaussian distribution function}, and $\theta_+>\theta_-$ are closed-form functions of $\beta$, $c$, and the signal-to-noise ratio (SNR). Unlike \cite{HAY09,POL10}, we do not obtain matching lower and upper bounds due to the presence of the non-ergodic random channel matrix. Nonetheless, it appears that the gap between both bounds is quite tight for SNR values of practical interest. Besides, numerical comparisons to LDPC codes reveal good similarities with theory in the slope of the error probability.

\subsection*{Notation and definitions}
The set of nonnegative integers is denoted by $\mathds{N}$, the real and complex fields by $\mathds{R}$ and $\mathds{C}$, respectively. Boldface letters $\mb{x}$ and upper-case letters $\mb{X}$ are used to denote vectors and matrices, respectively. The transpose, complex conjugate, and complex conjugate  (Hermitian) transpose are denoted by $(\cdot)\tp$, $(\cdot)^*$, and $(\cdot)\htp$, respectively. The trace and determinant of a square matrix $\Xm$ are written $\trace\Xm$ and $\det(\Xm)$, respectively. The spectral norm of a square matrix $\Xm$, i.e., the absolute largest eigenvalue, is denoted by $\Vert\Xm\Vert$. The Frobenius norm of a matrix $\Xm$ is denoted by $\Vert \Xm \Vert_F$. The $(i,j)$-element of $\Xm$ is denoted by $\Xm_{ij}$ or $\LSB \Xm\RSB_{ij}$. Random vectors and matrix variables are denoted by lowercase letters $x$ and uppercase letters $X$, respectively. The symbol $\Pr[\cdot]$ denotes the probability of the bracketed random argument. For a set $\mathcal S$, we define by $\mathcal P(\mathcal S)$ the set of probability measures with support a subset of $\mathcal S$. We also denote by ${\rm supp}(\mathbb{P})$ the support of $\mathbb{P}$.

For random matrices $X,Y$ in $\mathds{C}^{K\times n}$ and $\mathds{C}^{N\times n}$, let $\mathbb{P}_{X}\in \mathcal{P}(\mathds{C}^{K\times n})$ and let $\mb{X} \mapsto  \mathbb{P}_{Y|X}(\,\cdot\, | \mb{X}) $ be any Borel measurable mapping. We define the probability measure $\mathbb{P}_{XY}$ by $\mathbb{P}_{XY}(\mathcal{A}\times\mathcal{B})=\int_{\mathcal{A}} \mathbb{P}_{Y|X}(\mathcal{B} | \mb{X}) \mathbb{P}_{X}(d\mb{X})$ where $\mathcal{A},\mathcal{B}$ are Borel sets of $ \mathds{C}^{K\times n}$ and $ \mathds{C}^{N\times n}$, respectively. Similarly, we define the distribution $\mathbb{P}_{Y}$ as $\mathbb{P}_{Y}(\mathcal{B})=\int  \mathbb{P}_{Y|X}(\mathcal{B} | \mb{X}) \mathbb{P}_{X}(d\mb{X})$ for any Borel subset $\mathcal{B}\subset  \mathds{C}^{N\times n}$, where the integral is understood to be taken over $\mathds{C}^{K\times n}$. We also define, for a $\mathbb{P}_X$-measurable functional $f$, its mean $\EE[f(X)] = \int f(\Xm) \mathbb{P}_{X}(d\Xm)$ and variance $\mathbb{V}{\rm ar}[f(X)] = \EE[ |f(X)-\EE[f(X)]|^2]$.

Let $\mathbb{P}$ and $\mathbb{Q}$ be two measures on (the Borel $\sigma$-field of) $\mathds{C}^{K\times n}$. Then $\mathbb{P}$ is said to be {\it absolutely continuous} with respect to $\mathbb{Q}$ if $\mathbb{P}(\mathcal{A})=0$ for every Borel set $\mathcal{A}$ for which $\mathbb{Q}(\mathcal{A})=0$. This is written as $\mathbb{P} \ll\mathbb{Q}$. For such measures $\mathbb{P}$ and $\mathbb{Q}$, we denote $	\frac{d\mathbb{P}}{d\mathbb{Q}}(\Xm) = \frac{\mathbb{P}(d\Xm)}{\mathbb{Q}(d\Xm)}$ the Radon--Nykodym derivative \cite[Theorem~32.2]{BIL95} of $\mathbb{P}$ with respect to $\mathbb{Q}$ at position $\Xm$, i.e., for any Borel set $\mathcal A$, $\mathbb{P}(\mathcal A)=\int_{\mathcal A} \frac{d\mathbb{P}}{d\mathbb{Q}} d\mathbb{Q}=\int_{\mathcal A} \frac{\mathbb{P}(d\Xm)}{\mathbb{Q}(d\Xm)} \mathbb{Q}(d\Xm).$ The notation $\mathbb{P}(d\Xm) \leq \mathbb{Q}(d\Xm)$ will then be understood as $\frac{d\mathbb{P}}{d\mathbb{Q}}(\Xm)=\frac{\mathbb{P}(d\Xm)}{\mathbb{Q}(d\Xm)}\leq 1.$ If $\mathbb{P}$ is not absolutely continuous with respect to $\mathbb{Q}$, we set $d\mathbb{P}/d\mathbb{Q} \defines \infty$ and $\mathbb{P}(d\Xm) \leq \mathbb{Q}(d\Xm)$ is understood as an always false statement.

We denote $\mathcal{CN}(0,\sigma^2)$ the complex circularly symmetric normal distribution with zero mean and variance $\sigma^2$. We call $\Phi$ the distribution function of the real standard normal distribution, given by $\Phi(x) \defines \frac1{\sqrt{2\pi}}\int_{-\infty}^x \exp\left(-\frac{t^2}{2}\right)dt$. The weak convergence of the sequence of probability measures $\{\mu_n\}_{n=1}^\infty$ to $\mu$ is denoted by $\mu_n\Rightarrow \mu$; ``$\xrightarrow[]{\text{a.s.}}$'' stands for almost sure convergence.

The notation $f_n(t)=\Oc(t^\alpha n^{-\beta})$ means that there exists $C>0$ independent of $t$ and $n$ such that, for all $t>0$ and $n\in\mathds{N}$, $|f_n(t)|\leq Ct^\alpha n^{-\beta}$.

\section{Channel model and problem statement}
Consider the following MIMO memoryless Gaussian quasi-static fading channel:
\begin{align}
 \yv_t = \frac1{\sqrt{K}}\Hm^n\xv_t + \sigma\wv_t,\qquad t=\{1,\dots, n\}
 \label{eq-channel-definition}
\end{align}
where
$\yv_t\in \mathds{C}^N$ is the channel output at time $t$, ${\Hm^n\in \mathds{C}^{N\times K}}$ is a realization of the random channel matrix $H^n\in \mathds{C}^{N\times K}$ whose entries are independent and identically distributed (i.i.d.) $\Cc\Nc\LB0,1\RB$ and the index $n$ reminds that $\Hm^n$ is constant for the duration of $n$ channel uses, ${\xv_t \in  \mathds{C}^{K\times 1}}$ is the realization of the random channel input $x_t\in  \mathds{C}^{K\times 1}$ at time $t$, and $\sigma\wv_t$ is the realization of the random noise vector $\sigma w_t$ at time $t$ whose entries are i.i.d.\@ $\Cc\Nc\LB 0, \sigma^2\RB$. The transmitter end has only statistical knowledge about $H^n$ while the receiver end knows $H^n$ perfectly. In particular, we will assume $H^n$, $x_t$, and $w_t$ to be independent for each $t$.
We define the following matrices: $\Xm^n = ( \xv_1,\ldots,\xv_n)\in\mathds{C}^{K\times n}$, $\Wm^n = ( \wv_1,\ldots \wv_n) \in \mathds{C}^{N \times n}$, and $\Ym^n = ( \yv_1,\ldots, \yv_n) \in \mathds{C}^{N \times n}$. Associated to these matrices, we define the random matrices $X^n=(x_1,\ldots,x_n)\in\mathds{C}^{K\times n}$, $W^n=(w_1,\ldots,w_n)\in \mathds{C}^{N \times n}$, and $Y^n=(y_1,\ldots,y_n)\in \mathds{C}^{N \times n}$. 

We denote  the sets of admissible inputs $\Xm^n$ with unit maximal and exact energy constraint, respectively, by
\begin{align}\label{eq:power}
\mathcal{S}^n &\defines\left\{\mb{X}^n\in  \mathds{C}^{K\times n}\, \Big | \,   \frac1{nK} \trace\mb{X}^n(\mb{X}^n)\htp\le   1 \right \} \\
\label{eq:power_equal}
\mathcal{S}_=^n &\defines\left\{\mb{X}^n\in  \mathds{C}^{K\times n}\, \Big | \,   \frac1{nK} \trace\mb{X}^n(\mb{X}^n)\htp =  1 \right \}.
\end{align}

The {\it mutual information density} of $\mathbb{P}_{Y^n|X^n,H^n}$, i.e.,\@ the probability measure of $Y^n$ conditioned on $X^n$ and $H^n$, is defined by (see e.g. \cite{HAN03} for the AWGN definition):
\begin{equation}
I_{N,K}^{(n)} \defines\frac1{nK}\log\frac{\mathbb{P}_{Y^n|X^n,H^n}(dY^n|X^n,H^n)}{\mathbb{P}_{{Y}^n|H^n}(dY^n|H^n)}
\label{eq-information density-general}
\end{equation}
where the ratio ${\mathbb{P}_{Y^n|X^n,H^n}(\cdot|\Xm^n,\Hm^n)}/{\mathbb{P}_{{Y}^n|H^n}}(\cdot | \Hm^n)$, for given $\Xm^n,\Hm^n$, denotes the Radon--Nykodym derivative of the measure $\mathbb{P}_{Y^n|X^n,H^n}(\cdot |\Xm^n,\Hm^n)$ with respect to $\mathbb{P}_{Y^n|H^n}(\cdot | \Hm^n)$ whenever $\mathbb{P}_{Y^n|X^n,H^n}(\cdot |\Xm^n,\Hm^n)\ll \mathbb{P}_{Y^n | H^n} (\cdot | \Hm^n)$ and is set to $\infty$ otherwise. 

\begin{definition}[Code and average error probability]\label{def-code}
	A $({P}_e^{(n)},M_{n})$-code $\mathcal C_n$ for the channel model \eqref{eq-channel-definition} with power constraint \eqref{eq:power} consists of the following mappings:
\begin{itemize}
\item An encoder mapping: 
\begin{equation}
	\varphi:  \mathcal{M}_n  \longmapsto  \mathds{C}^{K \times n}.
\end{equation}
The transmitted symbols are $ \Xm_m^n = \varphi(m)\in\mathcal S^n$ for every message $m$ uniformly distributed over the set $\mathcal{M}_n=\{1,\dots,M_n\}$ of messages.
\item A set of decoder mappings $\{\phi_{\Hm^n}\}_{\Hm^n\in\mathds{C}^{N\times K}}$ with:
	\begin{equation}
	\phi_{\Hm^n}:  \mathds{C}^{N \times n} \longmapsto  \mathcal{M}_n \cup \{e\}
\end{equation}
which produces the decoder's decision $\hat{m}= \phi_{\Hm^n}(\Ym_m^n)$, $\Ym_m^n=\frac1{\sqrt{K}}\Hm^n\varphi(m)+\sigma\Wm^n$, on the transmitted message $m$, or the error event $e$. 
\end{itemize}
For a code $\mathcal{C}_{n}$ with block-length $n$, codebook size $M_n$, encoder $\varphi$, and decoder $\{\phi_{\Hm^n}\}_{\Hm^n\in\mathds{C}^{N\times K}}$, the {\it average error probability} is defined as
\begin{equation}
	P_e^{(n)}= {P}_{e}^{(n)}(\mathcal{C}_{n} ) \defines  \Pr\left[\hat{m}\neq m\right],
\end{equation}
where the probability is taken over the random variables $W^n$, $H^n$ and $m$.
\end{definition}

Let  $\textrm{supp}(\mathcal{C}_{n} )$ denote the {\it codebook} $\{\varphi(1),\dots,\varphi(M_n)\}$. The {\it optimal average error probability for the rate $R$} is defined as
\begin{align}
\mathbb{P}_{e}^{(n)}(R) \defines  \inf_{ \mathcal{C}_{n}: \textrm{supp}(\mathcal{C}_{n})\subseteq \mathcal{S}^n} \Big\{P_{e}^{(n)}(\mathcal{C}_{n} ) \, \Big| \,\frac{1}{nK}\log M_n\geq R \Big\}.
\label{errorprob_def2}   
\end{align}

The exact characterization of $\mathbb{P}_{e}^{(n)}(R)$ for fixed $n$, $K$, and $N$ is generally intractable. 
As mentioned in the introduction, a classical approach consists in considering rates within $\mathcal{O}(1/\sqrt{n})$ of the ergodic capacity with block-lengths $n$ growing to infinity (i.e., second-order coding rates). This leads to tractable limiting error probabilities, referred to as {\it optimal average error probabilities for the second-order coding rates} \cite{HAY09,POL10}. However, as the capacity of the quasi-static Rayleigh fading channel is zero, we assume here that the system dimensions $K$ and $N$ grow large. This induces ergodicity in the channel and entails a new definition of the second-order coding rate and the optimal average error probability for the quasi-static fading MIMO channel.
Precisely, we assume that $K$, $N$, and $n$ are large but of the same order of magnitude. This is expressed mathematically via the relations 
\begin{align}
	n\to \infty\ ,\qquad \frac{n}K &= \beta\ , \qquad 	\frac{N}K = c 
\end{align}
for some constants $\beta,c>0$.\footnote{This assumption can be relaxed to $\frac{n}K = \beta+ o(n^{-2})$ and $\frac{N}K = c  + o(n^{-2})$. However, it is easy to see that these constraints impose $c$ and $\beta$ to be rational numbers and the sequences $\{N/K\}_{n=1}^\infty$ and $\{n/K\}_{n=1}^\infty$ to be constant for all large $n$.} 
These relations will be denoted by $n \xrightarrow{(\beta,c)}\infty$ in the remainder of the article. For an infinite block-length, the per-antenna capacity of the channel converges for almost every channel realization to an asymptotic limit $C$ \cite{VER99}:
\begin{theorem}[{\cite[Eq. (9)]{VER99},\cite[Thm. 1]{hachem2008new}}]\label{thm:detequ_mutinf}
Let $\{ H^n\}_{n=1}^\infty$, where $H^n\in\mathds{C}^{N\times K}$ has i.i.d.\@ entries $H^n_{ij}\sim\Cc\Nc\LB 0,1\RB$. Let $\sigma^2>0$ and define
\begin{align}
\label{def-CNK}
C_{N,K} &\defines \frac{1}{K} \log\det\LB\Id_N+\frac{1}{\sigma^2K}H^n(H^n)\htp\RB.
\end{align}
Then, as $n\xrightarrow[]{(\beta,c)}\infty$,
\begin{itemize}
 \item[$(i)$]$C_{N,K}  \xrightarrow[]{\text{a.s.}}C \LB\sigma^2\RB$
 \item[$(ii)$]$\mathbb{E}\LSB C_{N,K}\RSB = C \LB\sigma^2\RB + \Oc\LB \frac1{n^2}\RB$
\end{itemize}
where, for $x>0$,
\begin{align}\nonumber
C\LB x\RB =& \log\LB 1+\delta_0\LB x\RB\RB + c\log\LB 1+\frac1{x\LB1+\delta_0\LB x\RB\RB}\RB \\
& - \frac{\delta_0\LB x\RB}{1+\delta_0\LB x\RB} 
\end{align}
and 
\begin{align}
		\label{def-delta0}
		\delta_0(x) \defines \frac{c-1}{2x} - \frac12 + \frac{\sqrt{(1-c+x)^2+4cx}}{2x}>0.
	\end{align}
\end{theorem}

Based on this observation, we can characterize the error probability in the second-order coding rate, i.e., when the coding rate is within $\Oc(1/\sqrt{nK})$ of the limiting capacity $C=C(\sigma^2)$, and estimate $\mathbb{P}_{e}^{(n)}(R)$ via the following limiting error probability:

\begin{definition}
The {\it optimal average error probability for the second-order coding rate $r$} is 
\begin{align}
&\mathbb{P}_{e}(r|\beta,c) \defines \inf_{\{\mathcal{C}_{n}: \textrm{supp}(\mathcal{C}_n )\subseteq \mathcal{S}^n\}_{n=1}^\infty} \left\{\limsup\limits_{n \xrightarrow{(\beta,c)} \infty}{P}_{e}^{(n)}(\mathcal{C}_{n} ) \, \Big| \right.  \nonumber\\   
&\quad\left. \liminf\limits_{n \xrightarrow{(\beta,c)}  \infty} \sqrt{nK} \,\left(\frac{1}{nK} \log M_n - C \right)\geq r \right\}.
\label{errorprob_def3}   
\end{align} 
\end{definition}

\begin{remark}[Fluctuation around ergodic capacity]
	For the channel model \eqref{eq-channel-definition}, the optimal average error probability may be alternatively written as
\begin{align}\label{eq:aveErrDef2}
	&\mathbb{P}_{e}(r|\beta,c) = \inf_{\substack{ \{\mathcal{C}_{n}\}_{n=1}^\infty \\ \textrm{supp}(\mathcal{C}_n )\subseteq \mathcal{S}^n} } \left\{\limsup\limits_{n \xrightarrow{(\beta,c)} \infty}P_e^{(n)}(\mathcal{C}_{n} ) \, \Big| \right.  \nonumber\\  
&\quad\left.\liminf\limits_{n \xrightarrow{(\beta,c)}  \infty} \sqrt{nK} \,\left(\frac{1}{nK} \log M_n - \EE[C_{N,K}] \right)\geq r \right\}
\end{align}
since 
\begin{equation}\label{eq:Gauss1storder}
	\sqrt{nK}\left( \EE[C_{N,K}] - C\right) \to 0
\end{equation}
as $n \xrightarrow{(\beta,c)}\infty$ by Theorem~\ref{thm:detequ_mutinf}~$(ii)$. In the finite $N,K,n$-regime, we may therefore see the optimal average error probability as an approximation of the optimal achievable error under the rate constraint
\begin{equation}
	\frac{1}{nK} \log M_n \geq \EE[C_{N,K}] + \frac{r}{\sqrt{nK}}.
\end{equation}
Note that the relation \eqref{eq:Gauss1storder} is fundamentally dependent on the Gaussianity of $H^n$. It was indeed shown in \cite[Theorem 4.4]{HAC08} that, whenever the entries of $H^n$ have a non-zero fourth order cumulant $\kappa = \mathbb{E}\LSB \left|H^n_{11}\right|^4\RSB-2$, a bias term $\mathcal{B}$ proportional to $\kappa$ arises such that \eqref{eq:Gauss1storder} must be modified to  $\sqrt{nK}\left( \EE[C_{N,K}] - C\right)\to \mathcal{B}$ as $n \xrightarrow{(\beta,c)}\infty$. In this case the equivalence of \eqref{eq:aveErrDef2} and \eqref{errorprob_def3} does not hold. For Gaussian channels (since $\kappa=0$ and then $\mathcal{B}=0$), however, the 
asymptotic mutual information is reached at the sufficiently fast rate of $\Oc(n^{-2})$ (as confirmed by Theorem~\ref{thm:detequ_mutinf}~$(ii)$).
\end{remark}

Instead of the optimal average error probability, we may consider the {\it second-order outage probability} $\mathbb{P}_{\rm out}(r|\beta,c)$ for the rate $r$, which we define as follows:
\begin{definition}
	\label{def:Pout}
	The {\it second-order outage probability for the second-order coding rate $r$} is
\begin{align}
&\mathbb{P}_{\rm out}(r|\beta,c) \defines \inf_{\{\mathcal{C}_{n}: \textrm{supp}(\mathcal{C}_n )\subseteq \mathcal{S}^n\}_{n=1}^\infty} \left\{\limsup\limits_{n \xrightarrow{(\beta,c)} \infty}P_e^{(n)}(\mathcal{C}_{n} ) \, \Big| \right.  \nonumber\\ 
&\quad\left. \liminf\limits_{n \xrightarrow{(\beta,c)}  \infty} K \,\left(\frac{1}{nK} \log M_n - C \right)\geq r \right\}.
\label{def-outage}   
\end{align}  
\end{definition}
The second-order outage probability and the optimal average error probability are related by $\mathbb{P}_{\rm out}(r|\beta,c)=\mathbb{P}_{e}(r\sqrt{\beta}|\beta,c)$. Definition~\ref{def:Pout} allows us to study the behavior of the second-order outage probability for growing $\beta$. In the finite dimensional setting, this corresponds to increasing the block-length while maintaining $N$ and $K$ (and thus the capacity $KC$) fixed. This cannot be performed on $\mathbb{P}_{e}(r|\beta,c)$ since, by growing $n$, $\sqrt{nK}C$ grows as well, therefore not maintaining the capacity fixed as $n$ grows alone.

The main objective of this article is to characterize $\mathbb{P}_{e}(r|\beta,c)$ (which will in turn characterize $\mathbb{P}_{\rm out}(r|\beta,c)$).

\section{Main result}\label{sec:main_results}
To determine the optimal average error probability, one ideally needs to determine the asymptotic fluctuations of the mutual information density $I_{N,K}^{(n)}$ for all codes $\mathcal C_n$. Since this is intractable, we shall resort to upper and lower bounds, which shall both rely on establishing the fluctuations of the random quantity $I_{N,K}^{X^n}$ as defined, for $\mathbb{P}_{X^n}\in\mathcal P(\mathcal S^n)$, in \eqref{def-INK_Xn} on the top of the next page.
\begin{figure*}[!t]
\begin{align}\label{def-INK_Xn}
	I_{N,K}^{X^n} \defines& \frac{1}{K} \log\det\LB\Id_N+\frac{1}{\sigma^2}\frac{H^n(H^n)\htp}{K}\RB\nonumber\\
	&+ \frac{1}{nK} \trace \LSB\LB \frac{ H^n(H^n)\htp}{K} + \sigma^2\Id_N\RB^{-1}\LB \frac{H^n}{\sqrt{K}}X^n+\sigma W^n\RB\LB \frac{H^n}{\sqrt{K}}X^n+\sigma W^n\RB\htp - W^n(W^n)\htp\RSB
\end{align}
\hrulefill
\end{figure*}

These fluctuations are provided in the following theorem.
\begin{theorem}\label{thm:CLT}
Let $\{ X^n\}_{n=1}^\infty$ be a sequence of random variables with probability $\mathbb{P}_{{X}^n}\in\mathcal P(\mathcal S_=^n)$ and, for ${A}^n=\Id_K-\frac1nX^n(X^n)\htp$, define $\theta_n>0$ the random variable given by
\begin{align}
	\label{eq:def-thetan}
	\theta_n^2 =&  -\beta \log\left( 1 - \frac1c \frac{\delta_0\LB\sigma^2\RB^2}{\LB1+\delta_0\LB\sigma^2\RB\RB^2} \right) + c + \sigma^4\delta_0'\LB\sigma^2\RB\nonumber\\
	&\qquad - \beta \frac{\delta_0'(\sigma^2)}{(1+\delta_0(\sigma^2))^{4}} \frac1K\trace\left[(A^n)^2\right] 
\end{align}
where the function $\delta_0(x)$ is defined in \eqref{def-delta0}.
Then, for any real $z$, as $n \xrightarrow{(\beta,c)} \infty$,
\begin{align}
	\Pr\left[\frac{\sqrt{nK}}{\theta_n}\left( I_{N,K}^{X^n} - C\right)\leq z\right] \to \Phi(z).
\end{align}
\end{theorem}
\begin{IEEEproof}
The proof is provided in Appendix~\ref{proof:CLT}.
\end{IEEEproof} \smallskip

Based on this result, we can determine the following lower and upper bounds on the optimal average error probability for the second-order coding rate.
\begin{theorem}\label{th-main}
	The optimal average error probability $\mathbb{P}_e(r|\beta,c)$ for the second-order coding rate~$r$ satisfies:
			\begin{itemize}
				\item If $r\leq 0$,
					\begin{align}
						\Phi\left( \frac{r}{\theta_-} \right) \leq \mathbb{P}_e(r|\beta,c) \leq \Phi\left( \frac{r}{\theta_+} \right)
					\end{align}
				\item If $r>0$,
					\begin{align}
						\frac12 \leq \mathbb{P}_e(r|\beta,c) \leq \Phi\left( \frac{r}{\theta_+} \right)
					\end{align}
			\end{itemize}
	where $\theta_- >0$ and $\theta_+ >0$ are defined by
	\begin{align}
		\label{def-theta_-}
		\theta_-^2 &\defines -\beta \log\left( 1 - \frac1c \frac{\delta_0\LB\sigma^2\RB^2}{\LB1+\delta_0\LB\sigma^2\RB\RB^2} \right)\nonumber\\
		&\qquad + c + \sigma^4\delta_0'\LB\sigma^2\RB  \\ 
		\label{def-theta_+}
		\theta_+^2 &\defines  -\beta \log\left( 1 - \frac1c \frac{\delta_0\LB\sigma^2\RB^2}{\LB1+\delta_0\LB\sigma^2\RB\RB^2} \right) \nonumber\\
		&\qquad + c + \sigma^4\delta_0'\LB\sigma^2\RB - \frac{\delta_0'\LB\sigma^2\RB}{(1+\delta_0(\sigma^2))^4} 
	\end{align}
	and $\delta_0(x)$ is defined in \eqref{def-delta0} with derivative, for $x>0$,
	\begin{align}
		\delta_0'(x) = - \frac{\delta_0(x)\left( 1+ \delta_0(x)\right)}{1-c+x + 2 x\delta_0(x)}<0.
	\end{align}	
\end{theorem}
\begin{IEEEproof}
The details of this proof are provided in Appendix~\ref{app:proof_th-main}.
\end{IEEEproof}

Theorem~\ref{th-main} shows that, for sufficiently large channel dimensions and block-length, the optimal error probability for a coding rate close to the asymptotic capacity, i.e., $(nK)^{-1} \log M_n = C+ (nK)^{-1/2} \,r$, is comprised between two explicit bounds which depend only on $c$, $\beta$, and $\sigma^2$. This is to be compared with the AWGN scenario of \cite{HAY09,POL10} where the corresponding bounds were found to depend only on $\sigma^2$. However, as opposed to Theorem~\ref{th-main}, the lower and upper bounds in these works were shown to be equal. We discuss in Remark~\ref{rem:tightness} below the technical reasons for this important difference. Note that, for rates above the capacity limit (i.e., for $r>0$), the lower bound is very loose and can be  far from its associated upper bound. In contrast, the more interesting case $r<0$ (corresponding to coding rates below the asymptotic capacity) features two bounds which are numerically shown to be quite close to one-another. 

\begin{remark}[On the quantity $\delta_0\LB\sigma^2\RB$]\label{rem:MPrelation}
	The function $c^{-1}\delta_0(\sigma^2)$ coincides with the Stieltjes transform $m_{\mu_c}(z)$ of the Mar\u{c}enko--Pastur measure $\mu_c$ with parameter $c$ \cite{MAR67} evaluated at position $z=-\sigma^2$, which is defined by $m_{\mu_c}(z)=\int (t-z)^{-1}\mu_c(dt)$ for all $z\in\mathds{C}\setminus {\rm supp}\LB \mu_c\RB$. This measure is the limiting distribution of the eigenvalues of $K^{-1}H^n(H^n)\htp$ as $N,K\to\infty$ and $N/K\to c$. For this reason, the quantities $C$, $\theta_-$, and $\theta_+$ of Theorem~\ref{th-main} naturally appear as functionals of $\mu_c$. 
\end{remark}

\begin{remark}[Tightness of the bounds]\label{rem:tightness}
The case $r=0$ set aside, the lower and upper bounds on the optimal average error probability are never equal. This unfolds from the presence of the random channel $H^n$ which induces a dependence of the second order statistics of $I_{N,K}^{(n)}$ on the ``fourth order moment'' $\EE[K^{-1}\trace (n^{-1}X^n(X^n)\htp)^2]$ of $\mathbb{P}_{X^n}$. The weak lower bound $1/2$ for $r>0$ is a consequence of the impossibility in the proof to bound the fourth order moment of $\mathbb{P}_{X^n}$ from above under the sole constraint \eqref{eq:power}; see Appendix~\ref{app:proof_th-main}. By contrast, in \cite{HAY09,POL10}, only (scalar) second order moments of $\mathbb{P}_{X^n}$ play a role in the second order statistics of $I_{N,K}^{(n)}$. These are easily controlled by \eqref{eq:power}. 
\end{remark}

\begin{remark}[High SNR-regime]
	In the high-SNR regime, we have the following result:
	\begin{align}
		\lim_{\sigma^2\to 0} \theta_-^2 &= \left\{ 
			\begin{array}{ll}
				- \beta\log \left( 1 - c \right) + c &,~c<1 \\ 
				\infty &,~c=1 \\
				- \beta\log \left( 1 - \frac1c \right) + 1 &,~c>1\,,
			\end{array}
			\right. \\
			\lim_{\sigma^2\to 0} \theta_+^2 &= \left\{ 
				\begin{array}{ll}
					- \beta\log \left( 1 - c \right) + c(2-c) &,~c<1 \\ 
					\infty &,~c=1 \\
					- \beta\log \left( 1 - \frac1c \right) + 1 &,~c>1\,.
				\end{array}
				\right. 
			\end{align}
			This follows from the definition of $\delta_0(x)$ and $\delta_0'(x)$ in Theorem~\ref{thm:detequ_mutinf} which brings, for $c<1$, $\delta_0(x)\to c(1-c)^{-1}$ and $\delta_0'(\sigma^2)\to -c(1-c)^3$ as $x\downarrow 0$, while, for $c>1$, $x\delta_0(x)\to c-1$ and $x^2\delta_0'(x)\to 1-c$ as $x\downarrow 0$.
\end{remark}

\begin{remark}[Low SNR-regime] Both $\theta^2_+$ and $\theta^2_-$ converge to $0$ as $\sigma^2\to\infty$. Thus, for $r<0$, the upper and lower bounds on $\mathbb{P}_e(r|\beta,c)$ are equal to zero and, for $r>0$, the upper bound tends to one. However, also the asymptotic capacity $C$ is zero. First order approximations of $C$ and $\theta_-^2,\theta_+^2$ for $\sigma^2\to \infty$ are thus meaningful and are given by
	\begin{align}
		C &= \frac{c}{\sigma^2} + \Oc(\sigma^{-4}) \\
		\theta_+^2 &=  \frac{2c}{\sigma^2} + \Oc(\sigma^{-4})\\
		\theta_-^2 &= \frac{2c}{\sigma^2} + \Oc(\sigma^{-4}).
	\end{align}
	This shows in particular that $(\theta_+^2-\theta_-^2)/\theta_+^2= \Oc(\sigma^{-2})$, implying the asymptotic closeness of the upper and lower bounds in the low SNR regime. Note additionally that, for $c=1$, the approximate standard deviation $\frac{2c}{\sigma^2}$ coincides with the low-SNR channel dispersion reported in \cite{POL10} for SISO AWGN channels.
\end{remark}

\begin{figure*}
\centering
\begin{tikzpicture}[scale=0.9]
	\pgfplotsset{every axis/.append style={mark options=solid}}
    \renewcommand{\axisdefaulttryminticks}{4}
    \pgfplotsset{every major grid/.append style={densely dashed}}
    \tikzstyle{every axis y label}+=[yshift=-10pt]
    \tikzstyle{every axis x label}+=[yshift=5pt]
    \pgfplotsset{every axis legend/.append style={cells={anchor=west},fill=white, at={(0.02,0.98)}, anchor=north west}}
    \pgfplotsset{
    tick label style={font=\normalsize},
    label style={font=\normalsize},
    legend style={font=\normalsize}
    }
    \tikzset{
	    every pin/.style={fill=yellow!50!white,rectangle,rounded corners=3pt,font=\small},
	    small dot/.style={fill=black,circle,scale=0.3}
    }
    \begin{axis}[
width=\textwidth,
      xmin=-4,
      ymin=0,
      xmax=4,
      ymax=1,    
      bar width=3pt,
      grid=major,    
      scaled ticks=true,
      ylabel={Bounds on $\mathbb{P}_{e}(r|\beta,c)$},
      xlabel={$r'=r/C(\sigma^2)$},
      ]
      \addplot[blue,smooth] plot coordinates{	     
	      (-3.981072,0.450082)(-3.162278,0.460310)(-2.511886,0.468454)(-1.995262,0.474932)(-1.584893,0.480083)(-1.258925,0.484177)(-1.000000,0.487430)(-0.794328,0.490015)(-0.630957,0.492068)(-0.501187,0.493699)(-0.398107,0.494995)(-0.316228,0.496024)(-0.251189,0.496842)(-0.199526,0.497492)(-0.158489,0.498008)(-0.125893,0.498417)(-0.100000,0.498743)(-0.079433,0.499001)(-0.063096,0.499207)(-0.050119,0.499370)(-0.039811,0.499500)(-0.031623,0.499602)(-0.025119,0.499684)(-0.019953,0.499749)(-0.015849,0.499801)(-0.012589,0.499842)(-0.010000,0.499874)(-0.007943,0.499900)(-0.006310,0.499921)(-0.005012,0.499937)(-0.003981,0.499950)(-0.003162,0.499960)(-0.002512,0.499968)(-0.001995,0.499975)(-0.001585,0.499980)(-0.001259,0.499984)(-0.001000,0.499987)(0.001000,0.500013)(0.001259,0.500016)(0.001585,0.500020)(0.001995,0.500025)(0.002512,0.500032)(0.003162,0.500040)(0.003981,0.500050)(0.005012,0.500063)(0.006310,0.500079)(0.007943,0.500100)(0.010000,0.500126)(0.012589,0.500158)(0.015849,0.500199)(0.019953,0.500251)(0.025119,0.500316)(0.031623,0.500398)(0.039811,0.500500)(0.050119,0.500630)(0.063096,0.500793)(0.079433,0.500999)(0.100000,0.501257)(0.125893,0.501583)(0.158489,0.501992)(0.199526,0.502508)(0.251189,0.503158)(0.316228,0.503976)(0.398107,0.505005)(0.501187,0.506301)(0.630957,0.507932)(0.794328,0.509985)(1.000000,0.512570)(1.258925,0.515823)(1.584893,0.519917)(1.995262,0.525068)(2.511886,0.531546)(3.162278,0.539690)(3.981072,0.549918)
};
      \addplot[red,densely dashed,smooth] plot coordinates{
	      (-3.981072,0.450069)(-3.162278,0.460300)(-2.511886,0.468446)(-1.995262,0.474926)(-1.584893,0.480078)(-1.258925,0.484173)(-1.000000,0.487427)(-0.794328,0.490012)(-0.630957,0.492066)(-0.501187,0.493698)(-0.398107,0.494994)(-0.316228,0.496024)(-0.251189,0.496841)(-0.199526,0.497491)(-0.158489,0.498007)(-0.125893,0.498417)(-0.100000,0.498743)(-0.079433,0.499001)(-0.063096,0.499207)(-0.050119,0.499370)(-0.039811,0.499499)(-0.031623,0.499602)(-0.025119,0.499684)(-0.019953,0.499749)(-0.015849,0.499801)(-0.012589,0.499842)(-0.010000,0.499874)(-0.007943,0.499900)(-0.006310,0.499921)(-0.005012,0.499937)(-0.003981,0.499950)(-0.003162,0.499960)(-0.002512,0.499968)(-0.001995,0.499975)(-0.001585,0.499980)(-0.001259,0.499984)(-0.001000,0.499987)(0.001000,0.500000)(0.001259,0.500000)(0.001585,0.500000)(0.001995,0.500000)(0.002512,0.500000)(0.003162,0.500000)(0.003981,0.500000)(0.005012,0.500000)(0.006310,0.500000)(0.007943,0.500000)(0.010000,0.500000)(0.012589,0.500000)(0.015849,0.500000)(0.019953,0.500000)(0.025119,0.500000)(0.031623,0.500000)(0.039811,0.500000)(0.050119,0.500000)(0.063096,0.500000)(0.079433,0.500000)(0.100000,0.500000)(0.125893,0.500000)(0.158489,0.500000)(0.199526,0.500000)(0.251189,0.500000)(0.316228,0.500000)(0.398107,0.500000)(0.501187,0.500000)(0.630957,0.500000)(0.794328,0.500000)(1.000000,0.500000)(1.258925,0.500000)(1.584893,0.500000)(1.995262,0.500000)(2.511886,0.500000)(3.162278,0.500000)(3.981072,0.500000)
};
      \addplot[blue,smooth] plot coordinates{	     
	      (-3.981072,0.156988)(-3.162278,0.211908)(-2.511886,0.262610)(-1.995262,0.306902)(-1.584893,0.344262)(-1.258925,0.375086)(-1.000000,0.400163)(-0.794328,0.420386)(-0.630957,0.436604)(-0.501187,0.449564)(-0.398107,0.459898)(-0.316228,0.468126)(-0.251189,0.474671)(-0.199526,0.479876)(-0.158489,0.484012)(-0.125893,0.487299)(-0.100000,0.489911)(-0.079433,0.491986)(-0.063096,0.493634)(-0.050119,0.494943)(-0.039811,0.495983)(-0.031623,0.496809)(-0.025119,0.497465)(-0.019953,0.497987)(-0.015849,0.498401)(-0.012589,0.498730)(-0.010000,0.498991)(-0.007943,0.499199)(-0.006310,0.499363)(-0.005012,0.499494)(-0.003981,0.499598)(-0.003162,0.499681)(-0.002512,0.499747)(-0.001995,0.499799)(-0.001585,0.499840)(-0.001259,0.499873)(-0.001000,0.499899)(0.001000,0.500101)(0.001259,0.500127)(0.001585,0.500160)(0.001995,0.500201)(0.002512,0.500253)(0.003162,0.500319)(0.003981,0.500402)(0.005012,0.500506)(0.006310,0.500637)(0.007943,0.500801)(0.010000,0.501009)(0.012589,0.501270)(0.015849,0.501599)(0.019953,0.502013)(0.025119,0.502535)(0.031623,0.503191)(0.039811,0.504017)(0.050119,0.505057)(0.063096,0.506366)(0.079433,0.508014)(0.100000,0.510089)(0.125893,0.512701)(0.158489,0.515988)(0.199526,0.520124)(0.251189,0.525329)(0.316228,0.531874)(0.398107,0.540102)(0.501187,0.550436)(0.630957,0.563396)(0.794328,0.579614)(1.000000,0.599837)(1.258925,0.624914)(1.584893,0.655738)(1.995262,0.693098)(2.511886,0.737390)(3.162278,0.788092)(3.981072,0.843012)
};
      \addplot[red,densely dashed,smooth] plot coordinates{
	      (-3.981072,0.154774)(-3.162278,0.209784)(-2.511886,0.260709)(-1.995262,0.305274)(-1.584893,0.342906)(-1.258925,0.373976)(-1.000000,0.399265)(-0.794328,0.419664)(-0.630957,0.436026)(-0.501187,0.449103)(-0.398107,0.459530)(-0.316228,0.467833)(-0.251189,0.474439)(-0.199526,0.479691)(-0.158489,0.483865)(-0.125893,0.487183)(-0.100000,0.489818)(-0.079433,0.491912)(-0.063096,0.493575)(-0.050119,0.494897)(-0.039811,0.495946)(-0.031623,0.496780)(-0.025119,0.497442)(-0.019953,0.497968)(-0.015849,0.498386)(-0.012589,0.498718)(-0.010000,0.498982)(-0.007943,0.499191)(-0.006310,0.499357)(-0.005012,0.499490)(-0.003981,0.499595)(-0.003162,0.499678)(-0.002512,0.499744)(-0.001995,0.499797)(-0.001585,0.499839)(-0.001259,0.499872)(-0.001000,0.499898)(0.001000,0.500000)(0.001259,0.500000)(0.001585,0.500000)(0.001995,0.500000)(0.002512,0.500000)(0.003162,0.500000)(0.003981,0.500000)(0.005012,0.500000)(0.006310,0.500000)(0.007943,0.500000)(0.010000,0.500000)(0.012589,0.500000)(0.015849,0.500000)(0.019953,0.500000)(0.025119,0.500000)(0.031623,0.500000)(0.039811,0.500000)(0.050119,0.500000)(0.063096,0.500000)(0.079433,0.500000)(0.100000,0.500000)(0.125893,0.500000)(0.158489,0.500000)(0.199526,0.500000)(0.251189,0.500000)(0.316228,0.500000)(0.398107,0.500000)(0.501187,0.500000)(0.630957,0.500000)(0.794328,0.500000)(1.000000,0.500000)(1.258925,0.500000)(1.584893,0.500000)(1.995262,0.500000)(2.511886,0.500000)(3.162278,0.500000)(3.981072,0.500000)
};
      \addplot[blue,smooth] plot coordinates{
	      (-3.981072,0.022285)(-3.162278,0.055294)(-2.511886,0.102507)(-1.995262,0.157033)(-1.584893,0.211951)(-1.258925,0.262649)(-1.000000,0.306935)(-0.794328,0.344289)(-0.630957,0.375108)(-0.501187,0.400181)(-0.398107,0.420401)(-0.316228,0.436616)(-0.251189,0.449573)(-0.199526,0.459905)(-0.158489,0.468132)(-0.125893,0.474676)(-0.100000,0.479880)(-0.079433,0.484015)(-0.063096,0.487302)(-0.050119,0.489913)(-0.039811,0.491987)(-0.031623,0.493635)(-0.025119,0.494944)(-0.019953,0.495984)(-0.015849,0.496810)(-0.012589,0.497466)(-0.010000,0.497987)(-0.007943,0.498401)(-0.006310,0.498730)(-0.005012,0.498991)(-0.003981,0.499199)(-0.003162,0.499363)(-0.002512,0.499494)(-0.001995,0.499598)(-0.001585,0.499681)(-0.001259,0.499747)(-0.001000,0.499799)(0.001000,0.500201)(0.001259,0.500253)(0.001585,0.500319)(0.001995,0.500402)(0.002512,0.500506)(0.003162,0.500637)(0.003981,0.500801)(0.005012,0.501009)(0.006310,0.501270)(0.007943,0.501599)(0.010000,0.502013)(0.012589,0.502534)(0.015849,0.503190)(0.019953,0.504016)(0.025119,0.505056)(0.031623,0.506365)(0.039811,0.508013)(0.050119,0.510087)(0.063096,0.512698)(0.079433,0.515985)(0.100000,0.520120)(0.125893,0.525324)(0.158489,0.531868)(0.199526,0.540095)(0.251189,0.550427)(0.316228,0.563384)(0.398107,0.579599)(0.501187,0.599819)(0.630957,0.624892)(0.794328,0.655711)(1.000000,0.693065)(1.258925,0.737351)(1.584893,0.788049)(1.995262,0.842967)(2.511886,0.897493)(3.162278,0.944706)(3.981072,0.977715)
};
      \addplot[red,densely dashed,smooth] plot coordinates{
	      (-3.981072,0.021794)(-3.162278,0.054469)(-2.511886,0.101456)(-1.995262,0.155909)(-1.584893,0.210873)(-1.258925,0.261685)(-1.000000,0.306109)(-0.794328,0.343602)(-0.630957,0.374546)(-0.501187,0.399726)(-0.398107,0.420035)(-0.316228,0.436323)(-0.251189,0.449340)(-0.199526,0.459719)(-0.158489,0.467984)(-0.125893,0.474558)(-0.100000,0.479786)(-0.079433,0.483941)(-0.063096,0.487242)(-0.050119,0.489866)(-0.039811,0.491950)(-0.031623,0.493605)(-0.025119,0.494920)(-0.019953,0.495965)(-0.015849,0.496795)(-0.012589,0.497454)(-0.010000,0.497978)(-0.007943,0.498394)(-0.006310,0.498724)(-0.005012,0.498986)(-0.003981,0.499195)(-0.003162,0.499361)(-0.002512,0.499492)(-0.001995,0.499597)(-0.001585,0.499679)(-0.001259,0.499745)(-0.001000,0.499798)(0.001000,0.500000)(0.001259,0.500000)(0.001585,0.500000)(0.001995,0.500000)(0.002512,0.500000)(0.003162,0.500000)(0.003981,0.500000)(0.005012,0.500000)(0.006310,0.500000)(0.007943,0.500000)(0.010000,0.500000)(0.012589,0.500000)(0.015849,0.500000)(0.019953,0.500000)(0.025119,0.500000)(0.031623,0.500000)(0.039811,0.500000)(0.050119,0.500000)(0.063096,0.500000)(0.079433,0.500000)(0.100000,0.500000)(0.125893,0.500000)(0.158489,0.500000)(0.199526,0.500000)(0.251189,0.500000)(0.316228,0.500000)(0.398107,0.500000)(0.501187,0.500000)(0.630957,0.500000)(0.794328,0.500000)(1.000000,0.500000)(1.258925,0.500000)(1.584893,0.500000)(1.995262,0.500000)(2.511886,0.500000)(3.162278,0.500000)(3.981072,0.500000)
};
      \addplot[blue,smooth] plot coordinates{
	     (-3.981072,0.000188)(-3.162278,0.002361)(-2.511886,0.012408)(-1.995262,0.037319)(-1.584893,0.078382)(-1.258925,0.130337)(-1.000000,0.185806)(-0.794328,0.238945)(-0.630957,0.286467)(-0.501187,0.327152)(-0.398107,0.361035)(-0.316228,0.388766)(-0.251189,0.411213)(-0.199526,0.429256)(-0.158489,0.443697)(-0.125893,0.455221)(-0.100000,0.464404)(-0.079433,0.471711)(-0.063096,0.477522)(-0.050119,0.482142)(-0.039811,0.485813)(-0.031623,0.488730)(-0.025119,0.491047)(-0.019953,0.492888)(-0.015849,0.494351)(-0.012589,0.495513)(-0.010000,0.496436)(-0.007943,0.497169)(-0.006310,0.497751)(-0.005012,0.498214)(-0.003981,0.498581)(-0.003162,0.498873)(-0.002512,0.499105)(-0.001995,0.499289)(-0.001585,0.499435)(-0.001259,0.499551)(-0.001000,0.499644)(0.001000,0.500356)(0.001259,0.500449)(0.001585,0.500565)(0.001995,0.500711)(0.002512,0.500895)(0.003162,0.501127)(0.003981,0.501419)(0.005012,0.501786)(0.006310,0.502249)(0.007943,0.502831)(0.010000,0.503564)(0.012589,0.504487)(0.015849,0.505649)(0.019953,0.507112)(0.025119,0.508953)(0.031623,0.511270)(0.039811,0.514187)(0.050119,0.517858)(0.063096,0.522478)(0.079433,0.528289)(0.100000,0.535596)(0.125893,0.544779)(0.158489,0.556303)(0.199526,0.570744)(0.251189,0.588787)(0.316228,0.611234)(0.398107,0.638965)(0.501187,0.672848)(0.630957,0.713533)(0.794328,0.761055)(1.000000,0.814194)(1.258925,0.869663)(1.584893,0.921618)(1.995262,0.962681)(2.511886,0.987592)(3.162278,0.997639)(3.981072,0.999812)
};
      \addplot[red,densely dashed,smooth] plot coordinates{
	      (-3.981072,0.000187)(-3.162278,0.002356)(-2.511886,0.012388)(-1.995262,0.037280)(-1.584893,0.078327)(-1.258925,0.130274)(-1.000000,0.185742)(-0.794328,0.238887)(-0.630957,0.286416)(-0.501187,0.327109)(-0.398107,0.361000)(-0.316228,0.388737)(-0.251189,0.411189)(-0.199526,0.429237)(-0.158489,0.443682)(-0.125893,0.455210)(-0.100000,0.464394)(-0.079433,0.471703)(-0.063096,0.477516)(-0.050119,0.482137)(-0.039811,0.485809)(-0.031623,0.488727)(-0.025119,0.491045)(-0.019953,0.492887)(-0.015849,0.494350)(-0.012589,0.495512)(-0.010000,0.496435)(-0.007943,0.497168)(-0.006310,0.497750)(-0.005012,0.498213)(-0.003981,0.498581)(-0.003162,0.498873)(-0.002512,0.499104)(-0.001995,0.499289)(-0.001585,0.499435)(-0.001259,0.499551)(-0.001000,0.499643)(0.001000,0.500000)(0.001259,0.500000)(0.001585,0.500000)(0.001995,0.500000)(0.002512,0.500000)(0.003162,0.500000)(0.003981,0.500000)(0.005012,0.500000)(0.006310,0.500000)(0.007943,0.500000)(0.010000,0.500000)(0.012589,0.500000)(0.015849,0.500000)(0.019953,0.500000)(0.025119,0.500000)(0.031623,0.500000)(0.039811,0.500000)(0.050119,0.500000)(0.063096,0.500000)(0.079433,0.500000)(0.100000,0.500000)(0.125893,0.500000)(0.158489,0.500000)(0.199526,0.500000)(0.251189,0.500000)(0.316228,0.500000)(0.398107,0.500000)(0.501187,0.500000)(0.630957,0.500000)(0.794328,0.500000)(1.000000,0.500000)(1.258925,0.500000)(1.584893,0.500000)(1.995262,0.500000)(2.511886,0.500000)(3.162278,0.500000)(3.981072,0.500000)
};
      \addplot[blue,smooth] plot coordinates{
	      (-3.981072,0.000000)(-3.162278,0.000000)(-2.511886,0.000000)(-1.995262,0.000000)(-1.584893,0.000000)(-1.258925,0.000000)(-1.000000,0.000000)(-0.794328,0.000000)(-0.630957,0.000011)(-0.501187,0.000370)(-0.398107,0.003675)(-0.316228,0.016618)(-0.251189,0.045389)(-0.199526,0.089563)(-0.158489,0.142954)(-0.125893,0.198314)(-0.100000,0.250372)(-0.079433,0.296381)(-0.063096,0.335478)(-0.050119,0.367885)(-0.039811,0.394329)(-0.031623,0.415694)(-0.025119,0.432847)(-0.019953,0.446565)(-0.015849,0.457508)(-0.012589,0.466224)(-0.010000,0.473159)(-0.007943,0.478673)(-0.006310,0.483057)(-0.005012,0.486540)(-0.003981,0.489307)(-0.003162,0.491506)(-0.002512,0.493253)(-0.001995,0.494641)(-0.001585,0.495743)(-0.001259,0.496618)(-0.001000,0.497314)(0.001000,0.502686)(0.001259,0.503382)(0.001585,0.504257)(0.001995,0.505359)(0.002512,0.506747)(0.003162,0.508494)(0.003981,0.510693)(0.005012,0.513460)(0.006310,0.516943)(0.007943,0.521327)(0.010000,0.526841)(0.012589,0.533776)(0.015849,0.542492)(0.019953,0.553435)(0.025119,0.567153)(0.031623,0.584306)(0.039811,0.605671)(0.050119,0.632115)(0.063096,0.664522)(0.079433,0.703619)(0.100000,0.749628)(0.125893,0.801686)(0.158489,0.857046)(0.199526,0.910437)(0.251189,0.954611)(0.316228,0.983382)(0.398107,0.996325)(0.501187,0.999630)(0.630957,0.999989)(0.794328,1.000000)(1.000000,1.000000)(1.258925,1.000000)(1.584893,1.000000)(1.995262,1.000000)(2.511886,1.000000)(3.162278,1.000000)(3.981072,1.000000)
};
      \addplot[red,densely dashed,smooth] plot coordinates{
(-3.981072,0.000000)(-3.162278,0.000000)(-2.511886,0.000000)(-1.995262,0.000000)(-1.584893,0.000000)(-1.258925,0.000000)(-1.000000,0.000000)(-0.794328,0.000000)(-0.630957,0.000011)(-0.501187,0.000370)(-0.398107,0.003675)(-0.316228,0.016618)(-0.251189,0.045389)(-0.199526,0.089563)(-0.158489,0.142954)(-0.125893,0.198314)(-0.100000,0.250372)(-0.079433,0.296381)(-0.063096,0.335478)(-0.050119,0.367885)(-0.039811,0.394329)(-0.031623,0.415694)(-0.025119,0.432847)(-0.019953,0.446565)(-0.015849,0.457508)(-0.012589,0.466224)(-0.010000,0.473159)(-0.007943,0.478673)(-0.006310,0.483057)(-0.005012,0.486540)(-0.003981,0.489307)(-0.003162,0.491506)(-0.002512,0.493253)(-0.001995,0.494641)(-0.001585,0.495743)(-0.001259,0.496618)(-0.001000,0.497314)(0.001000,0.500000)(0.001259,0.500000)(0.001585,0.500000)(0.001995,0.500000)(0.002512,0.500000)(0.003162,0.500000)(0.003981,0.500000)(0.005012,0.500000)(0.006310,0.500000)(0.007943,0.500000)(0.010000,0.500000)(0.012589,0.500000)(0.015849,0.500000)(0.019953,0.500000)(0.025119,0.500000)(0.031623,0.500000)(0.039811,0.500000)(0.050119,0.500000)(0.063096,0.500000)(0.079433,0.500000)(0.100000,0.500000)(0.125893,0.500000)(0.158489,0.500000)(0.199526,0.500000)(0.251189,0.500000)(0.316228,0.500000)(0.398107,0.500000)(0.501187,0.500000)(0.630957,0.500000)(0.794328,0.500000)(1.000000,0.500000)(1.258925,0.500000)(1.584893,0.500000)(1.995262,0.500000)(2.511886,0.500000)(3.162278,0.500000)(3.981072,0.500000)
};

\legend{ {$\Phi(r/\theta_+)$} , {$\Phi(r/\theta_-)$} }
\node[small dot,pin=90:{ ${\rm SNR}=-30\,$dB}] at (axis cs:-3.2,0.46) {};
\node[small dot,pin=90:{ ${\rm SNR}=-10\,$dB}] at (axis cs:-3.2,0.21) {};
\node[small dot,pin=0:{ ${\rm SNR}=0\,$dB}] at (axis cs:-0.6,0.38) {};
\node[small dot,pin=0:{ ${\rm SNR}=10\,$dB}] at (axis cs:-0.76,0.25) {};
\node[small dot,pin=0:{ ${\rm SNR}=100\,$dB}] at (axis cs:-0.2,0.1) {};
    \end{axis}
  \end{tikzpicture}
  \caption{Bounds on the optimal average error probability as a function of the second-order coding rate $r=r'C(\sigma^2)$ for different SNRs and the parameters $c=2$ and $\beta=16$.}
\label{fig:Pe}
\end{figure*}
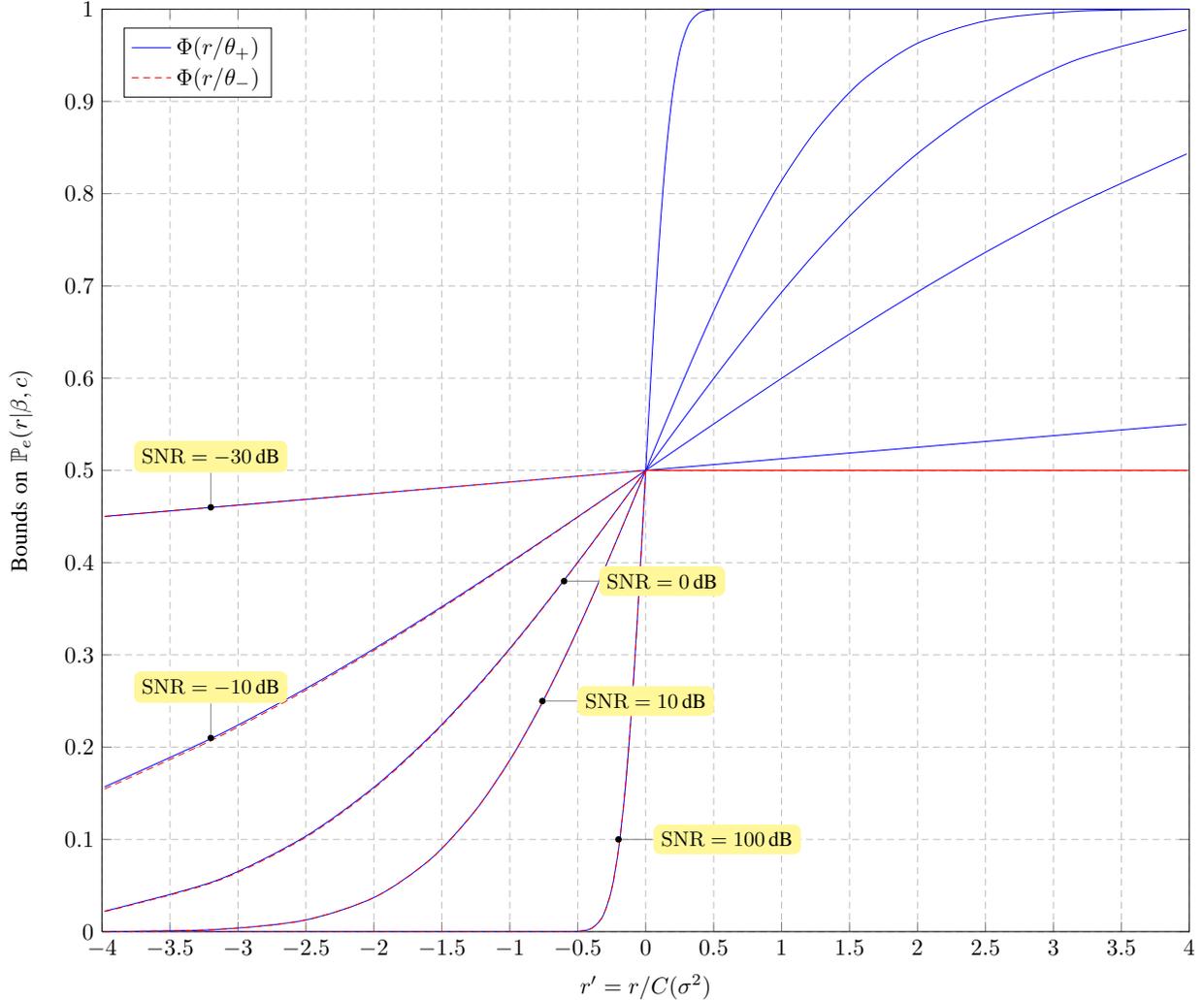 

Figure~\ref{fig:Pe} depicts the bounds on the optimal average error probability for varying second-order coding rates $r$ and for different SNR values (defined as ${\rm SNR}=\sigma^{-2}$). We choose $c=2$ and $\beta=16$. For fair comparison between the various SNR regimes, $r$ is taken to be proportional to $C(\sigma^2)$. For finite but large $N,K,n$ values, Figure~\ref{fig:Pe} therefore provides approximate error probability bounds when coding at rate $R=C(\sigma^2)(1+r'/\sqrt{nK})$ for various values of $r'$. We observe that, for negative second-order coding rates, the gap between the upper- and lower-bound is barely visible. 

\begin{figure*}
\centering
\begin{tikzpicture}[scale=0.9]
	\pgfplotsset{every axis/.append style={mark options=solid}}
    \renewcommand{\axisdefaulttryminticks}{4}
    \pgfplotsset{every major grid/.append style={densely dashed}}
    \pgfplotsset{every axis legend/.append style={cells={anchor=west},fill=white, at={(0.02,0.02)}, anchor=south west}}
    \pgfplotsset{
    tick label style={font=\normalsize},
    label style={font=\normalsize},
    legend style={font=\normalsize}
    }
    \tikzset{
	    every pin/.style={fill=yellow!50!white,rectangle,rounded corners=3pt,font=\small},
	    small dot/.style={fill=black,circle,scale=0.3}
    }
    \begin{semilogyaxis}[
width=\textwidth,
      xmin=-4,
      ymin=1e-4,
      xmax=5,
      ymax=1,   
      grid=major,    
      scaled ticks=true,
      ylabel={Error probability and bounds},
      xlabel={SNR [dB]},
      ]
      \addplot[blue,mark=*] plot coordinates{
	      (-0.25,0.998812)(0.00,0.995)(0.25,0.983062)(0.50,0.955562)(0.75,0.903438)(1.00,0.821562)(1.25,0.707)(1.50,0.5735)(1.75,0.42425)(2.00,0.295938)(2.25,0.185396)(2.50,0.10725)(2.75,0.0562083)(3.00,0.0275771)(3.25,0.0124577)(3.50,0.00522886)(3.75,0.00189055)(4.00,0.000676617)(4.25,0.000238806)
};
      \addplot[blue,smooth] plot coordinates{	     
	      (-4.0,9.957736e-01)(-3.9,9.930561e-01)(-3.8,9.889189e-01)(-3.7,9.828142e-01)(-3.6,9.740805e-01)(-3.5,9.619610e-01)(-3.4,9.456426e-01)(-3.3,9.243157e-01)(-3.2,8.972522e-01)(-3.1,8.638951e-01)(-3.0,8.239485e-01)(-2.9,7.774551e-01)(-2.8,7.248468e-01)(-2.7,6.669581e-01)(-2.6,6.049952e-01)(-2.5,5.404614e-01)(-2.4,4.750473e-01)(-2.3,4.104970e-01)(-2.2,3.484713e-01)(-2.1,2.904221e-01)(-2.0,2.374958e-01)(-1.9,1.904746e-01)(-1.8,1.497594e-01)(-1.7,1.153917e-01)(-1.6,8.710639e-02)(-1.5,6.440397e-02)(-1.4,4.663077e-02)(-1.3,3.305641e-02)(-1.2,2.294043e-02)(-1.1,1.558337e-02)(-1.0,1.036087e-02)(-0.9,6.741806e-03)(-0.8,4.293174e-03)(-0.7,2.675385e-03)(-0.6,1.631505e-03)(-0.5,9.735935e-04)(-0.4,5.685269e-04)(-0.3,3.248688e-04)(-0.2,1.816565e-04)(-0.1,9.939953e-05)(0.0,5.322475e-05)(0.1,2.788985e-05)(0.2,1.430177e-05)(0.3,7.177171e-06)(0.4,3.524890e-06)(0.5,1.694244e-06)(0.6,7.969910e-07)(0.7,3.669334e-07)(0.8,1.653431e-07)(0.9,7.292194e-08)(1.0,3.147831e-08)(1.1,1.330003e-08)(1.2,5.500327e-09)(1.3,2.226509e-09)(1.4,8.821960e-10)(1.5,3.421481e-10)(1.6,1.298896e-10)(1.7,4.826671e-11)(1.8,1.755636e-11)(1.9,6.250778e-12)(2.0,2.178437e-12)(2.1,7.431290e-13)(2.2,2.481348e-13)(2.3,8.109798e-14)(2.4,2.594326e-14)(2.5,8.123130e-15)(2.6,2.489414e-15)(2.7,7.466815e-16)(2.8,2.191927e-16)(2.9,6.297320e-17)(3.0,1.770557e-17)(3.1,4.871611e-18)(3.2,1.311675e-18)(3.3,3.455825e-19)(3.4,8.909003e-20)(3.5,2.247173e-20)(3.6,5.545633e-21)(3.7,1.338900e-21)(3.8,3.162289e-22)(3.9,7.306077e-23)(4.0,1.651077e-23)(4.1,3.649391e-24)(4.2,7.888821e-25)(4.3,1.667667e-25)(4.4,3.447286e-26)(4.5,6.967558e-27)(4.6,1.376834e-27)(4.7,2.659757e-28)(4.8,5.022518e-29)(4.9,9.269978e-30)(5.0,1.672133e-30)(5.1,2.947501e-31)(5.2,5.076711e-32)(5.3,8.542983e-33)(5.4,1.404388e-33)(5.5,2.255102e-34)(5.6,3.536681e-35)(5.7,5.416569e-36)(5.8,8.100248e-37)(5.9,1.182671e-37)(6.0,1.685642e-38)
	     };
      \addplot[blue,densely dashed,smooth] plot coordinates{
	      (-4.0,0.996268)(-3.9,0.993776)(-3.8,0.989929)(-3.7,0.984179)(-3.6,0.975857)(-3.5,0.964185)(-3.4,0.948318)(-3.3,0.927404)(-3.2,0.900665)(-3.1,0.867492)(-3.0,0.827546)(-2.9,0.780838)(-2.8,0.727793)(-2.7,0.669261)(-2.6,0.60649)(-2.5,0.541045)(-2.4,0.47469)(-2.3,0.409247)(-2.2,0.346445)(-2.1,0.287789)(-2.0,0.234456)(-1.9,0.187234)(-1.8,0.146511)(-1.7,0.112297)(-1.6,0.0842862)(-1.5,0.0619344)(-1.4,0.0445465)(-1.3,0.0313571)(-1.2,0.0215996)(-1.1,0.0145582)(-1.0,0.00960034)(-0.9,0.00619395)(-0.8,0.00390964)(-0.7,0.00241429)(-0.6,0.00145855)(-0.5,0.000862066)(-0.4,0.000498485)(-0.3,0.000282011)(-0.2,0.000156098)(-0.1,8.45392e-05)(0.0,4.47988e-05)(0.1,2.32294e-05)(0.2,1.17867e-05)(0.3,5.85259e-06)(0.4,2.84395e-06)(0.5,1.35249e-06)(0.6,6.29506e-07)(0.7,2.86774e-07)(0.8,1.27871e-07)(0.9,5.58097e-08)(1.0,2.38435e-08)(1.1,9.9717e-09)(1.2,4.08243e-09)(1.3,1.63618e-09)(1.4,6.41978e-10)(1.5,2.466e-10)(1.6,9.27387e-11)(1.7,3.41452e-11)(1.8,1.23085e-11)(1.9,4.34402e-12)(2.0,1.50104e-12)(2.1,5.07822e-13)(2.2,1.68207e-13)(2.3,5.45499e-14)(2.4,1.73202e-14)(2.5,5.38418e-15)(2.6,1.63865e-15)(2.7,4.88253e-16)(2.8,1.42425e-16)(2.9,4.06722e-17)(3.0,1.13702e-17)(3.1,3.11158e-18)(3.2,8.33531e-19)(3.3,2.1856e-19)(3.4,5.60932e-20)(3.5,1.40902e-20)(3.6,3.46393e-21)(3.7,8.33376e-22)(3.8,1.96204e-22)(3.9,4.52002e-23)(4.0,1.01885e-23)(4.1,2.2469e-24)(4.2,4.84766e-25)(4.3,1.0231e-25)(4.4,2.11206e-26)(4.5,4.26441e-27)(4.6,8.42049e-28)(4.7,1.62593e-28)(4.8,3.06978e-29)(4.9,5.66647e-30)(5.0,1.02252e-30)(5.1,1.8036e-31)(5.2,3.10933e-32)(5.3,5.23845e-33)(5.4,8.62379e-34)(5.5,1.38708e-34)(5.6,2.1795e-35)(5.7,3.34511e-36)(5.8,5.01426e-37)(5.9,7.33987e-38)(6.0,1.04905e-38)	      
	     };
      \addplot[red,mark=*] plot coordinates{
	      (-0.25,0.999938)(0.00,0.999875)(0.25,0.998563)(0.50,0.993)(0.75,0.971812)(1.00,0.920312)(1.25,0.820312)(1.50,0.656312)(1.75,0.460938)(2.00,0.267688)(2.25,0.12726)(2.50,0.0526042)(2.75,0.0204688)(3.00,0.00486567)(3.25,0.00116915)(3.50,0.00021393)(3.75,4.47761e-05)
};
      \addplot[red,smooth] plot coordinates{	     
	      (-4.0,9.998669e-01)(-3.9,9.996595e-01)(-3.8,9.991848e-01)(-3.7,9.981694e-01)(-3.6,9.961363e-01)(-3.5,9.923193e-01)(-3.4,9.855880e-01)(-3.3,9.744199e-01)(-3.2,9.569605e-01)(-3.1,9.312020e-01)(-3.0,8.952865e-01)(-2.9,8.478918e-01)(-2.8,7.886185e-01)(-2.7,7.182718e-01)(-2.6,6.389413e-01)(-2.5,5.538306e-01)(-2.4,4.668557e-01)(-2.3,3.820997e-01)(-2.2,3.032507e-01)(-2.1,2.331474e-01)(-2.0,1.735200e-01)(-1.9,1.249517e-01)(-1.8,8.703071e-02)(-1.7,5.862360e-02)(-1.6,3.818810e-02)(-1.5,2.405853e-02)(-1.4,1.466090e-02)(-1.3,8.643618e-03)(-1.2,4.931595e-03)(-1.1,2.723746e-03)(-1.0,1.456718e-03)(-0.9,7.546845e-04)(-0.8,3.788727e-04)(-0.7,1.843827e-04)(-0.6,8.701791e-05)(-0.5,3.984022e-05)(-0.4,1.770197e-05)(-0.3,7.636045e-06)(-0.2,3.199042e-06)(-0.1,1.302058e-06)(0.0,5.150508e-07)(0.1,1.980730e-07)(0.2,7.407940e-08)(0.3,2.695277e-08)(0.4,9.542781e-09)(0.5,3.288809e-09)(0.6,1.103610e-09)(0.7,3.606801e-10)(0.8,1.148335e-10)(0.9,3.562545e-11)(1.0,1.077202e-11)(1.1,3.175219e-12)(1.2,9.125970e-13)(1.3,2.557980e-13)(1.4,6.993693e-14)(1.5,1.865439e-14)(1.6,4.854996e-15)(1.7,1.233086e-15)(1.8,3.056704e-16)(1.9,7.396430e-17)(2.0,1.747224e-17)(2.1,4.029735e-18)(2.2,9.074978e-19)(2.3,1.995679e-19)(2.4,4.285899e-20)(2.5,8.989293e-21)(2.6,1.841463e-21)(2.7,3.684445e-22)(2.8,7.200550e-23)(2.9,1.374527e-23)(3.0,2.562947e-24)(3.1,4.667953e-25)(3.2,8.304453e-26)(3.3,1.443068e-26)(3.4,2.449312e-27)(3.5,4.060402e-28)(3.6,6.574220e-29)(3.7,1.039559e-29)(3.8,1.605313e-30)(3.9,2.420740e-31)(4.0,3.564380e-32)(4.1,5.124286e-33)(4.2,7.192155e-34)(4.3,9.854188e-35)(4.4,1.317880e-35)(4.5,1.720195e-36)(4.6,2.191177e-37)(4.7,2.723477e-38)(4.8,3.302644e-39)(4.9,3.906919e-40)(5.0,4.507980e-41)(5.1,5.072735e-42)(5.2,5.566090e-43)(5.3,5.954412e-44)(5.4,6.209243e-45)(5.5,6.310703e-46)(5.6,6.250029e-47)(5.7,6.030807e-48)(5.8,5.668649e-49)(5.9,5.189365e-50)(6.0,4.625908e-51)
};
      \addplot[red,densely dashed,smooth] plot coordinates{
	      (-4.0,0.99988)(-3.9,0.99969)(-3.8,0.999247)(-3.7,0.99829)(-3.6,0.996352)(-3.5,0.992679)(-3.4,0.986145)(-3.3,0.975222)(-3.2,0.958035)(-3.1,0.932537)(-3.0,0.896817)(-2.9,0.8495)(-2.8,0.790144)(-2.7,0.71954)(-2.6,0.639797)(-2.5,0.55417)(-2.4,0.466649)(-2.3,0.381394)(-2.2,0.302156)(-2.1,0.231812)(-2.0,0.172098)(-1.9,0.123577)(-1.8,0.0858006)(-1.7,0.0575939)(-1.6,0.0373756)(-1.5,0.0234512)(-1.4,0.0142293)(-1.3,0.00835106)(-1.2,0.00474204)(-1.1,0.00260612)(-1.0,0.00138669)(-0.9,0.000714631)(-0.8,0.000356835)(-0.7,0.000172705)(-0.6,8.10528e-05)(-0.5,3.68999e-05)(-0.4,1.63022e-05)(-0.3,6.99199e-06)(-0.2,2.91241e-06)(-0.1,1.17859e-06)(0.0,4.63539e-07)(0.1,1.77247e-07)(0.2,6.5915e-08)(0.3,2.38477e-08)(0.4,8.3966e-09)(0.5,2.87796e-09)(0.6,9.60544e-10)(0.7,3.12263e-10)(0.8,9.89029e-11)(0.9,3.05275e-11)(1.0,9.18484e-12)(1.1,2.69431e-12)(1.2,7.70742e-13)(1.3,2.15053e-13)(1.4,5.85376e-14)(1.5,1.55473e-14)(1.6,4.02973e-15)(1.7,1.01944e-15)(1.8,2.51752e-16)(1.9,6.06967e-17)(2.0,1.42885e-17)(2.1,3.2846e-18)(2.2,7.37385e-19)(2.3,1.6168e-19)(2.4,3.46257e-20)(2.5,7.24348e-21)(2.6,1.48022e-21)(2.7,2.95495e-22)(2.8,5.7628e-23)(2.9,1.09796e-23)(3.0,2.04366e-24)(3.1,3.71626e-25)(3.2,6.60198e-26)(3.3,1.14579e-26)(3.4,1.94262e-27)(3.5,3.21741e-28)(3.6,5.20529e-29)(3.7,8.22584e-30)(3.8,1.26966e-30)(3.9,1.91398e-31)(4.0,2.81773e-32)(4.1,4.05076e-33)(4.2,5.68604e-34)(4.3,7.79255e-35)(4.4,1.04256e-35)(4.5,1.36152e-36)(4.6,1.7354e-37)(4.7,2.15861e-38)(4.8,2.61994e-39)(4.9,3.10237e-40)(5.0,3.5836e-41)(5.1,4.03742e-42)(5.2,4.43589e-43)(5.3,4.75205e-44)(5.4,4.96287e-45)(5.5,5.05201e-46)(5.6,5.01185e-47)(5.7,4.8446e-48)(5.8,4.56207e-49)(5.9,4.18437e-50)(6.0,3.73746e-51)
};
\node[small dot,pin={[pin distance=0.3cm] 0:{$n=36$}}] at (axis cs:3.3,0.01) {};
\node[small dot,pin={[pin distance=0.3cm] 180:{$n=144$}}] at (axis cs:2.87,0.01) {};
\node[small dot,pin={[pin distance=0.3cm] 0:{$n=36$}}] at (axis cs:-1,0.01) {};
\node[small dot,pin={[pin distance=0.3cm] 180:{$n=144$}}] at (axis cs:-1.32,0.01) {};
\legend{ {LDPC code}, {$\Phi(r/\theta_+)$},{$\Phi(r/\theta_-)$} }
    \end{semilogyaxis}
  \end{tikzpicture}
  \caption{Approximate bounds on the error probability for finite $n$, as a function of the $\text{SNR}=1/\sigma^2$, $r=K(R-C)$ for $K=8$, $N=16$, $R=\log(2)$, $n\in\{36,144\}$, $C$ being evaluated with $c=N/K$, $\beta=n/K$ and for different SNR values. Theoretical curves are compared to a rate $1/2$ LDPC QPSK code (giving $R=\log(2)$).}
\label{fig:Pe_n_simu}
\end{figure*}
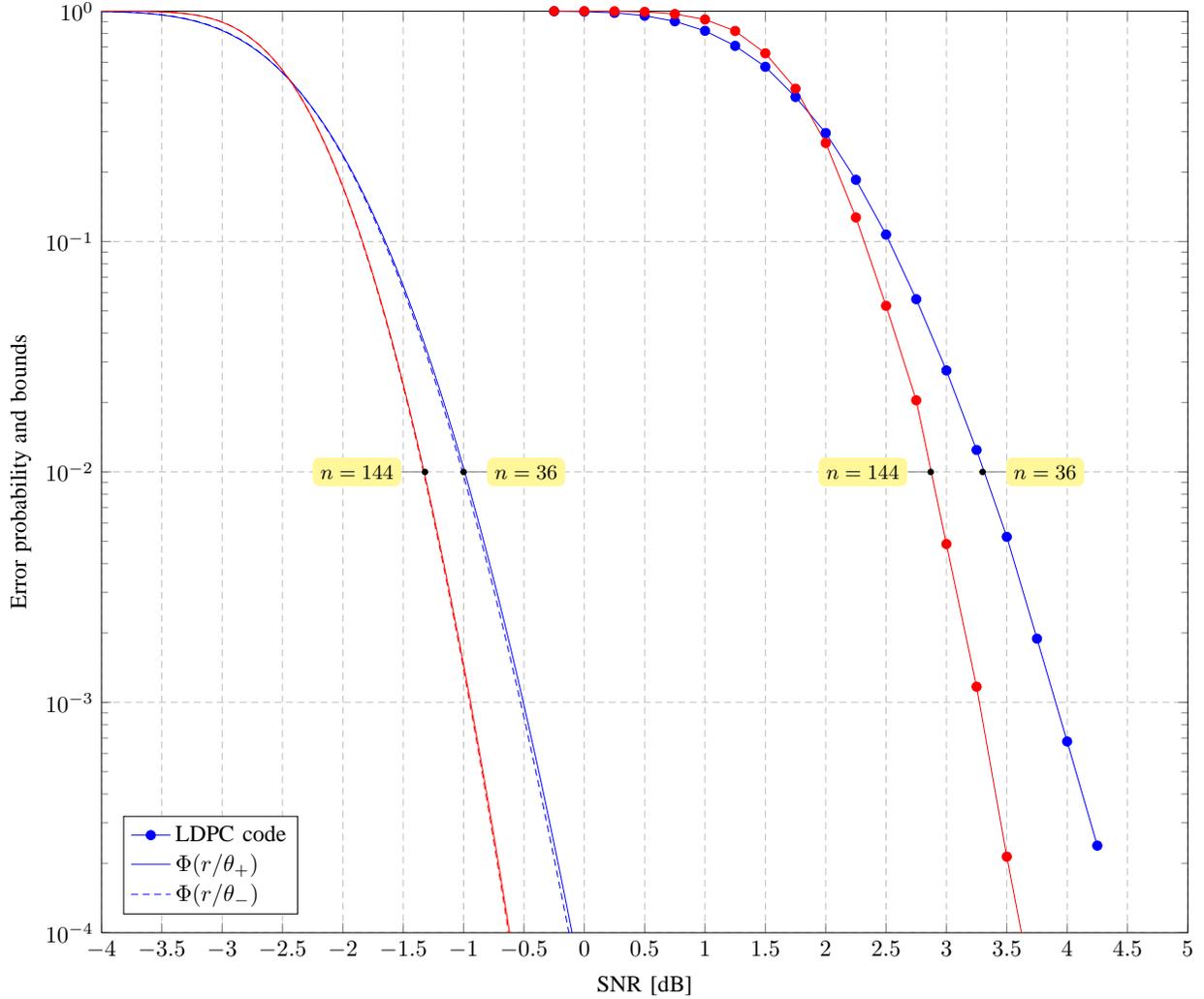

\begin{remark}[Relation to second-order outage probability]\label{rem:outage}
	Recalling Definition~\ref{def:Pout}, we have 
\begin{align}\label{eq:outbounds}
	\min\left\{\Phi\left( \frac{r}{\theta^{\rm out}_-} \right),\frac12 \right\} \leq \mathbb{P}_{\rm out}(r|\beta,c) \leq \Phi\left( \frac{r}{\theta^{\rm out}_+} \right)
\end{align}
where $\theta^{\rm out}_->0$ and $\theta^{\rm out}_+>0$ are defined by
\begin{align}
	\LB\theta^{\rm out}_-\RB^2 \defines& -\log\left( 1 - \frac1c \frac{\delta_0\LB\sigma^2\RB^2}{\LB1+\delta_0\LB\sigma^2\RB\RB^2} \right) \nonumber\\
	&\quad+ \frac1{\beta}\left(c + \sigma^4\delta_0'\LB\sigma^2\RB\right) \\	
	\LB\theta^{\rm out}_+\RB^2 \defines& -\log\left( 1 - \frac1c \frac{\delta_0\LB\sigma^2\RB^2}{\LB1+\delta_0\LB\sigma^2\RB\RB^2} \right)\nonumber\\
	& \quad+ \frac{2}{\beta}\left(c + \sigma^4\delta_0'\LB\sigma^2\RB - \frac{\delta_0'\LB\sigma^2\RB}{(1+\delta_0(\sigma^2))^4} \right).
\end{align}
Interestingly, for $r\leq 0$, as $\beta\to \infty$, we recover the limiting outage probability of MIMO Gaussian fading channels \cite{bai2004clt,HAC08},
\begin{align}\label{eq:limoutage}
	\lim_{\beta\to \infty} \mathbb{P}_{\rm out}(r|\beta,c) = \Phi\left( \frac{r}{\theta^{\rm out}} \right)
\end{align}
with $\theta^{\rm out}>0$ defined by
\begin{align}
	\LB\theta^{\rm out}\RB^2 \defines-\log\left( 1 - \frac1c \frac{\delta_0(\sigma^2)^2}{(1+\delta_0(\sigma^2))^2} \right).
\end{align}
Although both results coincide, there is a fundamental difference in the way they are obtained. In \cite{bai2004clt,HAC08}, the block-length is assumed to be infinitely large from the start and then the limit is taken in $N$ and $K$. By contrast, we have obtained \eqref{eq:limoutage} by changing the order of both limits. 
Note also that, while $\Phi\LB {r}/{\theta_-^{\text{out}}}\RB$ and $\Phi\LB {r}/{\theta_+^{\text{out}}}\RB$ are decreasing functions of $\beta$ for $r<0$, $\Phi\LB {r}/{\theta_+^{\text{out}}}\RB$ is increasing in $\beta$ for $r>0$. Although no tight lower bound was derived for $r>0$, this strongly suggests the existence of a crossing point for the optimal average error probability for an error rate of $1/2$. We will see a practical example of this crossing point effect in Figure~\ref{fig:Pe_n_simu}.
\end{remark}\smallskip

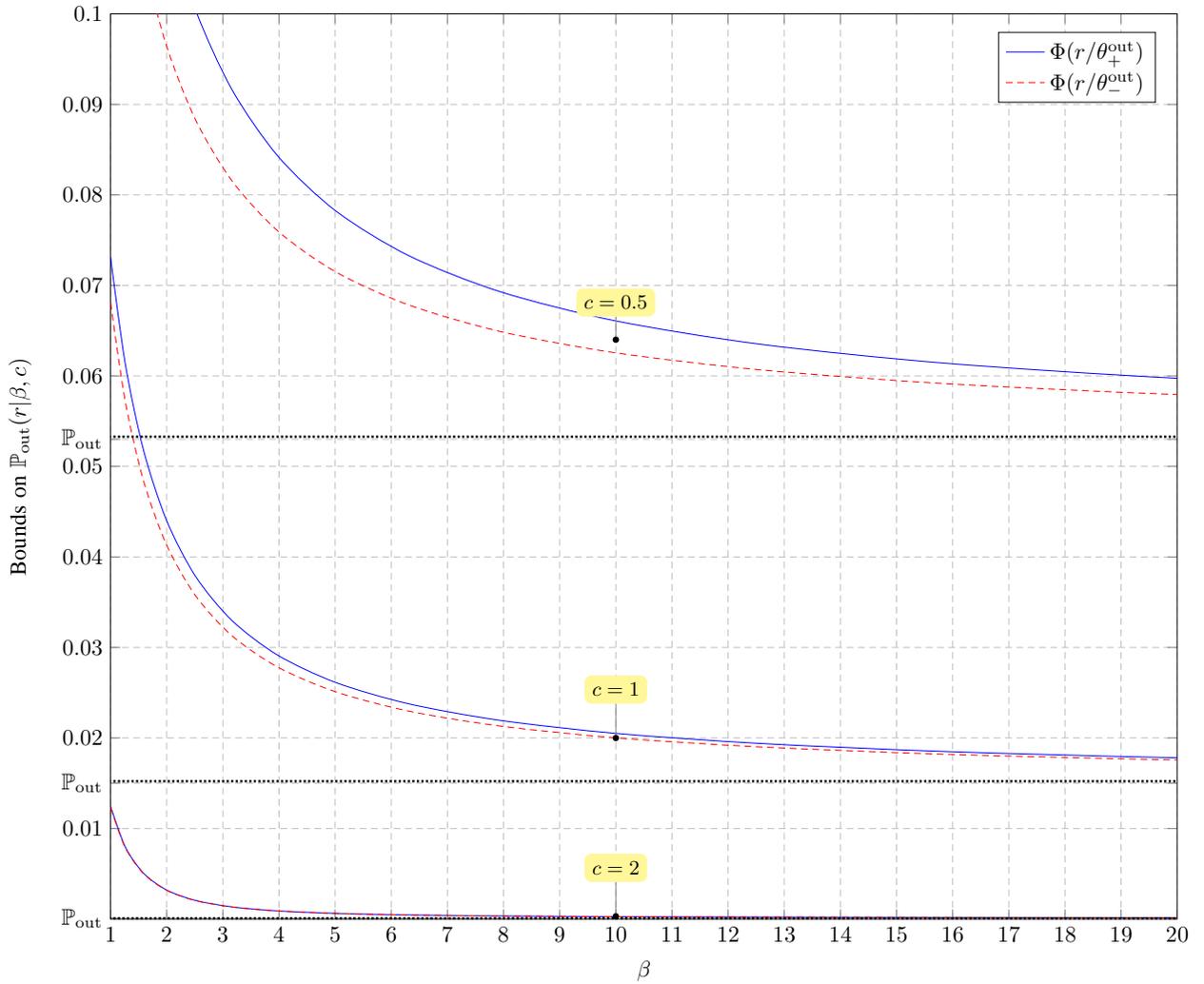
\begin{figure*}
\centering
\begin{tikzpicture}[scale=0.9]
	\pgfplotsset{every axis/.append style={mark options=solid}}
    \renewcommand{\axisdefaulttryminticks}{4}
    \pgfplotsset{every major grid/.append style={densely dashed}} 
    \pgfplotsset{every axis legend/.append style={cells={anchor=west},fill=white, at={(0.98,0.98)}, anchor=north east}}
    \pgfplotsset{
    tick label style={font=\normalsize},
    label style={font=\normalsize},
    legend style={font=\normalsize}
    }
    \tikzset{
	    every pin/.style={fill=yellow!50!white,rectangle,rounded corners=3pt,font=\small},
	    small dot/.style={fill=black,circle,scale=0.3}
    }
    \begin{axis}[
width=\textwidth,
      xmin=1,
      ymin=0,
      xmax=20,
      ymax=0.1,  
      bar width=3pt,
      grid=major,
      ytick={0.00008,0.01,0.015,0.02,0.03,0.04,0.05,0.053,0.06,0.07,0.08,0.09,0.1},
      yticklabels={$\mathbb{P}_{\rm out}$,$0.01$,$\mathbb{P}_{\rm out}$,$0.02$,$0.03$,$0.04$,$0.05$,$\mathbb{P}_{\rm out}$,$0.06$,$0.07$,$0.08$,$0.09$,$0.1$},
      scaled ticks=true,
      ylabel={Bounds on $\mathbb{P}_{\rm out}(r|\beta,c)$},
      xlabel={$\beta$},
      ylabel style={yshift=5pt}
      ]
      \addplot[blue,smooth] plot coordinates{	     
	      (1.000000,0.153125)(1.258925,0.137302)(1.584893,0.12323)(1.995262,0.110966)(2.511886,0.100469)(3.162278,0.0916258)(3.981072,0.0842732)(5.011872,0.0782251)(6.309573,0.0732923)(7.943282,0.0692956)(10.000000,0.0660734)(12.589254,0.0634854)(15.848932,0.0614127)(19.952623,0.059756)(25.118864,0.0584339)(31.622777,0.05738)(39.810717,0.0565407)(50.118723,0.0558725)(63.095734,0.055341)(79.432823,0.0549183)(100.000000,0.0545822)
};
      \addplot[red,densely dashed,smooth] plot coordinates{	     
	      (1.000000,0.131091)(1.258925,0.117787)(1.584893,0.106285)(1.995262,0.0965103)(2.511886,0.0883235)(3.162278,0.0815496)(3.981072,0.0759992)(5.011872,0.071486)(6.309573,0.0678376)(7.943282,0.0649013)(10.000000,0.0625461)(12.589254,0.0606615)(15.848932,0.0591564)(19.952623,0.0579558)(25.118864,0.0569991)(31.622777,0.0562374)(39.810717,0.0556313)(50.118723,0.0551491)(63.095734,0.0547657)(79.432823,0.0544609)(100.000000,0.0542186)
};
      \addplot[blue,smooth] plot coordinates{	          
	      (1.000000,0.0731087)(1.258925,0.0614426)(1.584893,0.0518389)(1.995262,0.0440702)(2.511886,0.0378711)(3.162278,0.0329733)(3.981072,0.0291288)(5.011872,0.0261222)(6.309573,0.0237745)(7.943282,0.0219416)(10.000000,0.020509)(12.589254,0.0193877)(15.848932,0.0185085)(19.952623,0.0178178)(25.118864,0.0172743)(31.622777,0.016846)(39.810717,0.016508)(50.118723,0.0162409)(63.095734,0.0160297)(79.432823,0.0158625)(100.000000,0.0157301)
};
      \addplot[red,densely dashed,smooth] plot coordinates{	       
	      (1.000000,0.068084)(1.258925,0.0572834)(1.584893,0.0484596)(1.995262,0.0413647)(2.511886,0.0357287)(3.162278,0.0312892)(3.981072,0.0278108)(5.011872,0.0250929)(6.309573,0.022971)(7.943282,0.0213138)(10.000000,0.0200178)(12.589254,0.0190028)(15.848932,0.0182062)(19.952623,0.0175801)(25.118864,0.0170871)(31.622777,0.0166983)(39.810717,0.0163913)(50.118723,0.0161487)(63.095734,0.0159567)(79.432823,0.0158047)(100.000000,0.0156842)	      
};
      \addplot[blue,smooth] plot coordinates{	           
	      (1.000000,0.0124356)(1.258925,0.00799282)(1.584893,0.00507216)(1.995262,0.00321165)(2.511886,0.00205166)(3.162278,0.00133612)(3.981072,0.000894908)(5.011872,0.000620476)(6.309573,0.000447079)(7.943282,0.000335252)(10.000000,0.00026145)(12.589254,0.00021157)(15.848932,0.000177071)(19.952623,0.000152692)(25.118864,0.000135129)(31.622777,0.000122261)(39.810717,0.000112695)(50.118723,0.000105496)(63.095734,0.000100023)(79.432823,9.58262e-05)(100.000000,9.25866e-05)
};
      \addplot[red,densely dashed,smooth] plot coordinates{	      
(1.000000,0.0123149)(1.258925,0.00791206)(1.584893,0.00502005)(1.995262,0.00317891)(2.511886,0.0020314)(3.162278,0.00132365)(3.981072,0.00088719)(5.011872,0.000615639)(6.309573,0.000443989)(7.943282,0.000333235)(10.000000,0.000260101)(12.589254,0.000210647)(15.848932,0.000176425)(19.952623,0.000152231)(25.118864,0.000134794)(31.622777,0.000122013)(39.810717,0.000112509)(50.118723,0.000105355)(63.095734,9.99152e-05)(79.432823,9.57436e-05)(100.000000,9.25226e-05)
};
      \addplot[black,densely dotted,line width=1pt] plot coordinates{(0,0.0532815)(100,0.0532815)};
      \addplot[black,densely dotted,line width=1pt] plot coordinates{(0,0.0152218)(100,0.0152218)};
      \addplot[black,densely dotted,line width=1pt] plot coordinates{(0,8.08299e-05)(100,8.08299e-05)};
\legend{ {$\Phi(r/\theta_+^{\rm out})$} , {$\Phi(r/\theta_-^{\rm out})$} }
\node[small dot,pin={[pin distance=0.3cm] 90:{$c=0.5$}}] at (axis cs:10,0.064) {};
\node[small dot,pin=90:{$c=1$}] at (axis cs:10,0.02) {};
\node[small dot,pin=90:{$c=2$}] at (axis cs:10,0.0003) {};
    \end{axis}
  \end{tikzpicture}
 \caption{Bounds on the second-order outage probability as a function of $\beta$ for different values of $c$, $r=-C(\sigma^2)$, and ${\rm SNR=10\,}$dB. The limiting outage probability is $\mathbb{P}_{\rm out}\defines \mathbb{P}_{\rm out}(r|\infty,c)$.}
\label{fig:Pout}
\end{figure*}

Figure~\ref{fig:Pout} depicts the bounds on $\mathbb{P}_{\rm out}(r|\beta,c)$ in \eqref{eq:outbounds} as a function of $\beta$ for different values of $c$, assuming ${\rm SNR}=10\,$dB and $r=-C(\sigma^2)$ (for fair comparison since $C(\sigma^2)$ is implicitly a function of $c$). For each value of $c$ we also provide the limiting outage probability as given in \eqref{eq:limoutage}. The upper and lower bounds are seen to approach the outage probability at a rate $\Oc(\beta^{-1})$ as $\beta$ grows, which is easily confirmed by direct calculus. 

We conclude this section by a comparison in Figure~\ref{fig:Pe_n_simu} of the theoretical results against practical codes. We specifically consider a scenario with $K=8$ transmit and $N=16$ receive antennas employing QPSK modulation at each antenna. Coding and modulation are set up in a conventional bit-interleaved coded modulation (BICM) scheme, with a random interleaver separating the code and the modulation. At the receiver, we employ a non-iterative demodulation scheme, with a MAP MIMO demodulator based on a full code book enumeration. We consider short LDPC codes and take as an example the rate $1/2$ code used in the WiMAX standard \cite{wimax}, corresponding to a coding rate in nats $R=\log(2)$. This code is a quasi-cyclic irregular repeat-accumulate (IRA) LDPC code where the accumulator is slightly modified to ease the encoding circuit. 

We consider code blocks of $n'=576\,$ bit and $n'=2304\,$ bit, corresponding to  $n=n'/(2K)\in \{36,144\}$ channel uses. The error probability of the code described above for ${n\in\{36,144\}}$ is compared against the approximate upper and lower bounds (Theorem~\ref{th-main}) obtained when coding at second order rate $r=(R-C(\sigma^2))\sqrt{nK}$, for different SNR (i.e., $\sigma^{-2}$) values (corresponding to a span from $r\simeq 5.2$ for $-4\,$dB SNR to $r\simeq -10$ for $0\,$dB SNR, when $n=144$). We can make several interesting observations from this figure. For both block-lengths, the SNR-gap between the simulation results and the corresponding bounds by Theorem~\ref{th-main} is roughly constant (to about $4\,$dB) for a large range of SNR values. 

Also note that both theoretical and simulated curves exhibit a crossing point close to $1/2$ error probability, which goes in line with Remark~\ref{rem:outage}.

\section{Summary and directions for future work}
We have studied the second-order coding rate of the MIMO quasi-static Rayleigh fading channel using information-spectrum methods and Gaussian tools from random matrix theory. To this end, we derived a CLT for the asymptotic analysis of the ``information density" where the channel dimensions as well as the block-length grow infinitely large at the same speed and the coding rate is a perturbation within $\Oc(1/\sqrt{nK})$ of the asymptotic capacity. The derived CLT allowed us to characterize  closed-form upper and lower bounds on the optimal average error probability which depend only on the main system and channel parameters. The proposed approach to the study of the asymptotic statistics of the ``mutual information density" for MIMO channels is original and can be further applied  to other scenarios, such as the block-fading regime where coding is performed over multiple coherence blocks or, in a more practical context, the error performance achieved under linear receive filters.

\appendices
\section{Auxiliary results on information spectrum}
\label{app:information_spectrum}
The objective of this section is to prove Proposition~\ref{prop-bounds_2nd_order} below which provides analytical bounds on the optimal average error probability $\mathbb{P}_e(r|\beta,c)$ and constitutes the first step of the proof of Theorem~\ref{th-main}, developed in Appendix~\ref{app:proof_th-main}.

We first state a variation of Verd\'u--Han's lemma~\cite{VER94} which appears to be more adequate to characterize the second-order approximation of the error probability. 

\begin{lemma}[Variation on Verd\'u-Han's lemma]\label{lemma-Converse}
For any integer $n\ge1$, let $X^n$ be an arbitrary random variable uniformly distributed over the set of $M_n$ messages issued from $M_n$ realizations of $\mathbb{P}_{X^n} \in \mathcal{P}(\mathcal{S}^n)$, and let $Y^n$ be the output random variable of the channel $ \mathbb{P}_{Y^n|X^n,H^n}$ corresponding to the input $X^n$ and the random fading $H^n$. Then, the  average error probability of such a $({P}_e^{(n)},M_{n})$-code  $\mathcal{C}_{n}$ must satisfy
\begin{align}
	&P_e^{(n)}(\mathcal{C}_{n}) \geq \sup\limits_{\gamma>0}\, \sup\limits_{ \{\mathbb{Q}_n\}_{n=1}^\infty } \nonumber\\
	& \Bigg\{ \Pr\left[ \log\frac{\mathbb{P}_{Y^n|X^n,H^n}(dY^n| X^n,H^n)}{\mathbb{Q}_n(dY^n|H^n)} \leq \log \gamma \right]- \frac{\gamma}{M_{n}} \Bigg\}
\end{align}
where $\mathbb Q_n(\cdot|H^n)$ is an $H^n$-measurable random variable valued in $\mathcal{P}(\mathds{C}^{N\times n})$. 
\end{lemma}
\begin{IEEEproof}
The proof follows straightforwardly from that in~\cite{VER94} which itself is related to \cite{DBLP:journals/iandc/Wolfowitz68}. We remark that a similar result was already used in~\cite{HAY09} without an explicit proof and also follows from the same steps used to prove the ``meta-converse'' theorem in~\cite[Thm. 26, 27]{POL10}.
\end{IEEEproof}

\begin{lemma}[Variation of Feinstein's lemma]\label{lemma-Feinstein}
Let $n\ge 1$ be an integer and denote by $Y^n$ the output from the channel $\mathbb{P}_{Y^n|X^n,H^n}$ corresponding to an input distribution $\mathbb{P}_{X^n}$ and random fading $H^n$. We denote $\mathbb{P}_n(dY^n| H^n)$ the distribution of such $Y^n$ given $H^n$. Then, there exists a block-length $n$ codebook of size $M_{n}$ that, together with the maximum {\it a posteriori} (MAP) decoder, forms a code  $\mathcal{C}_{n}$ whose average error probability $P_e^{(n)}(\mathcal{C}_{n})$ satisfies:
\begin{align}
	\label{eq-lemma_Feinstein}
&P_e^{(n)}(\mathcal{C}_{n}) \leq \nonumber\\
&\inf\limits_{\gamma>0}\left\{ \Pr\left[ \log\frac{\mathbb{P}_{Y^n|X^n,H^n}(dY^n| X^n,H^n)}{\widetilde{\mathbb{P}}_{ n} (d{Y}^n|H^n)  } \leq \log \gamma\kappa \right] + \frac{M_{n}}{\gamma} \right\} \nonumber \\
& +\Pr\left( \frac{\mathbb{P}_n(dY^n| H^n)   }{ \widetilde{\mathbb{P}}_{ n} (d{Y}^n|H^n)  }>\kappa\right)
\end{align}
for any probability measure $\widetilde{\mathbb{P}}_{ n} (d{Y}^n|H^n)  \gg \mathbb{P}_n(dY^n| H^n) $ and positive value $\kappa$, where $\frac{\mathbb{P}_n(dY^n| H^n)}{\widetilde{\mathbb{P}}_{ n} (d{Y}^n|H^n)}$ denotes the Radon-Nikodym derivative. 
\end{lemma}
\begin{IEEEproof}
The proof simply follows from Feinstein's lemma~\cite{FEI54} and the introduction of the event
\begin{equation}
\mathcal{B}_n= \left\{ Y^n \in \mathds{C}^{N\times n} \,:\, \frac{\mathbb{P}_n(Y^n| H^n)}{\widetilde{\mathbb{P}}_{ n} ({Y}^n|H^n)  } >\kappa\right\}.  \label{eq-set-zero-defintion}
\end{equation}
\end{IEEEproof}

In order to obtain Part-{\emph (ii)} (Upper bound) of Proposition~\ref{prop-bounds_2nd_order} below, we need the following technical result.	
\begin{lemma}[A divergence result]
	\label{lemma:yuri}
	Let $\widetilde{\mathbb{P}}_n(d\widetilde{Y}^n|\Hm^n)$ and $\mathbb{P}_n(dY^n|\Hm^n)$ be the output distributions of the channels $\widetilde{Y}^n=\frac1{\sqrt{K}}\Hm^n\widetilde{X}^n+\sigma W^n$ and $Y^n=\frac1{\sqrt{K}}\Hm^nX^n+\sigma W^n$, respectively, where $\widetilde{X}^n$ is standard Gaussian (i.e., with independent $\mathcal{CN}(0,1)$ entries) and $X^n = \sqrt{nK}\widetilde{X}^n/\Vert \widetilde{X}^n \Vert_F$. Then, for any sequence $\kappa_n$ satisfying $\kappa_n \to \infty$,
	\begin{align}
		\label{def:eta}
\alpha_n \triangleq \Pr\left( \log \frac{\mathbb{P}_n(dY^n| H^n)  }{ \widetilde{\mathbb{P}}_{ n} (d{Y}^n|H^n)  }\geq \log \kappa_n\right) \to 0
	\end{align}
	where $\Pr(\cdot)$ is taken over $H^n$ standard Gaussian and ${Y^n\sim\mathbb{P}_n}$.
\end{lemma}
\begin{IEEEproof}
 For two distributions $P$ and $Q$, let $\beta_\alpha(P,Q)$ be defined as in \cite[Eq. (100)]{POL10}. Then, we have the following bounds on $\beta_\alpha(P,Q)$ \cite[Eqs. (154)-(157)]{POL10}:
 \begin{align}
  \beta_\alpha(P,Q) \ge \exp\LB - \frac{D\LB P \Vert Q\RB + h(\alpha)}{\alpha}\RB
 \end{align}
 where $D\LB P \Vert Q\RB$ is the Kullback-Leibler divergence and $h(x)$ the binary entropy function, 
and  \cite[Eq. (103)]{POL10}
 \begin{align}
  \beta_\alpha(P,Q) \le \frac{1}{\gamma_0}
 \end{align}
 for any $\gamma_0$ satisfying
 \begin{align}
    \Pr\LB \frac{dP}{dQ} \ge \gamma_0 \RB \ge \alpha.
 \end{align}
 Setting $P = \mathbb{P}_n(dY^n|\Hm^n)$, $Q = \widetilde{\mathbb{P}}_n(dY^n|\Hm^n)$, $\alpha=\alpha_n$, $\gamma_0 = \kappa_n$ and using the upper and lower bounds on $\beta_\alpha$, we conclude that
 \begin{align}
 \label{proof-step-1}
&  \exp\LB - \frac{D\LB\mathbb{P}_n(dY^n|H^n) \Vert \widetilde{\mathbb{P}}_n(dY^n|H^n)\RB + h(\alpha_n)}{\alpha_n}\RB \nonumber\\
 &  \leq \frac1{\kappa_n} \to 0.
 \end{align}
 To obtain \eqref{def:eta}, it is thus sufficient to prove $D\LB\mathbb{P}_n(dY^n|H^n) \Vert \widetilde{\mathbb{P}}_n(dY^n|H^n)\RB = \Oc(1)$.
 
 By the data-processing inequality for the Kullback-Leibler divergence \cite{csiszar1982information},
 \begin{align}
  &D\LB\mathbb{P}_n(dY^n|H^n) \Vert \widetilde{\mathbb{P}}_n(dY^n|H^n)\RB\nonumber\\
   &=  D\LB\mathbb{P}_{Y^n|H^n} \Vert \mathbb{P}_{\widetilde{Y}^n|H^n}\RB\\
 &\leq  D\LB\mathbb{P}_{Y^n\widetilde{X}^n|H^n} \Vert \mathbb{P}_{\widetilde{Y}^n\widetilde{X}^n|H^n}\RB\\
  &= D\LB\mathbb{P}_{Y^n|H^n\widetilde{X}^n} \Vert \mathbb{P}_{\widetilde{Y}^n|H^n\widetilde{X}^n}\RB\\
  \label{data-proc-1}
  &= \mathbb{E}_{H^n, \widetilde{X}^n}\LSB D\LB\mathbb{P}_{Y^n|\Hm^n=H^n,\widetilde{\Xm}^n=\widetilde{X}^n} \Vert \mathbb{P}_{\widetilde{Y}^n|\Hm^n=H^n,\widetilde{\Xm}^n=\widetilde{X}^n}\RB \RSB.    
 \end{align}
Note that, for given $\Hm^n, \widetilde{\Xm}^n$, the channel outputs $Y^n, \widetilde{Y}^n$ are Gaussian distributed, i.e.,
\begin{align}
\label{cond-dist-1}
\mathbb{P}_{Y^n|\Hm^n,\widetilde{\Xm}^n} &\sim \Cc\Nc\LB \text{vec}\LB\frac1{\sqrt{K}}\Hm^n\frac{\sqrt{nK}\widetilde{\Xm}^n}{\Vert \widetilde{\Xm}^n\Vert_F}\RB, \sigma^2\Id_{N\times n}\RB\\
\label{cond-dist-2}
\mathbb{P}_{\widetilde{Y}^n|\Hm^n,\widetilde{\Xm}^n}&\sim \Cc\Nc\LB \text{vec}\LB\frac1{\sqrt{K}}\Hm^n\widetilde{\Xm}^n\RB, \sigma^2\Id_{N\times n}\RB
\end{align}
where the function $\text{vec}\LB \Am\RB$ vectorizes the matrix $\Am$. Using $D\LB\Cc\Nc\LB \mv_1, \sigma^2\Id\RB\Vert\Cc\Nc\LB \mv_2, \sigma^2\Id\RB\RB = \Vert \mv_1 - \mv_2 \Vert^2/\sigma^2$ together with \eqref{cond-dist-1} and \eqref{cond-dist-2} in \eqref{data-proc-1}, we obtain from standard computations
\begin{align}
 &\mathbb{E}_{H^n, \widetilde{X}^n}\LSB D\LB\mathbb{P}_{Y^n|\Hm^n=H^n,\widetilde{\Xm}^n=\widetilde{X}^n} \Vert \mathbb{P}_{\widetilde{Y}^n|\Hm^n=H^n,\widetilde{\Xm}^n=\widetilde{X}^n}\RB \RSB\nonumber\\
 & =  \mathbb{E}_{H^n, \widetilde{X}^n}\LSB \frac1{\sigma^2K} \left| \frac{\sqrt{nK}}{\Vert\widetilde{X}^n \Vert_F} -1 \right|^2 \Vert H^n\widetilde{X}^n \Vert_F^2 \RSB\\
 & = \frac1{\sigma^2}\frac{N}{K}\mathbb{E}\LSB  \left| \sqrt{nK} - \Vert\widetilde{X}^n \Vert_F\right|^2\RSB\\
 \label{mean-result}
  & = \frac1{\sigma^2}\frac{N}{K}\LB 2nK - 2\sqrt{nK}  \mathbb{E}\LSB \Vert\widetilde{X}^n \Vert_F \RSB \RB.
\end{align}
Now, since $\widetilde{X}^n$ is Gaussian, $\sqrt{2}\Vert\widetilde{X}^n \Vert_F$ is $\chi_{2nK}$-distributed, so that
\begin{align}
 \mathbb{E}\LSB \Vert\widetilde{X}^n \Vert_F \RSB = \frac{\Gamma(nK + 1/2)}{\Gamma(nK)}.
\end{align}
Using this result in \eqref{mean-result} leads to
\begin{align}
 \text{\eqref{mean-result}} & =  \frac{2nN}{\sigma^2}\LB1 - \frac{\Gamma(nK + 1/2)}{\sqrt{nK}\Gamma(nK)}\RB\\
 \label{gamma-inequ}
 & \leq \frac{2nN}{\sigma^2}\LB1 - \sqrt{\frac{nK}{nK+1/2}}\RB\\
  & = \frac{2nN}{\sigma^2}\LB1 - \sqrt{1 - \frac{1/2}{nK + 1/2}}\RB\\
   \label{sqrt-order}
  & = \frac{2nN}{\sigma^2}\LB1 - 1 + \Oc(1/(nK))\RB = \Oc(1)
\end{align}
where we used in \eqref{gamma-inequ} that for $a\in(0,1)$ and $x>0$ \cite{wendel1948note}
\begin{align}
 1\geq \frac{\Gamma(x+a)}{\Gamma(x)x^a} \geq \LB \frac{x}{x+a}\RB^{1-a}
\end{align}
and  \eqref{sqrt-order} follows because $\sqrt{1+x}=1+\Oc(x)$ as $x\to 0$. 
\end{IEEEproof}

With this result at hand, we can prove the following result.

\begin{prop}[Bounds on the average error probability]\label{prop-bounds_2nd_order}
The following two statements hold:\\
$(i)$ {\it Lower bound:}
	Let $Y_+^n\in \CC^{N\times (n+1)}$ denote the random variable associated to the output of the channel $ \mathbb{P}_{Y_+^n|X_+^n,H^n}$  corresponding to the input $X_+^n\in\CC^{K\times (n+1)}$ and fading $H^n\in\CC^{K\times N}$. Then, \eqref{eq-2orderFuncLow} on the top of the next page holds, 
	\begin{figure*}
	\begin{align}
	\mathbb{P}_e(r|\beta,c) &\geq  \mathbb{F}(r| \beta,c)\nonumber\\
	&\defines  \inf_{ \substack{ \{ \mathbb{P}_{X_+^n}\}_{n=1}^{\infty} \\ \mathbb{P}_{X_+^n} \in \mathcal{P}(\mathcal{S}^{n+1}_{=} ) }}\, \sup_{\left\{ \mathbb{Q}_{n,+} \right\}_{n=1}^\infty} 
  \lim_{\xi\downarrow 0}\limsup\limits_{n \xrightarrow{(\beta,c)}  \infty}\Pr\left[ \sqrt{nK}\left(\frac1{nK}\log\frac{\mathbb{P}_{Y_+^n|X_+^n,H^n}(d{Y}_+^n| {X}_+^n,H^n)}{\mathbb{Q}_{n,+}(d{Y}_+^n|H^n)} - C \right) \leq r-\xi \right]\label{eq-2orderFuncLow}
\end{align}
\hrulefill
	\end{figure*}
where $\mathbb Q_{n,+}(\cdot|H^n)$ is an $H^n$-measurable random variable taking values in $\mathcal{P}(\mathds{C}^{N\times (n+1)})$ and 
\begin{align}
	\label{def-equalS+}
	&\mathcal{S}^{n,+}_{=}=\left\{ \Xm_+^n\in \mathds{C}^{K\times (n+1)}\ \Big|\ \frac1{(n+1)K} \trace \Xm_+^n(\Xm_+^n)\htp = 1 \right\}.
\end{align}
	$(ii)$ {\it Upper bound:}  There exists a codebook of size $M_{n}$ with codewords of block-length $n$ that together with the ML decoder form a $({P}_e^{(n)},M_{n})$-code $\mathcal{C}_{n}$ such that, for all real $r$, \eqref{eq-2orderFuncUp} on the top of the next page holds, 
	\begin{figure*}
\begin{align}
&\mathbb{P}_e(r|\beta,c) \leq  \mathbb{G}(r| \beta,c) \defines \lim_{\xi\downarrow 0}  \limsup\limits_{n \xrightarrow{(\beta,c)}  \infty}\Pr\left[ \sqrt{nK}\left(\frac1{nK}\log\frac{\mathbb{P}_{Y^n|X^n,H^n}(d{Y}^n| {X}^n,H^n)}{  \widetilde{\mathbb{P}}_{n} (d{Y}^n|H^n)} - C \right) \leq r+\xi \right] \label{eq-2orderFuncUp} 
\end{align}
\hrulefill
	\end{figure*}
which is computed from the probability measure induced by inputs uniformly distributed over the power shell: 
\begin{equation}
\mathbb{P}_{X^n}(\Xm^n)=\frac{\mathbbm{1}\left[\trace \Xm^n(\Xm^n)\htp = nK \right]}{S_{2nK}(\sqrt{nK})}\label{eq-PD-uniform}
\end{equation}
which satisfy $ \mathbb{P}_{X^n} ( \mathcal{S}^n_=)=1$, and where $S_{2nK}(r)=\displaystyle{2\pi^{nK}\Gamma(nK)^{-1}r^{2nK-1}}$ is the surface area of a $2nK$-dimensional sphere of radius $r$,  and $\widetilde{\mathbb{P}}_{n} $ is the output distribution of the channel $\mathbb{P}_{Y^n|X^n,H^n}$ induced by a complex Gaussian input distribution with zero mean and covariance $\Id_{Kn}$.  

\end{prop}
\begin{IEEEproof}
This proof is segmented in two parts. We first derive error probability bounds for each $N,K,n$, based on the established slight variations on the Verd{\'u}--Han's Lemma~\ref{lemma-Converse} and the modified  Feinstein's Lemma~\ref{lemma-Feinstein} and then  bringing $N,K,n$ to infinity leads to Proposition~\ref{prop-bounds_2nd_order}.


We first start with the proof of the lower bound \eqref{eq-2orderFuncLow}.
Let $\mathcal{C}_{n}$ be a $({P}_e^{(n)},M_{n})$-code whose probability measure satisfies $\mathbb{P}_{X^n} \in \mathcal{P}(\mathcal{S}^n)$. From this code, following the approach in~\cite{POL10}, we define the code $\mathcal C_{n,+}$ with codewords $\{\Xm^n_{i,+}=[\Xm^n_i,\xv_i],~i=1,\ldots,M_n\}$, where $\{\Xm^n_i,i=1,\ldots,M_n\}=\supp (\mathcal C_n)$ and $\xv_i$ satisfies $\Vert \xv_i\Vert^2=(n+1)K-\trace \Xm^n_i(\Xm^n_i)\htp$, and with the same decision region as for $\mathcal C_n$ discarding the last channel output (corresponding to input $\xv_i$). Note that the probability measure $\mathbb{P}_{X_+^n}$ of the code $\mathcal C_{n,+}$ satisfies $\mathbb{P}_{X_+^n} \in \mathcal{P}(\mathcal{S}_{=}^{n,+})$ and that $P_e^{(n)}(\mathcal C_{n,+})=P_e^{(n)}(\mathcal C_n)$.

From Lemma~\ref{lemma-Converse}, the average error probability must satisfy
\begin{align}
	\label{eq-Pe1}
	&P_e^{(n)}(\mathcal{C}_{n}) = P_e^{(n)}(\mathcal C_{+,n}) \nonumber\\
	&\geq\Pr\left[ \log\frac{\mathbb{P}_{Y_+^n|X_+^n,H^n}(dY_+^n| X_+^n,H^n)}{\mathbb{Q}_{+,n}(dY_+^n|H^n)} \leq \log \gamma \right] - \frac{\gamma}{M_{n}} 
\end{align}
for each $n=1,2,\dots$, $\gamma>0$, where $\mathbb{Q}_{+,n}(\cdot|H^n)$ is $H^n$-measurable and takes values in $\mathcal{P}( \mathds{C}^{N\times (n+1)} )$, with $Y_+^n=\frac1{\sqrt{K}}H^nX_+^n+\sigma W_+^n$, $W_+^n\in\mathds{C}^{N\times (n+1)}$ with i.i.d. $\mathcal{CN}(0,1)$ entries. Let us choose $\gamma$ as 
\begin{equation}
	\label{eq-Pe2}
	\frac1{nK} \log \gamma= \frac1{nK} \log M_n - \frac{\xi}{\sqrt{nK}}
\end{equation}
for some $\xi>0$. We now set the coding rate
\begin{equation}
	\label{eq-Pe3}
	\frac1{nK} \log M_n =  C + \frac{r}{\sqrt{nK}}
\end{equation}
for some real $r$. Then, combining \eqref{eq-Pe1}--\eqref{eq-Pe3}, we obtain
\begin{align}
	&P_e^{(n)}(\mathcal{C}_{n}) \geq \nonumber\\
	&\Pr\Bigg[ \sqrt{nK}\left(\frac1{nK}\log\frac{\mathbb{P}_{Y_+^n|X_+^n,H^n}(dY^n| X_+^n,H^n)}{\mathbb{Q}_{+,n}(dY_+^n|H^n)} - C\right)\nonumber\\
	& \qquad\qquad\leq r - \xi \Bigg] - \exp(-\sqrt{nK}\xi).
\end{align}
Taking the limit superior over $n$ on the last equation, we obtain
\begin{align}\label{eq:petmp}
	&\limsup\limits_{n \xrightarrow{(\beta,c)}  \infty} P_e^{(n)}(\mathcal{C}_{n}) \geq  \limsup\limits_{n \xrightarrow{(\beta,c)}  \infty} \Pr\Bigg[ \sqrt{nK} \nonumber\\
	&\times\left(\frac1{nK}\log\frac{\mathbb{P}_{Y_+^n|X_+^n,H^n}(dY_+^n| X_+^n,H^n)}{\mathbb{Q}_{+,n}(dY_+^n|H^n)} - C\right) \leq r - \xi \Bigg].
\end{align}
As this is true for each $\xi>0$ and $\mathbb{Q}_{+,n}$ as defined above, we can take $\xi\downarrow 0$ followed by the supremum over $\mathbb{Q}_{+,n}$ on the RHS of \eqref{eq:petmp}. Taking then the infimum over the codes on the RHS then LHS, we conclude that
\begin{align}
\mathbb{P}_e(r|\beta,c) &\geq  \mathbb{F}(r| \beta,c)
\end{align}
which proves part $(i)$ of the proposition.


We now prove part $(ii)$ for the upper bound in \eqref{eq-2orderFuncUp}. From Lemma~\ref{lemma-Feinstein}, we know that there exists a $({P}_e^{(n)},M_{n})$-code  $\mathcal{C}_{n}$  whose average error probability satisfies 
\begin{align}
P_e^{(n)}(\mathcal{C}_{n}) & \leq  \,\inf\limits_{\gamma>0}\,  \Bigg\{ \Pr\Bigg[ \frac1{nK}\,\log\frac{\mathbb{P}_{Y^n|X^n,H^n}(dY^n| X^n,H^n)}{\widetilde{\mathbb{P}}_{n}(dY^n|H^n)}\nonumber\\
& \qquad\qquad\leq \frac1{nK}\log (\gamma \kappa_n) \Bigg] + \frac{M_{n}}{\gamma} \Bigg\} + \alpha_n
\label{eq-error-prob2}
\end{align}
for every $n=1,2,\dots$, where $\alpha_n$ is defined as in Lemma~\ref{lemma:yuri}.
Let us now set
\begin{equation}
	\frac1{nK} \, \log \gamma= \frac1{nK} \, \log M_n + \frac{\xi}{\sqrt{nK}} \label{eq-replacing2}
\end{equation}
for some $\xi>0$. Then, we have the following chain of inequalities:
\begin{align}
P_e^{(n)}(\mathcal{C}_{n})  &\leq \Pr\Bigg[ \frac1{nK}\,\log\frac{\mathbb{P}_{Y^n|X^n,H^n}(dY^n| X^n,H^n)}{ \widetilde{\mathbb{P}}_{ n} (d{Y}^n|H^n)  }  \nonumber\\
& \qquad\quad\leq \frac1{nK}\log \gamma +\frac1{nK}\log \kappa_n  \Bigg]  + \frac{M_{n}}{\gamma} +\alpha_n\\
  &= \Pr\Bigg[\frac1{nK}\,\log\frac{\mathbb{P}_{Y^n|X^n,H^n}(dY^n| X^n,H^n)}{ \widetilde{\mathbb{P}}_{ n} (d{Y}^n|H^n)  }\nonumber\\
  &  \qquad\quad\leq \frac1{nK} \, \log M_n + \frac{\xi}{\sqrt{nK}} +\frac1{nK}\log \kappa_n \Bigg]  \nonumber \\
&\qquad+ \exp(-\sqrt{nK}\xi) +\alpha_n 
\label{eq-replacing3}
\end{align}
which simply follows by replacing \eqref{eq-replacing2} in \eqref{eq-error-prob2}. For some $r$ real, we choose the coding rate 
\begin{equation}
\frac1{nK} \log M_n =  C + \frac{r}{\sqrt{nK}}.
\end{equation}
By combining \eqref{eq-error-prob2} and \eqref{eq-replacing3}, taking the superior limit on $n$, then $\xi\downarrow 0$ on the RHS, and the infimum over the codes on the LHS, we obtain
\begin{align}
&\mathbb{P}_e(r|\beta,c) \leq   
\lim_{\xi\downarrow 0}  \limsup\limits_{n \xrightarrow{(\beta,c)}  \infty}\Pr\Bigg[ \sqrt{nK} \nonumber\\
&\times\left(\frac1{nK}\log\frac{\mathbb{P}_{Y^n|X^n,H^n}(d{Y}^n| {X}^n,H^n)}{\widetilde{\mathbb{P}}_{ n} (d{Y}^n|H^n)} - C \right) \leq r+\xi \Bigg]\label{eq-2orderFuncUp2}
\end{align}
where we used $\alpha_n\to 0$ while $\kappa_n\to\infty$, such that $\frac1{\sqrt{nK}}\log\kappa_n\to 0$. This concludes the proof. 
\end{IEEEproof}

\section{Proof of Theorem~\ref{th-main}}\label{app:proof_th-main}
The proof relies on information spectrum methods \cite{HAN03} and is more exactly related to Hayashi's proof-techniques used in \cite{HAY09}. Our starting point is Proposition~\ref{prop-bounds_2nd_order} in Appendix~\ref{app:information_spectrum} which relates the optimal average error probability $\mathbb{P}_e(r|\beta,c)$ to the statistics of the mutual information density. 

The main problem in studying the optimal average error probability lies in the difficulty to perform any analytical calculus on the information spectrum of $\mathbb{P}_{Y^n|X^n,H^n}$, unless the underlying distributions (of $X^n$, $Y^n|\Xm^n,H^n$, or $Y^n|H^n$) are Gaussian. Proposition~\ref{prop-bounds_2nd_order} precisely handles this difficulty. Indeed, first note that the lower bound \eqref{eq-2orderFuncLow} can be further bounded by the same expression with $\mathbb{Q}_{n,+}$ chosen to be Gaussian with appropriate mean and variance. As for \eqref{eq-2orderFuncUp}, it already features an information spectrum of Gaussian distributions. Both lower and upper bounds will thus rely on exploiting Theorem~\ref{thm:CLT} with the major difference that, while the upper bound from \eqref{eq-2orderFuncUp} provides a definite choice for $\mathbb{P}_{X^n}$ that allows for an accurate control of the variance $\theta_n$ of Theorem~\ref{thm:CLT}, \eqref{eq-2orderFuncLow} does not and will force us to consider the worst case scenario where $\frac1K\trace (\Am^n)^2=0$, with $\Am^n=\Id_N-\frac1n\Xm^n(\Xm^n)\htp$. As briefly discussed in Section~\ref{sec:main_results}, the term $(\Am^n)^2$ appears due to the randomness in the channel $H^n$, leaving the problem of non-matching upper and lower bounds; this is unlike the previously studied AWGN scenarios (e.g., \cite{HAY09,POL10}) where $H^n=\Id_N$ and only terms in $\Am^n$ but not $(\Am^n)^2$ account for the second-order statistics.

\subsection{Proof of the lower bound on the optimal average error probability}\label{app:proof-LB}
From \eqref{eq-2orderFuncLow},
\begin{align}	
	&\mathbb{P}_e(r|\beta,c) \geq \inf_{ \substack{ \{ \mathbb{P}_{X_+^n}\}_{n=1}^{\infty} \\ \mathbb{P}_{X_+^n} \in \mathcal{P}(\mathcal{S}^{n+1}_{=}) }}\lim_{\xi\downarrow 0}\limsup\limits_{n \xrightarrow{(\beta,c)}  \infty}\Pr\Bigg[ \sqrt{nK}\nonumber\\
	&\times\left(\frac1{nK}\log\frac{\mathbb{P}_{Y_+^n|X_+^n,H^n}(d{Y}_+^n| {X}_+^n,H^n)}{\mathbb{Q}_{n,+}(d{Y}_+^n)} - C \right) \leq r-\xi \Bigg]
\end{align}
where, for fixed $\Hm^n$, $\mathbb{Q}_{n,+}$ is taken to be complex Gaussian with zero mean and covariance matrix $\frac1K\Hm^n(\Hm^n)\htp + \sigma^2\Id_N$. 
Thus,
\begin{align}
	\label{eq-Pe_inf}
	&\mathbb{P}_e(r|\beta,c) \geq \inf_{ \{\mathbb{P}_{X_+^n}\in \mathcal P(\mathcal S_=^{n,+})\}_{n=1}^\infty  } \lim_{\xi\downarrow 0}\nonumber\\
	&\qquad  \limsup_{n \xrightarrow{(\beta,c)}  \infty} \Pr\left[ \sqrt{nK}\left( I_{N,K}^{X_+^n} - C\right) \leq r-\xi\right]	
\end{align} 
where $I_{N,K}^{X_+^n}$ is defined in \eqref{def-INK_1+} on the next page
\begin{figure*}
\begin{align}
	\label{def-INK_1+}
	I_{N,K}^{X_+^n} \defines& \frac{1}{K} \log\det\LB\Id_N+\frac{1}{\sigma^2}\frac{H^n(H^n)\htp}{K}\RB \nonumber\\
	&+ \frac{1}{nK} \trace \LSB\LB \frac{ H^n(H^n)\htp}{K} + \sigma^2\Id_N\RB^{-1}\LB \frac{H^n}{\sqrt{K}}X_+^n+\sigma W_+^n\RB\LB \frac{H^n}{\sqrt{K}}X_+^n+\sigma W_+^n\RB\htp - W_+^n(W_+^n)\htp\RSB
\end{align}
\hrulefill
\end{figure*}
and where $W_+^n\in\mathds{C}^{N\times(n+1)}$ is composed of i.i.d.\@ $\Cc\Nc(0,1)$ elements.

To proceed, we now call Theorem~\ref{thm:CLT} for the random variable $I_{N,K}^{X_+^n}$. Let $\{X_+^n\}_{n=1}^\infty$ be a sequence with $X_+^n$ random with support in $\mathcal S_=^{n,+}$ for each $n$. Denoting ${A}_+^n=\Id_K-\frac1{n+1}{X}_+^n({X}_+^n)\htp$, for any real $z$, as $n \xrightarrow{(\beta,c)} \infty$, we then have
\begin{align}
	\Pr\left[ \frac{\sqrt{nK}}{\tilde\theta_{n,+}}\left( I_{N,K}^{{X}_+^n} - C\right)\leq z\right] \to \Phi(z)
\end{align}
where $\tilde\theta_{n,+}^2\defines \frac1{1+n^{-1}}(\theta_-^2+\zeta \frac1K\trace (A_+^n)^2 - \frac1K\log (1-c^{-1}\delta_0(\sigma^2)(1+\delta_0(\sigma^2))^{-2}) - \frac1K \zeta\beta^{-1})$, in which the terms in $K^{-1}$ or $n^{-1}$ arise from accounting for the fact that $X_+^n\in\mathds{C}^{K\times (n+1)}$ and $W^n_+\in \mathds{C}^{N\times (n+1)}$. But since $- \frac1K\log (1-c^{-1}\delta_0(\sigma^2)(1+\delta_0(\sigma^2))^{-2}) - \frac1K \zeta\beta^{-1}\to 0$ as $n \xrightarrow{(\beta,c)} \infty$, we have more simply by Slutsky's lemma
\begin{align}
	\Pr\left[ \frac{\sqrt{nK}}{\theta_{n,+}}\left( I_{N,K}^{{X}_+^n} - C\right)\leq z\right] \to \Phi(z)
\end{align}
with $\theta_{n,+}^2\defines \theta_-^2+\zeta \frac1K\trace (A_+^n)^2$.

We can now write
\begin{align}
	&\Pr\left[ \sqrt{nK}\left( I_{N,K}^{X_+^n} - C\right) \leq r-\xi\right]\nonumber\\
	 &= \Pr\left[ \frac{\sqrt{nK}}{\theta_{n,+}}\left( I_{N,K}^{X_+^n} - C\right) \leq \frac{ r-\xi }{\theta_{n,+}}\right] \\
	&\overset{(a)}{\geq} 
	\begin{cases} 
		\Pr\left[\frac{\sqrt{nK}}{\theta_{n,+}}\left( I_{N,K}^{\bar{X}_+^n} - C\right) \leq \frac{r-\xi}{\theta_-}\right]~ &,\ r\leq 0\\
		\Pr\left[\frac{\sqrt{nK}}{\theta_{n,+}}\left( I_{N,K}^{\bar{X}_+^n} - C\right) \leq 0\right]~ &,\ r>0
	\end{cases}\\\label{eq:tmp_001}
	&=
	\begin{cases} 
		\Phi\left(\frac{r-\xi}{\theta_-} \right) + \ell_n~  &,\ r\leq 0 \\ 
		\frac12 + \ell_n~ &, \ r>0 
	\end{cases}
\end{align}
for some sequence $\ell_n\downarrow 0$, where $(a)$ holds since $\theta_{n,+}\geq \theta_->0$ and since we took $r-\xi>0$ for $r>0$. The term $1/2$ arises from $\Phi(0)=1/2$ which originates from $\theta_n$ not being bounded from above since $\frac1K\trace (A_+^n)^2$ can grow like $\mathcal O(n)$. 

Taking the limit superior as $n\xrightarrow{(\beta,c)} \infty$ of the above equation leads to
\begin{align}
	&\limsup\limits_{n \xrightarrow{(\beta,c)}  \infty} \Pr\left[ \sqrt{nK}\left( I_{N,K}^{X_+^n} - C\right) \leq r-\xi \right]\nonumber\\
	&\geq 
	\begin{cases} 
		\Phi\left(\frac{r-\xi}{\theta_{-}}\right) &,\ r\leq 0 \\ 
		\frac12  &, \ r> 0 .
	\end{cases}
\end{align}
By continuity of $\Phi$, we can freely take the limit $\xi\downarrow 0$ on the right- then left-hand sides to obtain 
\begin{align}
	\label{eq-lowerbound}
	&\lim_{\xi\downarrow 0}\limsup\limits_{n \xrightarrow{(\beta,c)}  \infty} \Pr\left[ \sqrt{nK}\left( I_{N,K}^{X_+^n} -C \right) \leq r-\xi \right]\nonumber\\
	 &\geq 
	\begin{cases} 
		\Phi\left(\frac{r}{\theta_{-}}\right) &,\ r<0 \\ 
		\frac12  &, \ r\geq 0 .
	\end{cases}
\end{align}

Equation \eqref{eq-lowerbound} is valid regardless of the choice of the sequence $\{\mathbb{P}_{X_+^n}\in \mathcal P(\mathcal S_=^{n,+}) \}_{n=1}^\infty$. This therefore implies
\begin{align}
	\mathbb{P}_e(r|\beta,c) \geq
	\begin{cases} 
		\Phi\left(\frac{r}{\theta_{-}}\right) &,\ r<0 \\ 
		\frac12  &, \ r\geq 0
	\end{cases}
\end{align}
which completes the proof.

\subsection{Proof of the upper bound on the optimal average error probability}\label{app:proof_th-main_up}
From \eqref{eq-2orderFuncUp}, we recall that
\begin{align}
	&\mathbb{P}_e(r|\beta,c) \leq 
\lim_{\xi\downarrow 0}  \limsup\limits_{n \xrightarrow{(\beta,c)}  \infty}\Pr\Bigg[ \sqrt{nK} \times\nonumber\\
&\left(\frac1{nK}\log\frac{\mathbb{P}_{Y^n|X^n,H^n}(d{Y}^n| {X}^n,H^n)}{  \widetilde{\mathbb{P}}_{n} (d{Y}^n|H^n)} - C \right) \leq r+\xi \Bigg] \label{eq:mainthmtmp1}
\end{align}
where $\widetilde{\mathbb{P}}_{n}(\cdot|H^n)$ is a Gaussian random variable with zero mean and covariance $H^n(H^n)\htp+\sigma^2\Id_N$ and the outer probability is taken over $H^n$ and over the random variable $X^n$ having uniform distribution ${\mathbb P}_{X^n}$ over the sphere $\mathcal S^n_=$, as per \eqref{eq-PD-uniform}. Denoting, similar to above,
\begin{align}
	I_{N,K}^{X^n} &= \frac1{nK}\log\frac{\mathbb{P}_{Y^n|X^n,H^n}(d{Y}^n| {X}^n,H^n)}{  \widetilde{\mathbb{P}}_{n} (d{Y}^n|H^n)}
\end{align}
we get from the Gaussianity of both $\mathbb{P}_{Y^n|X^n,H^n}$ and $\widetilde{\mathbb{P}}_{n}$ that $I_{N,K}^{X^n}$ is given by \eqref{def-INK_Xn} with $X^n$ of law ${\mathbb P}_{X^n}$, while $H^n$ and $W^n$ are zero mean Gaussian with (properly normalized) unit covariance.

Once again, we resort to Theorem~\ref{thm:CLT} to determine the limiting behavior of $I_{N,K}^{X^n}$. As opposed to the lower bound, where ${\mathbb P}_{X^n}\in \mathcal P(\mathcal S^n_=)$ was left undefined, ${\mathbb P}_{X^n}$ is now fixed and will allow for a more accurate control of the limiting variance of $I_{N,K}^{X^n}$. We first obtain
\begin{align}
	&\limsup\limits_{n \xrightarrow{(\beta,c)}  \infty} \Pr\left[ \sqrt{nK}\left( I_{N,K}^{X^n} - C \right) \leq r+\xi \right] \nonumber\\
	&= \limsup\limits_{n \xrightarrow{(\beta,c)}  \infty} \Pr\left[ \frac{\sqrt{nK}}{\theta_n}\left( I_{N,K}^{X^n} - C \right) \leq \frac{r+\xi}{\theta_n} \right]
\end{align}
where $\theta_n$ is defined in \eqref{eq:def-thetan} where we recall that $A^n=\Id_N-\frac1nX^n(X^n)\htp$. Now, it appears that 
\begin{align}
	\frac1K\trace \left[(A^n)^2\right] &\xrightarrow{(\beta,c)} \frac1{\beta}
\end{align}
almost surely. To obtain this result, it suffices to realize that $X^n=\bar{X}^n(\frac1{NK}\trace \bar{X}^n(\bar{X}^n)\htp)^{-\frac12}$ for $\bar{X}^n\in\CC^{K\times n}$ a standard Gaussian random matrix with entries of zero mean and unit variance; from classical random matrix results (that may be obtained by means of the Gaussian tools defined in Appendix~\ref{app:Gaussian_relResults}), we have that $\frac1{NK}\trace \bar{X}^n(\bar{X}^n)\htp\to 1$ while $\frac1{NK}\trace (\bar{X}^n(\bar{X}^n)\htp)^2\to 1+\beta^{-1}$, almost surely; plugging these results in the expression of $\frac1K\trace(A^n)^2$ gives the expected result. As such, we now have that $\theta_n \xrightarrow{(\beta,c)}\theta_+$ almost surely (and so in probability), with $\theta_+$ defined in Theorem~\ref{th-main}. By Slutsky's lemma and Theorem~\ref{thm:CLT}, we thus have
\begin{align}
	&\limsup\limits_{n \xrightarrow{(\beta,c)}  \infty} \Pr\left[ \sqrt{nK}\left( I_{N,K}^{X^n} - C \right) \leq r+\xi \right]\nonumber\\
	&= \limsup\limits_{n \xrightarrow{(\beta,c)}  \infty} \Pr\left[ \frac{\sqrt{nK}}{\theta_n}\left( I_{N,K}^{X^n} - C \right) \leq \frac{r+\xi}{\theta_+} \right] \\
	&= \Phi \left( \frac{r+\xi}{\theta_+} \right)
\end{align}
which, along with the fact that
\begin{align}
	\lim_{\xi\downarrow 0} \Phi\left( \frac{r+\xi}{\theta_+} \right) = \Phi\left( \frac{r}{\theta_+} \right)
\end{align}
concludes the proof.

\section{Gaussian tools and related results}\label{app:Gaussian_relResults}

The CLT, Theorem~\ref{thm:CLT}, relies on advanced tools from random matrix theory along with standard linear algebraic relations which are constantly called for. This section introduces the random matrix concepts and collects the aforementioned relations.

\begin{lemma}[Some matrix inequalities]\label{lem:matrixineq}
For two $N\times N$ matrices $\Am$ and $\Bm$, the following holds
\begin{align}
(i)\quad\left| \trace\Am\Bm\right| &\le \sqrt{\trace\Am\Am\htp \trace\Bm\Bm\htp}.
\end{align}
If $\Am$ is Hermitian nonnegative definite, it further holds that
\begin{align}
(ii)\quad\left|\trace \Am\Bm \right| &\le \left \lVert \Bm \right \rVert \trace \Am\\
(iii)\quad\frac1N\trace\Am &\le \left \lVert \Am \right \rVert.
\end{align}
\end{lemma}

\begin{lemma}[Cauchy-Schwarz inequality]\label{lem:cauchyschwarz}
For two complex random variables $x$ and $y$, 
\begin{align}
\left|\mathbb{E}\LSB x y\RSB \right| \le \sqrt{\mathbb{E}\LSB |x|^2\RSB} \sqrt{\mathbb{E}\LSB |y|^2\RSB}.
\end{align}
\end{lemma}

\begin{remark}[Application of the Cauchy-Schwarz inequality]\label{rem:cauchyschwarz}
Consider two random variables $x$ and $y$. By the Cauchy-Schwarz inequality, 
\begin{align}
\left|\mathbb{E}\LSB \LB x -\mathbb{E}\LSB x\RSB \RB\LB y - \mathbb{E}\LSB y\RSB\RB\RSB \right| \le \sqrt{\var[x]} \sqrt{\var[y]}.
\end{align}
 Thus,
\begin{align}
\left|\mathbb{E}\LSB xy\RSB\right| &= \left|\mathbb{E}\LSB x\RSB \mathbb{E}\LSB y\RSB  + \mathbb{E}\LSB \LB x -\mathbb{E}\LSB x\RSB \RB\LB y - \mathbb{E}\LSB y\RSB\RB\RSB \right|\\
& \le \left|\mathbb{E}\LSB x\RSB \mathbb{E}\LSB y\RSB \right| + \sqrt{\var[x]} \sqrt{\var[y]}. 
\end{align}
Moreover, it follows that
\begin{align}
&\var[x + y]\nonumber\\
&= \var[x] + \var[y] + 2 \Re \left\{\mathbb{E}\LSB\LB x -\mathbb{E}\LSB x\RSB \RB\LB y - \mathbb{E}\LSB y\RSB\RB\RSB \right\} \\
&\le \var[x] + \var[y] + 2\sqrt{\var[x]} \sqrt{\var[y]}\\
&= \LB\sqrt{\var[x]} + \sqrt{\var[y]}\RB^2.
\end{align}
\end{remark}

\begin{lemma}[{Integration by parts formula \cite[Equation~(2.1.42)]{pastur2011eigenvalue}}]\label{lem:IBP}
 Let $x=\LSB x_1,\dots, x_N\RSB\tp\sim\Cc\Nc\LB\zerov,\Rm\RB$ and let $f(x) = f\LB x_1,\dots x_N,x_1^*,\dots x_N^*\RB$
be a $\Cc^1$ complex function, polynomially bounded together with its derivatives. Then, 
\begin{align}
\mathbb{E}\LSB x_i f(x)\RSB = \sum_{j=1}^N \Rm_{ij}\mathbb{E}\LSB \frac{\partial f(x)}{\partial x_j^*}\RSB.
\end{align}
\end{lemma}

\begin{remark}[{Integration by parts formula for functionals of matrices with i.i.d. entries}]\label{rem:IBP}
Let $f\LB W \RB$ be a $\Cc^1$ complex function of the elements of $W$ and $W^*$, polynomially bounded together with its derivatives, where $W$ has i.i.d.\@ entries $W_{ij}\sim\Cc\Nc(0,1)$. Then
\begin{align}
\mathbb{E}\LSB W_{ij}f\LB W\RB\RSB = \mathbb{E}\LSB \frac{\partial f\LB W\RB}{\partial W^*_{ij}}\RSB.
\end{align}
\end{remark}

\begin{lemma}[{Poincar\'e-Nash Inequality \cite[Propostion~2.1.6]{pastur2011eigenvalue}}] \label{lem:PNI}
Let $x$ and $f(x)$ be defined as in Lemma~\ref{lem:IBP} and let $\nabla_x f(x) = \LSB \partial f(x)/\partial x_1,\dots,\partial f(x)/\partial x_N\RSB\tp$ and  $\nabla_{x^*} f(x) = \LSB \partial f(x)/\partial x^*_1,\dots,\partial f(x)/\partial x^*_N\RSB\tp$. Then,
\begin{align}
\var\LSB f(x)\RSB &\le \mathbb{E}\LSB\nabla_x f(x)\tp \Rm \nabla_x f(x)^* \RSB\nonumber\\
& \qquad+ \mathbb{E}\LSB\nabla_{x^*} f(x)\htp \Rm \nabla_{x^*} f(x) \RSB.
\end{align}
\end{lemma}

\begin{remark}[Poincar\'e-Nash Inequality for functionals of matrices with i.i.d. entries]\label{rem:PNI}
Let $f\LB W \RB$ be a function of the elements of $W$ and $W^*$ as in Remark~\ref{rem:IBP}, where $W\in\mathds{C}^{N\times n}$ has i.i.d.\@ entries $W_{ij}\sim\Cc\Nc(0,1)$. Then,
\begin{align}
\var\LSB f\LB W \RB\RSB \le \sum_{i=1}^N\sum_{j=1}^n \mathbb{E}\LSB \left| \frac{\partial f\LB W\RB}{\partial W_{ij}}\right|^2 + \left| \frac{\partial f\LB W\RB}{\partial W^*_{ij}}\right|^2\RSB.
\end{align}
\end{remark}

\begin{lemma}[Identities for Complex Derivatives]\label{lem:useful_derivatives}
Let $\Hm \in\mathds{C}^{N\times K}$. Then,
\begin{align}
\frac{\partial \Hm _{pq}}{\partial \Hm _{ij}^*} &= 0\\
\frac{\partial \Hm _{pq}}{\partial \Hm _{ij}} &= \delta_{ip}\delta_{jq}\\
\frac{\partial \LSB \Hm  \Hm \htp\RSB_{pq}}{\partial \Hm _{ij}^*} &= \delta_{iq}\Hm _{pj}\\
\frac{\partial \LSB \Hm \Hm \htp\RSB_{pq}}{\partial \Hm _{ij}} &= \delta_{ip}\Hm ^*_{qj}\\
\frac{\partial \LSB \Hm \htp \Hm \RSB_{pq}}{\partial \Hm _{ij}^*} &= \delta_{jp}\Hm _{iq}\\
\frac{\partial \LSB \Hm \htp \Hm \RSB_{pq}}{\partial \Hm _{ij}} &= \delta_{jq}\Hm ^*_{ip}.
\end{align}

Moreover, denote $\Qm   = \LB \frac1K \Hm \Hm \htp + x\Id_N\RB^{-1}$ and $\tilde{\Qm  } = \LB \frac1K \Hm \htp \Hm  + x\Id_K\RB^{-1}$ for some $x>0$. Then,
\begin{align}
\frac{\partial \Qm  _{pq}}{\partial \Hm ^*_{ij}} &= -\frac1K[ \Qm  \Hm ]_{pj}\Qm  _{iq}\\
\frac{\partial \Qm  _{pq}}{\partial \Hm _{ij}} &= -\frac1K[\Hm \htp \Qm  ]_{jq} \Qm  _{pi} \\
\frac{\partial \tilde \Qm  _{pq}}{\partial \Hm ^*_{ij}} &= -\frac1K \tilde \Qm  _{pj} [\Hm \tilde \Qm  ]_{iq}\\
\frac{\partial \tilde \Qm  _{pq}}{\partial \Hm _{ij}} &= -\frac1K \tilde{\Qm  }_{jq} [\tilde \Qm   \Hm \htp]_{pi}.
\end{align}
\end{lemma}

\begin{cor}\label{cor:derivatives}
Let $\Hm  \in\mathds{C}^{N\times K}$ and $\Cm\in\mathds{C}^{N\times N}$. Denote $\Qm   = \LB \frac1K \Hm  \Hm  \htp + x\Id_N\RB^{-1}$ for some $x>0$. Then,
\begin{align}
(i) \qquad\quad \trace \frac{\partial \Qm  }{\partial \Hm  _{ij}^*}\Cm &= -\frac{1}{K}\LSB  \Qm  \Cm \Qm   \Hm  \RSB_{ij}\\
(ii)\quad \trace\frac{\partial\LB \Hm  \Hm  \htp\RB}{\partial \Hm  _{ij}^*}\Cm &= \LSB \Cm \Hm  \RSB_{ij}.
\end{align}
\end{cor}
\begin{IEEEproof}
The proof follows directly from Lemma~\ref{lem:useful_derivatives} and some straightforward calculus.
\end{IEEEproof}

\begin{theorem}\label{thm:detequ_stieltjes}
Let $\{ H^n\}_{n=1}^\infty$, where $H^n\in\mathds{C}^{N\times K}$ has i.i.d.\@ entries $H^n_{ij}\sim\Cc\Nc\LB 0,1\RB$. For $u>0$, let $Q^n(u)=\LB \frac1K H^n\LB H^n\RB\htp + u\Id_N\RB^{-1}$ and $\tilde{Q}^n(u)=\LB \frac1K \LB H^n\RB\htp H^n + u\Id_K\RB^{-1}$. Then, as $n\xrightarrow[]{(\beta,c)}\infty$,
\begin{align}
\mathbb{E}\LSB\frac1K \trace Q^n(u)\RSB &= \delta_0\LB u\RB + \Oc\LB \frac{1}{u^4n^2}\RB\\
\mathbb{E}\LSB\frac1K \trace\tilde{Q}^n(u)\RSB &= \tilde{\delta}_0\LB u\RB + \Oc\LB \frac{1}{u^4n^2}\RB
\end{align}
where
\begin{align}
\delta_0(u) &= \frac{c-1}{2u}-\frac{1}{2} + \frac{\sqrt{(1-c+u)^2+4cu}}{2u}\\
\tilde{\delta}_0(u) &= \delta_0(u) - \frac{c-1}{u}.
\end{align}
\end{theorem}
\begin{IEEEproof}
	The proof follows from a direct adaption of \cite[Theorem~7.2.2]{pastur2011eigenvalue} (see also \cite[Theorem~3 and Proposition~5]{hachem2008new} for a more complex matrix model) along with a careful control of the dependence on $u$ in the bounds. 
\end{IEEEproof}

\begin{remark}\label{rem:stieltjes}
 The function $s(z)=\frac{\delta_0(-z)}{c}$ for $z\in\mathds{C}\setminus\mathds{R}_+$ corresponds to the Stieltjes transform of the Mar\u{c}enko-Pastur law, see e.g., \cite[Chapter 3.2]{COUbook}.
\end{remark}

\begin{property}[Some properties of $\delta_0(u)$]\label{properties:delta0}
The function $\delta_0(u)$, $u>0$, as defined in Theorem~\ref{thm:detequ_stieltjes} satisfies 
\begin{align}
	(i)\quad\quad\quad \delta_0(u) &> \frac{c}{(1+\sqrt{c})^2+u} > 0 \\
(ii)\quad\quad\quad \delta_0(u) &< \frac{c}{u}\\
(iii)\quad\quad\quad \delta_0(u) &= \frac{c}{1 - c + u\LB 1 + \delta_0(u)\RB}\\
(iv)\quad \frac{\delta_0(u)}{1+\delta_0(u)} &= c -u\delta_0(u)\\
(v)\quad \frac1{1+\delta_0(u)} &= 1-c + u\delta_0(u) \\
(vi)\quad\quad\quad \delta_0'(x) &= - \frac{\delta_0(x)(1+\delta_0(x))}{1-c+x(1+2\delta_0(x))}\,.
\end{align}
\end{property}
\begin{IEEEproof}
	Properties $(i)$--$(iii)$ are due to $\delta_0(u)=cm(-u)$, where $m(z)$ is the Stieltjes transform of the Mar\u{c}enko-Pastur law with support in $[(1-\sqrt{c})^2,(1+\sqrt{c})^2]\cup \{0\}$ (see Remark~\ref{rem:stieltjes} in Appendix~\ref{app:Gaussian_relResults}). Property $(iv)$ follows from $(iii)$ since
\begin{align}
 \delta_0(u) &= \frac{c}{1 - c + u\LB 1 + \delta_0(u)\RB}\\ 
\Longleftrightarrow\qquad \delta_0(u) &= \LB 1+\delta_0(u) \RB c - u\delta_0(u)\LB  1+\delta_0(u)\RB\\ 
\Longleftrightarrow \frac{\delta_0(u)}{1+\delta_0(u)} &= c -u\delta_0(u).
\end{align}
Property~$(v)$ follows from $(iii)$ and $(iv)$. Property~$(vi)$ is obtained from the differentiation of
\begin{equation}
c = \delta_0(x)(1-c+x)+x\delta_0(x)^2
\end{equation}
which follows from Property~$(iii)$.
\end{IEEEproof}

\begin{lemma}\label{lem:integrals} Let $\sigma^2,c>0$ and $\delta_m(x), m\ge 0$, be as defined in Proposition~\ref{prop:dettrace} in Appendix~\ref{app:aux_RMTaux}. Then,
  \begin{align}
	(i) & \int_{\sigma^2}^\infty c \frac{1-c+2u\delta_0(u)-\frac{u^2}c\delta_0(u)^2}{u(1-c+u(1+2\delta_0(u)))}du = \log(1+\delta_0(\sigma^2))\nonumber\\
	&\qquad-\frac{\delta_0(\sigma^2)}{1+\delta_0(\sigma^2)}+c\log\left(1+\frac1{\sigma^2}\frac1{1+\delta_0(\sigma^2)}\right)\\
	(ii) & \int_{\sigma^2}^\infty \frac{\delta_0(u)-\sigma^2\delta_1(u)}{1-c+u(1+2\delta_0(u))}du \nonumber\\
	&= -\log\left( 1 - \frac1c \frac{\delta_0(\sigma^2)^2}{(1+\delta_0(\sigma^2))^2} \right).
\end{align}
\end{lemma}
\begin{IEEEproof}
For the proof of part $(i)$, simply note that
\begin{align}
	&c \frac{1-c+2u\delta_0(u)-\frac{u^2}c\delta_0(u)^2}{u(1-c+u(1+2\delta_0(u)))}\nonumber\\
	&= \frac{c}u - \frac{u\delta_0(u)^2+c}{1-c+u(1+2\delta_0(u))} \\
	&= \frac{c}u - \frac{u\delta_0(u)^2+c}{u\delta_0(u)+\frac{c}{\delta_0(u)}} \\
	&= \frac{c}u - \delta_0(u)
\end{align}
where we used Property~\ref{properties:delta0}~$(iii)$ in the second equality. The result then unfolds from Theorem~\ref{thm:detequ_mutinf}.

For part $(ii)$, we start with the following calculus:
\begin{align}
	&\int_{\sigma^2}^\infty \frac{\delta_0(u)-\sigma^2\delta_1(u)}{1-c+u(1+2\delta_0(u))}du \nonumber\\
	&= \int_{\sigma^2}^\infty \Bigg[\frac{\delta_0(u)}{1-c+u(1+2\delta_0(u))}+\nonumber\\
	&\frac{\sigma^2\delta_0(u)(1+\delta_0(\sigma^2))}{(1-c+\sigma^2(1+\sigma^2)+u\delta_0(u))(1-c+u(1+2\delta_0(u)))} \Bigg]du \\
	&= \int_{\sigma^2}^\infty \Bigg[-\frac{\delta_0(u)\delta_0'(u)}{\delta_0(u)(1+\delta_0(u))}\nonumber\\
	&\qquad +\frac{\sigma^2\delta_0'(u)(1+\delta_0(\sigma^2))}{1+\sigma^2(1+\delta_0(\sigma^2))+\delta_0(u)\sigma^2(1+\delta_0(\sigma^2))} \Bigg]du
\end{align}
where in the first equality we developed the expression of $\delta_1(u)$ and in the second equality we introduced $\delta_0'(u)$ in both numerators and used the relation by iterating the relation $x\delta_0(x)^2=c-\delta_0(x)(1-c+x)$ (from Property~\ref{properties:delta0}~$(iii)$) in the second denominator in order to maintain a degree one polynomial in $\delta_0(u)$. Writing $\delta_0(u)\delta_0'(u) = [2\delta_0(u)\delta_0'(u)+\delta_0'(u)]-\delta_0'(u)(1+\delta_0(u))$ in the numerator of the first term, we then find
\begin{align}
	& \int_{\sigma^2}^\infty \frac{\delta_0(u)-\sigma^2\delta_1(u)}{1-c+u(1+2\delta_0(u))}du \nonumber\\
	&= \int_{\sigma^2}^\infty \Bigg[-\frac{2\delta_0(u)\delta_0'(u)+\delta_0'(u)}{\delta_0(u)(1+\delta_0(u))} + \frac{\delta_0'(u)}{\delta_0(u)} \nonumber\\
	&\quad +\frac{\sigma^2\delta_0'(u)(1+\delta_0(\sigma^2))}{1+\sigma^2(1+\delta_0(\sigma^2))+\delta_0(u)\sigma^2(1+\delta_0(\sigma^2))} \Bigg]du \\
	&= \Big[ -\log(1+\delta_0(u))  \nonumber\\
	&\qquad +\log(1+\sigma^2(1+\delta_0(\sigma^2))(1+\delta_0(u))) \Big]_{u=\sigma^2}^\infty \\
	&= \log(1+\delta_0(\sigma^2)) + \log(1+\sigma^2(1+\delta_0(\sigma^2)))\nonumber\\
	&\qquad -\log(1+\sigma^2(1+\delta_0(\sigma^2))^2) \\
	&= \log\left( \frac{(1+\delta_0(\sigma^2))(1+\sigma^2(1+\delta_0(\sigma^2)))}{1+\sigma^2(1+\delta_0(\sigma^2))^2} \right).
\end{align}

At this point, remark that
\begin{align}
	&\frac{(1+\delta_0(\sigma^2))(1+\sigma^2(1+\delta_0(\sigma^2)))}{1+\sigma^2(1+\delta_0(\sigma^2))^2}\nonumber\\
	&= 1 - \frac{\delta_0(\sigma^2)}{1+\sigma^2(1+\delta_0(\sigma^2))^2}
\end{align}
and that 
\begin{align}
1+\sigma^2(1+\delta_0(\sigma^2))^2 &= 1 + \sigma^2 + \sigma^2\delta_0(\sigma^2) + c + c\delta_0(\sigma^2) \\
&= \frac{c}{\delta_0(\sigma^2)} + 2c + c\delta_0(\sigma^2) \\
&= c\frac{(1+\delta_0(\sigma^2))^2}{\delta_0(\sigma^2)}
\end{align}
using Property~\ref{properties:delta0}~$(iii)$ in the second equality.

This allows us to finally conclude that
\begin{align}
  &\int_{\sigma^2}^\infty \frac{\delta_0(u)-\sigma^2\delta_1(u)}{1-c+u(1+2\delta_0(u))}du\nonumber\\
  &= -\log\left( 1 - \frac1c \frac{\delta_0(\sigma^2)^2}{(1+\delta_0(\sigma^2))^2} \right).
\end{align}
\end{IEEEproof}

\section{Proofs of the main random matrix results}
\label{app:RMT_results}
In the proof of Theorem~\ref{thm:CLT}, we fundamentally rely on the fact that the random matrices $W^n$ and $H^n$ are Gaussian by assumption. This allows us to use the powerful integration-by-parts and Poincar\'e--Nash inequalities (Lemma~\ref{lem:IBP} and Lemma~\ref{lem:PNI} in Appendix~\ref{app:Gaussian_relResults}) to compute the expectation and bound the variance of functionals of Gaussian variables. The derivation of Theorem~\ref{thm:CLT}  is specifically based on the characteristic function approach as explained in great detail in \cite{hachem2008new,pastur2011eigenvalue}.

This appendix is structured as follows: In Appendix~\ref{app:prelim}, we introduce some additional notations and useful identities. We then prove Theorem~\ref{thm:CLT} in Appendix~\ref{proof:CLT}. 

\subsection{Preliminaries}\label{app:prelim}
For readability, we often drop the index $n$ in matrix notations when there is no confusion, e.g., we write $H$ instead of $H^n$. 

We start with the definition of two matrices, the so-called ``resolvents" of $K^{-1}HH\htp$ and $K^{-1}H\htp H$, respectively, which will be of repeated use:
\begin{align}\label{eq:defQ}
 Q(x)&=\left(\frac1K HH\htp+x\Id_N\right)^{-1}\in\mathds{C}^{N\times N}\\\label{eq:defQtilde}
\tilde{Q}(x)&=\left(\frac1K H\htp H+x\Id_K\right)^{-1}\in\mathds{C}^{K\times K}
\end{align}
for $x>0$. One can easily verify that:
\begin{align}\label{eq:Qrelation}
 Q(x)\frac{HH\htp}K = \Id_N - xQ(x),\quad  \tilde{Q}(x)\frac{H\htp H}K = \Id_K - x \tilde{Q}(x).
\end{align}
We will also rely several times on the following identities:
\begin{align}
Q(x)H &= H\tilde{Q}(x), \qquad\ \ \tilde{Q}(x)H\htp = H\htp Q(x)\\
Q(x)\frac{HH\htp}{K} &= \frac{HH\htp}{K}Q(x),\quad \tilde{Q}(x)\frac{H\htp H}{K}=\frac{H\htp H}{K}\tilde{Q}(x)\\
Q(x)Q(y) &= Q(y)Q(x),\quad \tilde{Q}(x)\tilde{Q}(y) = \tilde{Q}(y)\tilde{Q}(x).
\end{align}
Using the above relations, it is easy to prove the following bounds on the spectral norm:
\begin{align}
\left \lVert Q(x) \right\rVert = \left \lVert \tilde{Q}(x) \right\rVert &\le \frac 1x\\\label{eq:Qbound1}
\left \lVert Q(x)\frac{HH\htp}{K} \right\rVert= \left \lVert \tilde{Q}(x)\frac{H\htp H}{K} \right\rVert &\le 1.
\end{align}

\subsection{Proof of Theorem~\ref{thm:CLT}}
\label{proof:CLT}

\subsection*{\underline{Outline of the proof:}}
The central object of Theorem~\ref{thm:CLT}  is the real quantity
\begin{align}
&\Gamma_{n} \triangleq \sqrt{nK}I_{N,K}^{X_+^n}\\
	 &= \sqrt{\frac{n}K}\log\det\left(\Id_N + \frac1{\sigma^2}\frac1K H H\htp \right)\nonumber\\
	 & + \frac1{\sqrt{nK}}\trace Q(\sigma^2) \left(\frac1{\sqrt{K}}HX+\sigma W \right)\left(\frac1{\sqrt{K}}HX+\sigma W \right)\htp\nonumber\\
	 & - \frac1{\sqrt{nK}}\trace WW\htp
\end{align}
where $I_{N,K}^{X_+^n}$ was defined in \eqref{def-INK_1+}. We also recall the dimensions $H\in\mathds{C}^{N\times K}$, $X\in\mathds{C}^{K\times n}$, and $W\in\mathds{C}^{N\times n}$. Moreover, $X\in\Sc^n_=$, where $\Sc^n_=$ was defined in \eqref{eq:power_equal}.

It is our goal to prove that, under the hypotheses of the theorem,
\begin{align}\label{eq:thildephiconv}
 \tilde{\phi}_n(t) \defines \mathbb{E}\LSB e^{\frac{\iv t}{\theta_n}\LB \Gamma_n - \mu_n\RB}\RSB \to e^{-\frac{t^2}{2}}
\end{align}
for $t\in\mathds{R}$ as $n\xrightarrow[]{(\beta,c)}\infty$, where $\mu_n \defines \sqrt{nK} C$.
This will imply, by L\'evy's continuity theorem \cite[Theorem~16.3]{BIL95}, that
\begin{align}
	\theta_n^{-1} \left( \Gamma_n-\mu_n \right) \Rightarrow \Nc(0,1)
\end{align}
which is equivalent to the statement of the theorem. The main difficulty arises from the evaluation of the expectation in \eqref{eq:thildephiconv} which must be taken with respect to the three random matrices $W$, $H$, and $X$. Since the direct computation of $\tilde{\phi}_n(t)$ is intractable, we calculate its derivative with respect to $t$, leading to a differential equation which must be integrated. In order to further simplify the analysis, we split the computation of the expectation in three steps by successively considering the conditional expectations with respect to each of the matrices. These expectations are developed by the integration by parts formula (Lemma~\ref{lem:IBP} in Appendix~\ref{app:Gaussian_relResults}) which yields terms that are either further developed or  shown to be asymptotically negligible by bounding their variance with the help of the Poincar\'{e}-Nash inequality (Lemma~\ref{lem:PNI} in Appendix~\ref{app:Gaussian_relResults}). The analysis makes use of several auxiliary results summarized in Appendix~\ref{app:Gaussian_relResults}. In more detail, the proof consists of the following three main steps:
\begin{enumerate}
	\item We first take the expectation over $W$ by fixing $\Xm\in \mathcal S^n_=$ and $\Hm\in\mathds{C}^{N\times K}$: we define the function $\phi^{\Xm^n,\Hm^n}_n(t)\defines \mathbb{E}\LSB e^{\iv t \Gamma_n^{\Xm^n,\Hm^n}}\RSB$, where $\Gamma_n^{\Xm^n,\Hm^n}$ is the random variable $\Gamma_n$ taken for fixed $H=\Hm_n$ and $X=\Xm^n$, and show that
\begin{align}\label{eq:eqdiffXH}
 &\frac{\partial \phi^{\Xm^n,\Hm^n}_n(t)}{\partial t}=\nonumber\\
  & \LB \iv \mu_n^{\Xm^n,\Hm^n} - t \LB \theta_n^{\Xm^n,\Hm^n}\RB^2 + \iv t^2 \kappa_n^{\Xm^n,\Hm^n}  \RB \phi^{\Xm^n,\Hm^n}_n(t)\nonumber\\
  &\qquad + \bar{\varepsilon}_n^{\Xm^n,\Hm^n}(t)
\end{align}
for some $\mu_n^{\Xm^n,\Hm^n}=\Oc(n)$, $\theta_n^{\Xm^n,\Hm^n}=\Oc(1)$, $\kappa_n^{\Xm^n,\Hm^n}=\Oc(n^{-1})$, and $\bar{\varepsilon}_n^{\Xm^n,\Hm^n}(t)=\Oc(n^{-2})$ which must be carefully controlled. This establishes a differential equation for $\phi^{\Xm^n,\Hm^n}_n(t)$ the solution of which allows us to obtain an estimate of $\phi^{\Xm^n,\Hm^n}_n(t)$ under the form $e^{f(t,\Xm,\Hm)}$ (i.e., with no expectation over $W$).\footnote{Note importantly that, although the term $\kappa_n^{\Xm^n,\Hm^n}$ is of order $\Oc(n^{-1})$ and will not play a role at the end of the calculus, it needs to be isolated and not contained into $\bar{\varepsilon}_n^{\Xm^n,\Hm^n}(t)$ as the estimation error $\phi^{\Xm^n,\Hm^n}_n(t)-e^{f(t,\Xm,\Hm)}$, which is of the same order of magnitude as $\bar{\varepsilon}_n^{\Xm^n,\Hm^n}(t)$, will increase by a factor $n$ when we take its expectation over $\Hm$ (this is due to $\mu_n^{\Xm^n,\Hm^n}$ being of order $\Oc(n)$).} 

\item We then compute the expectation over $H$: we introduce the function $\phi^{\Xm^n}_n(t)\defines \mathbb{E}\LSB \phi^{\Xm^n,H^n}_n(t)\RSB$. Working mainly with the tractable estimator $e^{f(t,\Xm,\Hm)}$ of $\phi^{\Xm^n,\Hm^n}_n(t)$ as developed in step~1), instead of $\phi^{\Xm^n,\Hm^n}_n(t)$ itself, we prove in a similar fashion that
\begin{align}\label{eq:eqdiffX}
  \frac{\partial \phi^{\Xm^n}_n(t)}{\partial t} = \LB \iv \mu_n^{\Xm^n} - t \LB\theta_n^{\Xm^n}\RB^2\RB \phi^{\Xm^n}_n(t) + \varepsilon_n^{\Xm^n}(t)
\end{align}
for some $\mu_n^{\Xm^n}=\Oc(n)$, $\theta_n^{\Xm^n}$, and $\varepsilon_n^{\Xm^n}(t)=\Oc(n^{-1})$. 
This establishes a second differential equation.

\item We finally integrate \eqref{eq:eqdiffX} and show that
\begin{align}\label{eq:phitildeconv}
	\tilde{\phi}_n^{\Xm^n}\defines \mathbb{E}\LSB e^{\iv \frac{t}{\theta_n^{\Xm^n}}\LB  \Gamma_n^{\Xm^n}-\mu_n^{\Xm^n}\RB}\RSB = e^{-\frac{t^2}{2}} + \Oc\left(n^{-\frac12}\right)
\end{align}
(as $n\xrightarrow[]{(\beta,c)}\infty$). Since \eqref{eq:phitildeconv} holds almost surely for any random matrix $X^n$ with law $\mathbb{P}_{X^n}\in\mathcal P(\mathcal S_=^n)$ for all $n$, it holds also for the function $\tilde{\phi}_n(t) =\mathbb{E}\LSB \tilde{\phi}_n^{X^n}(t)\RSB =  \mathbb{E}\LSB e^{\frac{\iv t}{\theta_n}\LB \Gamma_n - \mu_n\RB}\RSB$ which finally proves \eqref{eq:thildephiconv}. 
\end{enumerate}

We now detail all these steps rigorously.

\subsection*{\underline{Step 1:}}
In a first step, we consider the expectation over $W$ by treating $\Hm\in\mathds{C}^{N\times K}$ and $\Xm\in \mathcal S_=^n$ fixed. We define the function $\phi^{\Xm^n,\Hm^n}_n(t)\defines \mathbb{E}\LSB e^{\iv t \Gamma_n^{\Xm^n,\Hm^n}}\RSB$ which we would like to express as a differential equation of the form $\frac{\partial \phi^{\Xm^n,\Hm^n}_n(t)}{\partial t} = f\LB \Xm, \Hm, t\RB \phi^{\Xm^n,\Hm^n}_n(t) + \bar{\varepsilon}_n^{\Xm^n,\Hm^n}(t)$ for some functional $f$ and quantity $ \bar{\varepsilon}_n^{\Xm^n,\Hm^n}(t)$ which vanishes asymptotically. Since $\Gamma_n^{\Xm^n,\Hm^n}$ is real, $\phi^{\Xm^n,\Hm^n}_n(-t)=\phi^{\Xm^n,\Hm^n}_n(t)^*$, so that it is sufficient to consider $t\ge 0$ for the rest of the proof. 

With the help of \eqref{eq:Qrelation}, we can decompose $\Gamma_n$ in the following way:
\begin{align}\label{eq:gamma_decomposition}
\Gamma_{n} = \Gamma_{n,1} + \Gamma_{n,2} + \Gamma_{n,3} + \Gamma_{n,4}
\end{align}
where
\begin{align}
	\Gamma_{n,1} &= \sqrt{\frac{n}K}\log\det\left(\Id_N + \frac1{\sigma^2}\frac1K HH\htp \right)\nonumber\\
	&\qquad + \frac1{\sqrt{nK}}\trace Q\frac{HXX\htp H\htp}K \\
	\Gamma_{n,2} &= -\frac1{\sqrt{nK}}\trace Q \frac{HH\htp}K WW\htp \\
	\Gamma_{n,3} &= \frac{\sigma}{\sqrt{nK}} \trace Q \frac{HXW\htp}{\sqrt{K}} \\
	\Gamma_{n,4} &= \frac{\sigma}{\sqrt{nK}} \trace Q \frac{W X\htp H\htp}{\sqrt{K}}
\end{align}
and where we have defined $Q\triangleq Q(\sigma^2)$ to simplify the notations. 

By \eqref{eq:gamma_decomposition}, 
\begin{align}
	\frac{\partial \phi^{\Xm^n,\Hm^n}_n(t)}{\partial t}=\sum_{k=1}^4 \iv \EE\left[\Gamma^{\Xm^n,\Hm^n}_{n,k} e^{\iv t \Gamma_n^{\Xm^n,\Hm^n}}\right].
\end{align}

Since $\Gamma^{\Xm^n,\Hm^n}_{n,1}$ is independent of $W$,
\begin{align}
	\EE\left[ \Gamma^{\Xm^n,\Hm^n}_{n,1} e^{\iv t \Gamma_n^{\Xm^n,\Hm^n}} \right] &=  \Gamma^{\Xm^n,\Hm^n}_{n,1} \phi^{\Xm^n,\Hm^n}_n(t).
\end{align}

The term in $ \Gamma_{n,2}^{\Xm^n,\Hm^n}$ is studied as follows:
\begin{align}
&\mathbb{E}\LSB \Gamma_{n,2}^{\Xm^n,\Hm^n}\RSB\nonumber\\
&= -\frac1{\sqrt{nK}} \mathbb{E}\LSB\trace \Qm \frac{\Hm\Hm\htp}K WW\htp  e^{\iv t \Gamma_{n}^{\Xm^n,\Hm^n}}\RSB\\ \label{eq:Gamma_1}
&= -\frac1{\sqrt{nK}} \sum_{k=1}^N\sum_{i=1}^N \LSB \Qm \frac{\Hm\Hm\htp}K\RSB_{ki}\mathbb{E}\LSB\LSB WW\htp\RSB_{ik}e^{\iv t \Gamma_{n}^{\Xm^n,\Hm^n}}\RSB\\
&= -\frac1{\sqrt{nK}} \sum_{k=1}^N\sum_{i=1}^N \LSB \Qm \frac{\Hm\Hm\htp}K\RSB_{ki} \sum_{j=1}^{n}\mathbb{E}\LSB  W_{ij}W_{kj}^*e^{\iv t \Gamma_{n}^{\Xm^n,\Hm^n}}\RSB.
\end{align}
We now use the integration by parts formula (Lemma~\ref{lem:IBP} in Appendix~\ref{app:RMT_results}) to develop the individual terms $\EE\left[ W_{ij}W_{kj}^* e^{\iv t \Gamma_{n}^{\Xm^n,\Hm^n}} \right]$ as follows:
\begin{align}\label{eq:derWij}
	&\EE\left[ W_{ij}W_{kj}^* e^{\iv t \Gamma_{n}^{\Xm^n,\Hm^n}} \right]\nonumber\\
	&= \delta_{ik}\EE\left[  e^{\iv t \Gamma_{n}^{\Xm^n,\Hm^n}} \right] + \iv t \EE\left[ W_{kj}^* \frac{\partial \Gamma_{n}^{\Xm^n,\Hm^n}}{\partial W_{ij}^*}e^{\iv t \Gamma_{n}^{\Xm^n,\Hm^n}} \right].
\end{align}
The derivatives $\frac{\partial \Gamma_{n,k}^{\Xm^n,\Hm^n}}{\partial W_{ij}^*}$ and $\frac{\partial \Gamma_{n,k}^{\Xm^n,\Hm^n}}{\partial W_{ij}}$ can be computed by straightforward application of the derivation rules provided in Lemma~\ref{lem:useful_derivatives} in Appendix~\ref{app:Gaussian_relResults}:
\begin{align}\label{eq:gammader1}
	\frac{\partial \Gamma_{n,1}}{\partial W_{ij}^*} &= \frac{\partial \Gamma_{n,1}}{\partial W_{ij}} = \frac{\partial \Gamma_{n,3}}{\partial W_{ij}} = \frac{\partial \Gamma_{n,4}}{\partial W_{ij}^*} = 0 \\\label{eq:gammader2}
	\frac{\partial \Gamma_{n,2}}{\partial W_{ij}^*} &= -\frac1{\sqrt{nK}}\left[Q\frac{HH\htp}{K}W\right]_{ij} \\\label{eq:gammader3}
	\frac{\partial \Gamma_{n,2}}{\partial W_{ij}} &= -\frac1{\sqrt{nK}}\left[W\htp\frac{HH\htp}{K}Q\right]_{ji} \\\label{eq:gammader4}	
	\frac{\partial \Gamma_{n,3}}{\partial W_{ij}^*} &= \frac{\sigma}{\sqrt{nK}} \left[ Q\frac{H}{\sqrt{K}}X \right]_{ij} \\\label{eq:gammader5}
	\frac{\partial \Gamma_{n,4}}{\partial W_{ij}} &= \frac{\sigma}{\sqrt{nK}}\LSB X\htp\frac{H\htp}{\sqrt{K}}Q\RSB_{ji}.
\end{align}

Using \eqref{eq:gamma_decomposition} together with the derivatives \eqref{eq:gammader1}, \eqref{eq:gammader2}, \eqref{eq:gammader4} in \eqref{eq:derWij}, we obtain
\begin{align}
	&\EE\left[ [WW\htp]_{ik} e^{\iv t \Gamma_n^{\Xm^n,\Hm^n}} \right]\nonumber\\ 
	\label{eq:n+1_1}
 &= n \delta_{ik} \phi_n^{\Xm^n,\Hm^n}(t)\nonumber\\
 &\qquad + \iv t \sum_{j=1}^{n} \mathbb{E}\LSB  W_{kj}^*\LB\frac{\partial \Gamma_{n,2}^{\Xm^n,\Hm^n}}{\partial W_{ij}^*}+\frac{\partial \Gamma_{n,3}^{\Xm^n,\Hm^n}}{\partial W_{ij}^*}\RB e^{\iv t \Gamma_{n}^{\Xm^n,\Hm^n}}\RSB \\
	\label{eq:n+1_2}	
	&= n \delta_{ik} \phi_n^{\Xm^n,\Hm^n}(t) - \iv t \frac{1}{\sqrt{nK}}	\EE\Bigg[ \Bigg(\left[\Qm\frac{\Hm\Hm\htp}{K} W  W \htp\right]_{ik}\nonumber\\
	& \qquad\qquad- \sigma\left[\Qm\frac{\Hm}{\sqrt{K}}\Xm W \htp \right]_{ik} \Bigg) e^{\iv t \Gamma_{n}^{\Xm^n,\Hm^n}}\Bigg].
\end{align}
Replacing the last result in \eqref{eq:Gamma_1} yields 
\begin{align}
&\mathbb{E}\LSB \Gamma_{n,2}^{\Xm^n,\Hm^n} e^{\iv t \Gamma_{n}^{\Xm^n,\Hm^n}}\RSB\nonumber\\
&= - \frac{n}{\sqrt{nK}}\trace \Qm\frac{\Hm\Hm\htp}K  \phi^{\Xm^n,\Hm^n}_n(t)\nonumber \\
&\ \ \ + \iv t \mathbb{E}\Bigg[\Bigg( \frac{1}{nK}\trace \LB\Qm\frac{\Hm\Hm\htp}{K}\RB^2 WW\htp\nonumber\\
&\qquad\ \  - \frac{\sigma}{nK}\trace \Qm\frac{\Hm\Hm\htp}K\Qm\frac{\Hm}{\sqrt{K}}\Xm  W \htp \Bigg) e^{\iv t \Gamma_{n}^{\Xm^n,\Hm^n}}\Bigg].\label{eq:split1}
\end{align}

We will now individually treat the second and third terms on the RHS of the last equation. For the second term, using the same steps as above, we arrive at 
\begin{align}\nonumber
&\EE\LSB \frac{1}{nK}\trace \LB\Qm\frac{\Hm\Hm\htp}{K}\RB^2 W  W \htp e^{\iv t\Gamma_{n}^{\Xm^n,\Hm^n}}\RSB\\
&= \frac{1}{nK} \sum_{k=1}^N\sum_{i=1}^N \LSB\LB \Qm \frac{\Hm\Hm\htp}K\RB^2\RSB_{ki}\EE\LSB\LSB  W  W \htp\RSB_{ik}e^{\iv t \Gamma_{n}^{\Xm^n,\Hm^n}}\RSB\\
\label{eq:n+1_4}
&=  \frac{n}{nK}\trace \LB\Qm\frac{\Hm\Hm\htp}{K}\RB^2 \phi_n^{\Xm^n,\Hm^n}(t) \nonumber\\
&\quad -\frac{\iv t}{(nK)^{\frac 32}}\mathbb{E}\Bigg[\Bigg(\trace\LB\Qm\frac{\Hm\Hm\htp}{K}\RB^3 W  W \htp\nonumber\\
 &\qquad- \sigma\trace\LB\Qm\frac{\Hm\Hm\htp}{K}\RB^2\Qm\frac{\Hm}{\sqrt{K}}\Xm W \htp\Bigg) e^{\iv t\Gamma_{n}^{\Xm^n,\Hm^n}} \Bigg]\\
&= \frac{n}{nK}\trace \LB\Qm\frac{\Hm\Hm\htp}{K}\RB^2 \phi_n^{\Xm^n,\Hm^n}(t)\nonumber\\
\label{eq:n+1_5}
 &\quad - \iv t\frac{n}{\sqrt{n^3K^3}} \trace\LB\Qm\frac{\Hm\Hm\htp}{K}\RB^3 \phi_n^{\Xm^n,\Hm^n}(t) + \varepsilon^{\Xm^n,\Hm^n}_{n,1}(t)
\end{align}
where
\begin{align}\nonumber	
	&\varepsilon^{\Xm^n,\Hm^n}_{n,1}(t) = \nonumber\\
	 &-\iv t  \EE\left[ \frac{n}{\sqrt{n^3K^3}}\trace \left(\Qm\frac{\Hm\Hm\htp}K\right)^3\LB\frac{ W  W \htp}{n} - \Id_N \RB e^{\iv t \Gamma_{n}^{\Xm^n,\Hm^n}} \right]\nonumber \\\label{eq:epsilon_1}
	&+ \iv t \EE\left[ \frac{\sigma}{(nK)^{\frac32}}\trace \Qm \LB\Qm\frac{\Hm\Hm\htp}K\RB^2 \frac{\Hm}{\sqrt{K}}\Xm W \htp e^{\iv t \Gamma_{n}^{\Xm^n,\Hm^n}} \right].
\end{align} 

Consider now the third term on the RHS of \eqref{eq:split1} and define $\Tm=\Qm\frac{\Hm\Hm\htp}{K}\Qm\frac{\Hm}{\sqrt{K}}\Xm$. Then, 
\begin{align}	
&\frac{\sigma}{nK}\EE\LSB \trace \Tm W \htp e^{\iv t \Gamma_{n}^{\Xm^n,\Hm^n}}\RSB\nonumber\\
\label{eq:n+1_7}
&= \iv t\frac{\sigma}{nK}\sum_{i=1}^N\sum_{j=1}^{n} T_{ij} \EE\LSB \frac{\partial \Gamma_{n}^{\Xm^n,\Hm^n}}{\partial W_{ij}}e^{\iv t \Gamma_{n}^{\Xm^n,\Hm^n}}\RSB\\
&= \iv t \frac{\sigma^2}{\sqrt{n^3K^3}}\trace \Qm^2\frac{\Hm\Hm\htp}{K}\Qm\frac{\Hm \Xm\Xm\htp \Hm\htp}{K} \phi_n^{\Xm^n,\Hm^n}(t)\nonumber\\
&\quad + \varepsilon^{\Xm^n,\Hm^n}_{n,2}(t)
\end{align}
where
\begin{align}
&\varepsilon^{\Xm^n,\Hm^n}_{n,2}(t) =\nonumber\\
& - \iv t \frac{\sigma}{(nK)^{\frac32}}\mathbb{E}\LSB \trace\Qm \LB\Qm\frac{\Hm\Hm\htp}{\sqrt{K}}\RB^2\frac{\Hm}{\sqrt{K}}\Xm  W \htp e^{\iv t \Gamma_{n}^{\Xm^n,\Hm^n}}\RSB.
\end{align}

Combining the last results, we arrive at
\begin{align}	
&\EE\LSB \Gamma_{n,2}^{\Xm^n,\Hm^n} e^{\iv t \Gamma_{n}^{\Xm^n,\Hm^n}}\RSB =\nonumber\\\label{eq:n+1_8}
& - \frac{n}{\sqrt{nK}}\trace \Qm\frac{\Hm\Hm\htp}K \phi_n^{\Xm^n,\Hm^n}(t)\nonumber\\
&+ \iv t \frac{n}{nK}\trace \LB\Qm\frac{\Hm\Hm\htp}{K}\RB^2 \phi_n^{\Xm^n,\Hm^n}(t)\nonumber\\
&+ t^2\Bigg\{ \frac{n}{\sqrt{n^3K^3}} \trace\LB\Qm\frac{\Hm\Hm\htp}{K}\RB^3\nonumber\\
&\qquad + \frac{\sigma^2n}{\sqrt{n^3K^3}}\trace \Qm^2\frac{\Hm\Hm\htp}{K}\Qm\frac{\Hm \Xm\Xm\htp \Hm\htp}{nK} \Bigg\} \phi_n^{\Xm^n,\Hm^n}(t)\nonumber\\
&+ \iv t\left\{ \varepsilon^{\Xm^n,\Hm^n}_{n,1}(t) -\varepsilon^{\Xm^n,\Hm^n}_{n,2}(t) \right\}.
\end{align}

We now consider the terms in $\Gamma_{n,4}^{\Xm^n,\Hm^n}$ and $\Gamma_{n,3}^{\Xm^n,\Hm^n}$. Using similar calculus as above, 
\begin{align}\nonumber
&\EE\LSB \Gamma_{n,4}^{\Xm^n,\Hm^n} e^{it\Gamma_{n}^{\Xm^n,\Hm^n}}\RSB\\
 &= \mathbb{E}\LSB \frac{\sigma}{\sqrt{nK}}\trace\Qm W \Xm\htp\frac{\Hm\htp}{\sqrt{K}} e^{\iv t\Gamma_{n}^{\Xm^n,\Hm^n}}\RSB\\
 \label{eq:n+1_9}
& = \frac{\sigma}{\sqrt{nK}}\sum_{i=1}^N\sum_{j=1}^{n} \LSB \Xm\htp\frac{\Hm\htp}{\sqrt{K}}\Qm \RSB_{ji} \EE\LSB W_{ij}e^{\iv t\Gamma_{n}^{\Xm^n,\Hm^n}}\RSB\\
 \label{eq:n+1_10}
&= \iv t\frac{ \sigma}{\sqrt{nK}}\sum_{i=1}^N\sum_{j=1}^{n} \LSB \Xm\htp\frac{\Hm\htp}{\sqrt{K}}\Qm \RSB_{ji}\EE\LSB \frac{\partial\Gamma_{n}^{\Xm^n,\Hm^n}}{\partial W_{ij}^*}e^{\iv t \Gamma_{n}^{\Xm^n,\Hm^n}}\RSB\\
&=\iv t\frac{ \sigma}{\sqrt{nK}}\sum_{i=1}^N\sum_{j=1}^{n} \LSB \Xm\htp\frac{\Hm\htp}{\sqrt{K}}\Qm \RSB_{ji}\EE\Bigg[\Bigg( \frac{\sigma}{\sqrt{nK}}\LSB\Qm\frac{\Hm}{\sqrt{K}}\Xm\RSB_{ij}\nonumber\\
 \label{eq:n+1_11}
&\quad -\frac{1}{\sqrt{nK}}\LSB\Qm\frac{\Hm\Hm\htp}{K} W \RSB_{ij}   \Bigg) e^{\iv t \Gamma_{n}^{\Xm^n,\Hm^n}} \Bigg]\\
&= \iv t\frac{\sigma^2n}{nK}\trace \Qm^2\frac{\Hm\Xm\Xm\htp\Hm\htp}{(n+1)K} \phi_n^{\Xm^n,\Hm^n}(t)\nonumber\\
 \label{eq:n+1_12}
&\quad -\iv t \EE\LSB\frac{\sigma}{nK}\trace\Qm\frac{\Hm\Hm\htp}K  W  \Xm\htp\frac{\Hm\htp}{\sqrt{K}} \Qm e^{\iv t \Gamma_{n}^{\Xm^n,\Hm^n}} \RSB.
\end{align}

Doing the same calculus for the second term on the RHS of the last equation, one arrives at
\begin{align}
&\EE\LSB\frac{\sigma}{nK}\trace\Qm\frac{\Hm\Hm\htp}K  W  \Xm\htp\frac{\Hm\htp}{\sqrt{K}} \Qm e^{\iv t \Gamma_{n}^{\Xm^n,\Hm^n}} \RSB\nonumber\\
&= \iv t \frac{\sigma^2n}{\sqrt{n^3K^3}} \trace\Qm^2\frac{\Hm\Hm\htp}{K}\Qm\frac{\Hm\Xm\Xm\htp\Hm\htp}{nK} \phi_n^{\Xm^n,\Hm^n}(t)\nonumber\\
	\label{eq:n+1_13}
&\quad + \varepsilon^{\Xm^n,\Hm^n}_{n,4}(t)
\end{align}
where
\begin{align}
&\varepsilon^{\Xm^n,\Hm^n}_{n,4}(t)=\nonumber\\
& -\iv t\frac{\sigma}{(nK)^{\frac32}} \EE\LSB\trace\Qm\LB\Qm\frac{\Hm\Hm\htp}K\RB^2  W  \Xm\htp\frac{\Hm\htp}{\sqrt{K}}  e^{\iv t \Gamma_{n}^{\Xm^n,\Hm^n}} \RSB.
\end{align} 
Thus,
\begin{align}\nonumber
&\EE\LSB \Gamma_{n,4}^{\Xm^n,\Hm^n} e^{it\Gamma_{n}^{\Xm^n,\Hm^n}}\RSB\\
 &= \iv t\frac{\sigma^2n}{nK}\trace \Qm^2\frac{\Hm\Xm\Xm\htp\Hm\htp}{nK} \phi_n^{\Xm^n,\Hm^n}(t)\nonumber\\
 &\quad + t^2\frac{\sigma^2n}{\sqrt{n^3K^3}} \trace\Qm^2\frac{\Hm\Hm\htp}{K}\Qm\frac{\Hm\Xm\Xm\htp\Hm\htp}{nK} \phi_n^{\Xm^n,\Hm^n}(t)\nonumber\\
  \label{eq:n+1_14}
 &\quad -\iv t\varepsilon^{\Xm^n,\Hm^n}_{n,4}(t).
\end{align} 

Since $\Gamma_{n,3}^{\Xm^n,\Hm^n}=\left(\Gamma_{n,4}^{\Xm^n,\Hm^n}\right)^*$, it follows that
\begin{align}
\EE\left[ \Gamma_{n,3}^{\Xm^n,\Hm^n} e^{it \Gamma_n^{\Xm^n,\Hm^n}}\right]=\EE\left[ \Gamma_{n,4}^{\Xm^n,\Hm^n} e^{-it \Gamma_n^{\Xm^n,\Hm^n}}\right]^*.
\end{align}

Gathering all pieces together as a polynomial in $t$, we obtain a first differential equation of $\phi_n^{\Xm^n,\Hm^n}(t)$ as given in \eqref{eq:clt_W_1}--\eqref{eq:epsilonW} on the top of the next page, where
\begin{align}
\Am\defines\Id_K-\frac{1}{n}\Xm\Xm\htp.
\end{align}
\begin{figure*}
\begin{align}\label{eq:clt_W_1}
	\frac{\partial \phi_n^{\Xm^n,\Hm^n}(t)}{\partial t} &= (\iv \mu_{n}^{\Xm^n,\Hm^n} - t\LB\theta_{n}^{\Xm^n,\Hm^n}\RB^2 + \iv t^2\kappa_{n}^{\Xm^n,\Hm^n})\phi_n^{\Xm^n,\Hm^n}(t) + \bar{\varepsilon}_{n}^{\Xm^n,\Hm^n}(t)\\
\label{eq:muforproof}
	\mu_{n}^{\Xm^n,\Hm^n} &= \sqrt{\frac{n}K}\log\det \left(\Id_N+\frac1{\sigma^2}\frac{\Hm\Hm\htp}K \right) - \frac{n}{\sqrt{nK}}\trace \Qm\frac{\Hm\Am\Hm\htp}{K} \\\label{eq:thetaforproof}
	\LB\theta_{n}^{\Xm^n,\Hm^n}\RB^2 &= \frac{n}{nK}\trace \left(\Qm\frac{\Hm\Hm\htp}K \right)^2 + \frac{2\sigma^2n}{nK} \trace \Qm^2 \frac{\Hm\Xm\Xm\htp\Hm\htp}{nK} \\
	\label{eq:n+1_17}
	\kappa_{n}^{\Xm^n,\Hm^n} &= \frac{n}{\sqrt{n^3K^3}}\trace\LB\Qm\frac{\Hm\Hm\htp}{K}\RB^3 + \frac{3\sigma^2n}{\sqrt{n^3 K^3}}\trace\Qm^2\frac{\Hm\Hm\htp}{K}\Qm\frac{\Hm\Xm\Xm\htp\Hm\htp}{nK}\\
	\bar{\varepsilon}_{n}^{\Xm^n,\Hm^n}(t)&= {\bf i} t^2 \EE\LSB\left\{ \frac{n}{\sqrt{n^3K^3}}\trace \left(\Qm\frac{\Hm\Hm\htp}K\right)^3\LB\frac{ W  W \htp}{n} - \Id_N \RB  - \frac{3\sigma}{\sqrt{n^3K^3}}\trace \Qm\LB\Qm\frac{\Hm\Hm\htp}K\RB^2 \frac{\Hm\Xm W \htp}{\sqrt{K}} \right. \right.\nonumber\\ 
&\qquad\qquad\quad\left. \left. -\frac{\sigma}{\sqrt{n^3K^3}} \trace\Qm\LB\Qm\frac{\Hm\Hm\htp}K\RB^2 \frac{ W  \Xm\htp\Hm\htp}{\sqrt{K}}
\right\} e^{{\bf i} t \Gamma}\RSB.\label{eq:epsilonW}
\end{align}
\hrulefill
\end{figure*}

Let us now have a closer look at the quantities $\theta^{\Xm^n,\Hm^n}_{n}$, $\kappa_{n}^{\Xm^n,\Hm^n}$, and $\bar{\varepsilon}_{n}^{\Xm^n,\Hm^n}(t)$ individually. Using the identities and bounds presented at the beginning of this proof, one can verify that
\begin{align}\label{eq:Wbnd1}
0\le\LB\theta_n^{\Xm^n,\Hm^n} \RB^2 &\le \frac{N}{K} + \frac{2}{nK}\trace\Xm\Xm\htp = \LB\frac NK + 2\RB \\\label{eq:Wbnd2}
0\le\kappa_{n}^{\Xm^n,\Hm^n} &\le \frac{N}{\sqrt{nK^3}} + \frac{3}{\sqrt{n^3K^3}}\trace\Xm\Xm\htp\nonumber\\
& = \frac1{\sqrt{nK}}\LB \frac NK + 3\RB. 
\end{align}
Based on Remark~\ref{rem:cauchyschwarz} in Appendix~\ref{app:Gaussian_relResults}, we can bound the absolute value of $\bar{\varepsilon}_{n}^{\Xm^n,\Hm^n}(t)$ as
\begin{align}
&\left|\bar{\varepsilon}_{n}^{\Xm^n,\Hm^n}(t) \right| \le \nonumber\\
& t^3 \LB \sqrt{\var\LSB\frac{1}{\sqrt{nK^3}}\trace \left(\Qm\frac{\Hm\Hm\htp}K\right)^3\LB\frac{ W  W \htp}{n} - \Id_N \RB\RSB} \right.\nonumber\\
\label{eq:epsilonWbnd}
 & \qquad \left. +4 \sqrt{\var\LSB \frac{\sigma}{\sqrt{n^3K^3}}\trace \Qm\LB\Qm\frac{\Hm\Hm\htp}K\RB^2 \frac{\Hm\Xm W \htp}{\sqrt{K}}\RSB} \RB.
\end{align}
By Lemma~\ref{lem:varbnds}~$(ii)$ in Appendix~\ref{app:aux_RMTaux}, it follows that
\begin{align}
&\var\LSB\frac1{\sqrt{nK^3}}\trace \left(\Qm\frac{\Hm\Hm\htp}K\right)^3\LB\frac{ W  W \htp}{n} - \Id_N \RB\RSB\nonumber\\
 &= \var\LSB\frac{n}{\sqrt{n^3K^3}}\trace \left(\Qm\frac{\Hm\Hm\htp}K\right)^3\frac{ W  W \htp}{n}\RSB\\
&\le \frac{2}{nK^3}\trace\LB \Qm\frac{\Hm\Hm\htp}{K}\RB^6\\\label{eq:epsvar1bnd}
&\le \frac{2N}{nK^3}.
\end{align}
Similarly, by Lemma~\ref{lem:varbnds}~$(i)$ in Appendix~\ref{app:aux_RMTaux}, it follows that
\begin{align}
&\var\LSB \frac{\sigma}{\sqrt{n^3K^3}}\trace \Qm\LB\Qm\frac{\Hm\Hm\htp}K\RB^2 \frac{\Hm\Xm W \htp}{\sqrt{K}}\RSB\nonumber\\
& = \frac{\sigma^2}{n^3K^3}\trace \LB\Qm\frac{\Hm\Hm\htp}K\RB^4\Qm\frac{\Hm\Xm\Xm\htp\Hm\htp}{K}\Qm\\
&\le  \frac{\sigma^2}{n^3K^3}\trace \tilde{\Qm}^2\frac{\Hm\htp\Hm}{K}\Xm\Xm\htp\\
& \le \frac1{n^3K^3}\trace\Xm\Xm\htp\\\label{eq:epsvar2bnd}
&= \frac1{n^2K^2}.
\end{align}
Replacing \eqref{eq:epsvar1bnd} and \eqref{eq:epsvar2bnd} in \eqref{eq:epsilonWbnd}, we then obtain
\begin{align}\label{eq:eqnbound}
\bar{\varepsilon}_{n}^{\Xm^n,\Hm^n}(t) = \Oc\LB t^3 n^{-2}\RB.
\end{align}
Similarly, from \eqref{eq:Wbnd1} and \eqref{eq:Wbnd2}, 
\begin{align}
\LB\theta_{n}^{\Xm^n,\Hm^n}\RB^2 &= \Oc(1)\\
\kappa_{n}^{\Xm^n,\Hm^n} &= \Oc\LB n^{-1}\RB. 
\end{align}

Two remarks are important at this point. First observe that the introduction of $\kappa_{n}^{\Xm^n,\Hm^n}$ allows one to gain at each step one order of precision on the estimation of $\phi_n^{\Xm^n,\Hm^n}$ (through refinements of the coefficients of its differential equation). The choice of the order to be used is mainly ruled by the subsequent averaging steps. For the present proof, we need the error (given by $\bar{\varepsilon}_{n}^{\Xm^n,\Hm^n}(t)$) to be within $\Oc(n^{-2})$.

Second, it is very important to keep the terms in $t$ in the various bounds derived here and below. The reason for this is twofold: (i) to solve the differential equations in $\phi_n^{\Xm^n,\Hm^n}$, then $\phi_n^{\Xm^n}$, it will be necessary to integrate these bounds and their integrability must be controlled, (ii) at the end of the calculus, the normalization of $\Gamma_n$ by (the estimate for) its standard deviation $\theta_n^{\Xm^n}$, used to ensure a limiting unit variance, will be performed via a change of variable $t\mapsto t/\theta_n^{\Xm^n}$ which requires a close inspection of the polynomials in $t$ and $n^{-1}$ in the bounds.

\subsection*{\underline{Step 2:}}
In this step, we first solve \eqref{eq:clt_W_1} to express $\phi_n^{\Xm^n,\Hm^n}(t)$ as a function of $\Xm$ and $\Hm$. We then proceed similar to Step~1  and express the function $\phi_n^{\Xm^n}(t)=\EE[\phi_n^{\Xm^n,H^n}(t)]$ as the solution of a differential equation. 

The solution of \eqref{eq:clt_W_1} reads
\begin{align}
	&\phi_n^{\Xm^n,\Hm^n}(t) = e^{\iv t \mu_{n}^{\Xm^n,\Hm^n} - \frac{t^2}2 \LB\theta_{n}^{\Xm^n,\Hm^n}\RB^2 +\iv \frac{t^3}{3}\kappa_{n}^{\Xm^n,\Hm^n} }\Bigg( 1 +\nonumber\\
	\label{eq:equ_diff1}
	&\ \int_0^t e^{-\iv x \mu_{n}^{\Xm^n,\Hm^n} + \frac{x^2}2 \LB\theta_{n}^{\Xm^n,\Hm^n}\RB^2 -\iv\frac{x^3}{3}\kappa_{n}^{\Xm^n,\Hm^n} } \bar{\varepsilon}_{n}^{\Xm^n,\Hm^n}(x) dx\Bigg).
\end{align}

Define the function $\phi_n^{\Xm^n}(t)=\EE[\phi_n^{\Xm^n,H^n}(t)]$. The equation \eqref{eq:tmpdiffequ2} on the top of the next page follows then from \eqref{eq:equ_diff1}. 
\begin{figure*}
\begin{align}
	\frac{\partial \phi_n^{\Xm^n}(t)}{\partial t} &= \EE \left[ \frac{\partial \phi_n^{\Xm^n,H^n}(t)}{\partial t} \right] \\
	&=  \EE\Bigg[ \left(\iv \mu_{n}^{\Xm^n,H^n} - t \LB\theta_{n}^{\Xm^n,H^n}\RB^2 + \iv t^2 \kappa_{n}^{\Xm^n,H^n}\right)e^{\iv t \mu_{n}^{\Xm^n,H^n} - \frac{t^2}2 \LB\theta_{n}^{\Xm^n,H^n}\RB^2+\iv\frac{t^3}{3}\kappa_{n}^{\Xm^n,H^n}} \Bigg] \nonumber\\
	&\quad + \EE \Bigg[\left(\iv \mu_{n}^{\Xm^n,H^n} - t \LB\theta_{n}^{\Xm^n,H^n}\RB^2 + \iv t^2 \kappa_{n}^{\Xm^n,H^n}\right)\LB \int_0^t e^{-\iv x \mu_{n}^{\Xm^n,H^n} + \frac{x^2}2 \LB\theta_{n}^{\Xm^n,H^n}\RB^2 -\iv \frac{x^3}{3}\kappa_{n}^{\Xm^n,H^n}} \bar{\varepsilon}_{n}^{\Xm^n,H^n}(x) dx \right)\nonumber\\
 &\qquad\qquad \times e^{\iv t \mu_{n}^{\Xm^n,H^n} - \frac{t^2}2 \LB\theta_{n}^{\Xm^n,H^n}\RB^2 +\iv \frac{t^3}{3}\kappa_{n}^{\Xm^n,H^n}}  \Bigg]  + \mathbb{E}\LSB \bar{\varepsilon}_{n}^{\Xm^n,H^n}(t)\RSB.\label{eq:tmpdiffequ2}
\end{align}
\hrulefill
\end{figure*}

We will now show that only the first term on the RHS of \eqref{eq:tmpdiffequ2} is asymptotically non-negligible. Let us first define
\begin{align}
 &\Theta = \nonumber\\
 & \LB \int_0^t e^{-\iv x \mu_{n}^{\Xm^n,H^n} + \frac{x^2}2 \LB\theta_{n}^{\Xm^n,H^n}\RB^2 -\iv \frac{x^3}{3}\kappa_{n}^{\Xm^n,H^n}} \bar{\varepsilon}_{n}^{\Xm^n,H^n}(x) dx \right)\nonumber\\
 &\qquad\times  e^{\iv t \mu_{n}^{\Xm^n,H^n} - \frac{t^2}2 \LB\theta_{n}^{\Xm^n,H^n}\RB^2 +\iv \frac{t^3}{3}\kappa_{n}^{\Xm^n,H^n}}. 
\end{align}
Since
\begin{align}
 \left| \Theta \right| &\le  e^{- \frac{t^2}2 \LB\theta_{n}^{\Xm^n,H^n}\RB^2 } \int_{0}^t e^{\frac{x^2}{2}\LB\theta_{n}^{\Xm^n,H^n}\RB^2}\left|\bar{\varepsilon}_{n}^{\Xm^n,H^n}(x) \right|dx\\
 & = \Oc\LB t^4n^{-2} \RB
\end{align}
it follows that $\EE[\Theta]=\Oc(t^4n^{-2})$ and $\var[\Theta] = \Oc(t^8n^{-4})$. Thus, by Remark~\ref{rem:cauchyschwarz} in Appendix~\ref{app:Gaussian_relResults}, 
\begin{align}
 &\left|\EE\LSB \mu_{n}^{\Xm^n,H^n} \Theta \RSB \right|\nonumber\\
 &\le \left|\EE\LSB \mu_{n}^{\Xm^n,H^n}\RSB \right|\left| \EE\LSB \Theta \RSB\right| + \sqrt{\var\LSB\mu_{n}^{\Xm^n,H^n}\RSB}\sqrt{\var\LSB \Theta\RSB}\\
 &\le \left|\EE\LSB \mu_{n}^{\Xm^n,H^n}\RSB \right| \Oc\LB t^4 n^{-2}\RB + \sqrt{\var\LSB\mu_{n}^{\Xm^n,H^n}\RSB} \Oc\LB t^4n^{-2}\RB.
\end{align}

Again, from Remark~\ref{rem:cauchyschwarz} in Appendix~\ref{app:Gaussian_relResults},
\begin{align}
\var\LSB \mu_{n}^{\Xm^n,H^n}\RSB &\le \Bigg( \sqrt{ \frac{n}{K} \var\LSB \log\det \left(\Id_N+\frac1{\sigma^2}\frac{ H H\htp}K \right) \RSB}\nonumber\\
\label{eq:varmuwn}
& \qquad\quad+ \sqrt{\frac{n}{K}\var\LSB \trace  Q\frac{ H\Am H\htp}{K}  \RSB} \Bigg)^2.
\end{align}
From Proposition~\ref{prop:vartrace}~$(iii)$ in Appendix~\ref{app:aux_RMTaux}, we know that $\var\LSB \trace  Q\frac{ H\Am H\htp}{K}  \RSB = \Oc\LB \frac1K\trace\Am^2\RB$. It remains to find a bound for the variance of the first term in \eqref{eq:varmuwn}. By Lemma~\ref{lem:PNI} in Appendix~\ref{app:Gaussian_relResults}, 
\begin{align}
&\var\LSB \log\det \left(\Id_N+\frac1{\sigma^2}\frac{ H H\htp}K \right) \RSB \nonumber\\
& \le \frac2{\sigma^4} \sum_{i,j}\mathbb{E}\LSB \left| \frac1K\trace Q \frac{\partial \LB H H\htp\RB}{\partial H^n_{ij}} \right|^2\RSB\\
&= \frac2{\sigma^4} \sum_{i,j}\mathbb{E}\LSB \left| \frac1K \sum_{p,q} \delta_{ip} \LB H_{qj}\RB^*Q_{qp} \right|^2\RSB\\
&=\frac2{\sigma^4} \sum_{i,j}\mathbb{E}\LSB \left| \frac1K \LSB  H\htp Q\RSB_{ji} \right|^2\RSB\\
&= \frac2{\sigma^4} \frac1K\trace Q^2\frac{ H  H\htp}{K}\\\label{eq:logdetvarbnd}
&= \Oc(1).
\end{align}
Using the fact that $\trace\Am^2 = \Oc(n^2)$, we conclude that $\var\LSB \mu_{n}^{\Xm^n,H^n}\RSB = \Oc\LB n\RB$.

Similarly, we have from Proposition~\ref{prop:dettrace}~$(i)$ in Appendix~\ref{app:aux_RMT}
\begin{align}
	\left|\EE\LSB \mu_{n}^{\Xm^n,H^n} \RSB \right| &= \left| \sqrt{\frac nK}\mathbb{E}\Bigg[\log\det\LB\Id_N + \frac1{\sigma^2}\frac{HH\htp}{K}\RB \right. \nonumber\\
	 &\left. \qquad\qquad-\sqrt{ \frac{n}{K} } \trace Q\frac{H\Am H\htp}{K}\Bigg] \right| \\
 & = \Oc\LB n + \sqrt{\frac1{n^3}\trace\Am^2}\RB\\
 &= \Oc\LB n\RB.
\end{align}

Combining the last results, we have shown that
\begin{align}\label{eq:result_11}
  \left|\EE\LSB \mu_{n}^{\Xm^n,H^n} \Theta \RSB \right| = \Oc\LB t^4n^{-1}\RB.
\end{align}

Similarly, one can show that
\begin{align}\label{eq:result_22}
 \left| \EE\LSB \LB\theta_{n}^{\Xm^n,H^n}\RB^2 \Theta\RSB \right|  &= \Oc\LB t^4n^{-2}\RB \\
  \left| \EE\LSB \kappa_{n}^{\Xm^n,H^n} \Theta\RSB \right | &= \Oc\LB t^4 n^{-3}\RB \label{eq:result_33}.
\end{align}

Using \eqref{eq:result_11}, \eqref{eq:result_22}, and \eqref{eq:result_33}, we can finally conclude that
\begin{align}\nonumber
 &\EE \left[\left(\iv \mu_{n}^{\Xm^n,H^n} - t \LB\theta_{n}^{\Xm^n,H^n}\RB^2 + \iv t^2 \kappa_{n}^{\Xm^n,H^n}\right) \Theta  \right] \nonumber \\
&=\Oc\LB t^4n^{-1}+t^5n^{-2} +t^6 n^{-3}\RB.
\end{align}
Since all bounds are clearly integrable over $t$, this now means that $e^{\iv t \mu_{n}^{\Xm^n,H^n} - \frac{t^2}2 \LB\theta_{n}^{\Xm^n,H^n}\RB^2+\iv\frac{t^3}{3}\kappa_{n}^{\Xm^n,H^n}}$ is an estimator of $\phi_n^{\Xm^n}$ within $\Oc(n^{-1})$. Note that this bound would be $\Oc(1)$ if we had only used an estimation of $\phi_n^{\Xm^n,\Hm^n}$ within $\Oc(n^{-1})$ in the previous step. Hence the fundamental importance of the term $\kappa_n^{\Xm^n,H^n}$.

We can therefore proceed to study $\phi_n^{\Xm^n}$ via the estimator $e^{\iv t \mu_{n}^{\Xm^n,H^n} - \frac{t^2}2 \LB\theta_{n}^{\Xm^n,H^n}\RB^2+\iv\frac{t^3}{3}\kappa_{n}^{\Xm^n,H^n}}$.

Starting back from \eqref{eq:tmpdiffequ2}, we first verify that
\begin{align}
&\left|\mathbb{E}\LSB \iv t^2 \kappa_{n}^{\Xm^n,H^n}e^{\iv t \mu_{n}^{\Xm^n,H^n} - \frac{t^2}2 \LB\theta_{n}^{\Xm^n,H^n}\RB^2+\iv\frac{t^3}{3}\kappa_{n}^{\Xm^n,H^n}} \RSB\right| \nonumber\\
& = \Oc(t^2 n^{-1})\mathbb{E}\LSB e^{-\frac{t^2}{2}\LB\theta_{n}^{\Xm^n,H^n}\RB^2}\RSB = \Oc(t^2n^{-1}).
\end{align}
Thus, we have 
\begin{align}
	\frac{\partial \phi_n^{\Xm^n}(t)}{\partial t} =&\ \EE\Bigg[ \left(\iv \mu_{n}^{\Xm^n,H^n} - t \LB\theta_{n}^{\Xm^n,H^n}\RB^2\right) \nonumber\\
	&\qquad \times e^{\iv t \mu_{n}^{\Xm^n,H^n} - \frac{t^2}2 \LB\theta_{n}^{\Xm^n,H^n}\RB^2+\iv\frac{t^3}{3}\kappa_{n}^{\Xm^n,H^n}} \Bigg]\nonumber\\ \label{eq:phiprime}
	&\ + \Oc\LB  \frac{t^2+t^4}{n} + \frac{t^3+t^5}{n^2} + \frac{t^6}{n^3}\RB.
\end{align}

We now develop the term in the expectation and express it under the form of $f\LB \Xm\RB \phi_n^{\Xm^n}(t) + \varepsilon^{\Xm^n}_n(t)$ for some functional $f$ and asymptotically negligible quantity $\varepsilon^{\Xm^n}_n(t)$. For better readability, we define the shorthand notation
\begin{align}\label{eq:defgamma}
\gamma_n^{\Xm^n,H^n} = \iv t\mu_{n}^{\Xm^n,H^n}-\frac{t^2}2\LB\theta_{n}^{\Xm^n,H^n}\RB^2 + \iv \frac{t^3}{3}\kappa_{n}^{\Xm^n,H^n}
\end{align}
and consider individually the terms
\begin{align*}
	\text{A:\ }\EE\left[ \mu_{n}^{\Xm^n,H^n}e^{\gamma_n^{\Xm^n,H^n}} \right], \qquad
	\text{B:\ }\EE\left[ \LB\theta_{n}^{\Xm^n,H^n}\RB^2e^{\gamma_n^{\Xm^n,H^n}} \right] .
\end{align*}

\subsubsection*{Term A}
The term A cannot be evaluated in a straightforward manner as the integration by parts formula (Lemma~\ref{lem:IBP} in Appendix~\ref{app:Gaussian_relResults}) cannot be applied to the log-term in $\mu_{n}^{\Xm^n,H^n}$ (as defined in \eqref{eq:muforproof}). To avert this difficulty, we use the identity
\begin{align}
 \log\det\LB \Id_N + \frac 1{\sigma^2}\frac{HH\htp}{K}\RB = \int_{\sigma^2}^\infty \frac1u \trace Q(u)\frac{HH\htp}{K} du
\end{align}
which, together with the Fubini theorem (using $\trace Q(u)HH\htp \leq u^{-1}\trace HH\htp$), gives for A:
\begin{align}
&\EE\LSB \mu_{n}^{\Xm^n,H^n}e^{\gamma_n^{\Xm^n,H^n}}\RSB \nonumber\\
&= \sqrt{\frac{n}{K}}\int_{\sigma^2}^{\infty} \frac1u \EE\LSB  \trace Q(u)\frac{HH\htp}K e^{{\gamma_n^{\Xm^n,H^n}}}\RSB du \nonumber\\
\label{eq:termA}
&\quad -  \sqrt{\frac{n}{K}}\EE\LSB\trace Q\frac{H\Am H\htp}{K}e^{{\gamma_n^{\Xm^n,H^n}}}\RSB.
\end{align}

Before we continue, we need the following result which is the cornerstone of the subsequent analysis:
\begin{prop}\label{prop:tracexp} 
Let $u\ge\sigma^2>0$ and $\gamma_n^{\Xm^n,H^n}$ be defined as in \eqref{eq:defgamma}. Then,
\begin{align}
(i)\quad &\EE\LSB \trace Q(u)\frac{HH\htp}{K} e^{\gamma_n^{\Xm^n,H^n}}\RSB \nonumber\\
&= N\frac{\LB 1-c + 2u\delta_0(u) - \frac{u^2}{c}\delta_0(u)^2\RB}{1-c +u\LB 1 +2\delta_0(u)\RB}\EE\LSB e^{\gamma_n^{\Xm^n,H^n}}\RSB\nonumber\\
&\quad +{\bf i}t\sqrt{\frac nK} u\frac{\delta_0(u) - \sigma^2\delta_1(u)}{1 - c + u\LB 1 + 2\delta_0(u)\RB}\EE\LSB e^{\gamma_n^{\Xm^n,H^n}}\RSB\nonumber\\
&\quad + \Oc\left( \frac{P(t)}{u\sqrt{K}} \right)\\
(ii)\quad &\EE\LSB \trace Q\frac{H\Am H\htp}{K} e^{\gamma_n^{\Xm^n,H^n}}\RSB\nonumber\\
&= -{\bf i}t \sqrt{\frac{n}{K}}\frac{\gamma_1\LB\sigma^2\RB\frac1K\trace\Am^2}{\LB 1+\delta_0(\sigma^2)\RB^2}\mathbb{E}\LSB e^{{\gamma_n^{\Xm^n,H^n}}}\RSB\nonumber\\
&\quad + \Oc\LB \frac{P_1(t)}{\sqrt{K}} + \frac{tP_2(t)}{\sqrt{K}}\frac1K\trace \Am^2 \RB
\end{align}
for some non-zero polynomials $P(t),P_1(t),P_2(t)$ in $t$ with nonnegative coefficients and with $\delta_m(x)$ and $\gamma_m(x)$ given by Proposition~\ref{prop:dettrace} in Appendix~\ref{app:aux_RMT}.
\end{prop}
\begin{IEEEproof}
The proof is provided in Appendix~\ref{proof:prop:tracexp}.
\end{IEEEproof}

Applying Proposition~\ref{prop:tracexp}~$(i)$ and $(ii)$ to the first and second terms of \eqref{eq:termA}, respectively, we obtain the result in \eqref{eq:termAapprox} on the top of the next page,
\begin{figure*}
\begin{align}\label{eq:termAapprox}
\EE\LSB \mu_{n}^{\Xm^n,H^n}e^{\gamma_n^{\Xm^n,H^n}}\RSB &= \sqrt{nK}\left\{ \int_{\sigma^2}^\infty \frac{c\LB 1-c + 2u\delta_0(u) - \frac{u^2}{c}\delta_0(u)^2\RB}{u\LB 1-c +u\LB 1 +2\delta_0(u)\RB\RB} du \right\}\EE\LSB e^{\gamma_n^{\Xm^n,H^n}}\RSB\nonumber\\
&\qquad   + \iv t \frac nK\left\{ \int_{\sigma^2}^\infty\frac{\delta_0(u) - \sigma^2\delta_1(u)}{1 - c + u\LB 1 + 2\delta_0(u)\RB} du+ \frac{\gamma_1\LB \sigma^2\RB\frac1K\trace\Am^2}{\LB 1+\delta_0(\sigma^2)\RB^2}\right\}\EE\LSB e^{\gamma_n^{\Xm^n,H^n}}\RSB\nonumber\\
& \qquad + \Oc\LB \frac{P_1(t)}{\sqrt{K}} + \frac{tP_2(t)}{\sqrt{K}} \frac1K\trace\Am^2 \RB
\end{align}
\hrulefill
\end{figure*}
where for the last RHS term, we used $\int_{\sigma^2}^\infty u^{-2}du<\infty$ and $P_1,P_2$ are non-zero polynomials with nonnegative coefficients, possibly different from those of Proposition~\ref{prop:tracexp}. Note in passing the fundamental importance of maintaining $1/u$ in the big-$\Oc$ term of Proposition~\ref{prop:tracexp}~$(i)$. The existence of the two integrals in \eqref{eq:termAapprox} can be proved via bounds on the $\delta_t(u)$ and $\gamma_t(u)$, essentially relying on their definitions in Proposition~\ref{prop:dettrace} and on controls similar to Property~\ref{properties:delta0}~(i) and (ii) in Appendix~\ref{app:Gaussian_relResults}. Nonetheless, a more immediate argument consists in remarking that, since the LHS of \eqref{eq:termAapprox} is finite, and so are all terms aside from the integrals on the RHS, so is the sum of the integrals. Taking $t=0$ then justifies with the same argument that the first integral is finite which, taking then $t\neq 0$, ensures the finiteness of the second 
integral.

Also note that the last RHS term of \eqref{eq:termAapprox} is {\it not} necessarily negligible in the large $n$ limit. Indeed, for $\Xm\in \mathcal{S}^n$, $\trace\Am^2$ can grow as $\Oc(K^2)$, so that the whole term may grow as $\Oc(\sqrt{K})$. It is therefore essential to keep track of the terms in $\Am$. The pre-factor $t$ in front of $\frac1K\trace\Am^2$ will play a significant role in controlling these terms at the end of the proof, which explains why we also need to keep track of $t$ in the various bounds.

\subsubsection*{Term B}
For the term B, we have from the identities in \eqref{eq:Qrelation}
\begin{align}\nonumber
&\EE\left[ \LB\theta_n^{\Xm^n,H^n}\RB^2e^{\gamma_n^{\Xm^n,H^n}} \right]\\
 & = \mathbb{E}\Bigg[\LB \frac1{K}\trace \left( Q \frac{ H  H \htp}K \right)^2 + \frac{2\sigma^2}{K} \trace \left( Q  \frac{ H \Xm\Xm\htp H \htp}{nK} Q \right) \RB\nonumber\\
 &\qquad\qquad \times e^{\gamma_n^{\Xm^n,H^n}} \Bigg]\\
&= \mathbb{E}\Bigg[ \Bigg( \frac1K\trace Q \frac{ H  H \htp}{K} + \frac{\sigma^2}{K}\trace Q ^2\frac{ H  H \htp}{K}\nonumber\\
&\quad -\frac{2\sigma^2}{K}\trace Q ^2\frac{ H \LB\Id_K-\frac{1}{n}\Xm\Xm\htp\RB H \htp}{K} \Bigg) e^{\gamma_n^{\Xm^n,H^n}}\Bigg] + \Oc\LB\frac1K\RB\\
&= \mathbb{E}\LSB \LB c - \frac{\sigma^4}{K}\trace Q ^2 -\frac{2\sigma^2}{K}\trace Q ^2\frac{ H \Am H \htp}{K} \RB e^{\gamma_n^{\Xm^n,H^n}}\RSB\nonumber\\
\label{eq:partB}
&\qquad + \Oc\LB\frac1K\RB.
\end{align}
To proceed with this term, which is essentially equal to the product of the expectations of the two arguments, we rely on Remark~\ref{rem:cauchyschwarz} in Appendix~\ref{app:Gaussian_relResults}. Using Proposition~\ref{prop:vartrace} in Appendix~\ref{app:aux_RMTaux} and Proposition~\ref{prop:dettrace} in Appendix~\ref{app:aux_RMTaux} to bound the variances of each term, we have
\begin{align}
	\EE\left[\LB\theta_n^{\Xm^n,H^n}\RB^2e^{\gamma_n^{\Xm^n,H^n}} \right] & = \LB c -\sigma^4\delta_1(\sigma^2) \RB \mathbb{E}\LSB e^{\gamma_n^{\Xm^n,H^n}}\RSB \nonumber\\
	\label{eq:termBapprox}
	&\quad + \Oc \left( \frac1{\sqrt{K}} \right) 
\end{align}
where we used in particular $\sqrt{K^{-3}\trace\Am^2}\leq 1/\sqrt{K}$.

Combining \eqref{eq:phiprime}, \eqref{eq:termAapprox}, and \eqref{eq:termBapprox} we finally obtain the differential equation \eqref{eq:phiprime2} on the next page, where $\mu_{n}^{\Xm^n}$, $\LB\theta_n^{\Xm^n}\RB^2$, and $\bar{\varepsilon}_n^{\Xm^n}(t)$ are defined in \eqref{eq:defmun}--\eqref{eq:tmpvareps} for some non-zero polynomials with nonnegative coefficients $P_1(t)$ and $P_2(t)$.
\begin{figure*}
\begin{align}\label{eq:phiprime2}
\frac{\partial \phi_n^{\Xm^n}(t)}{\partial t} &= \LB \iv \mu_{n}^{\Xm^n} - t\LB\theta_n^{\Xm^n}\RB^2\RB\mathbb{E}\LSB e^{\gamma_n^{\Xm^n,H^n}}\RSB + \bar{\varepsilon}_n^{\Xm^n}(t)\\
\label{eq:defmun}
  \mu_{n}^{\Xm^n}  &=  \sqrt{nK}\left\{ \int_{\sigma^2}^\infty \frac{c\LB 1-c + 2u\delta_0(u) - \frac{u^2}{c}\delta_0(u)^2\RB}{u\LB 1-c +u\LB 1 +2\delta_0(u)\RB\RB} du\right\}\\
 \LB\theta_n^{\Xm^n}\RB^2 &=  \frac nK\left\{ \int_{\sigma^2}^\infty\frac{\delta_0(u) - \sigma^2\delta_1(u)}{1 - c + u\LB 1 + 2\delta_0(u)\RB} du+ \frac{\gamma_1\LB \sigma^2\RB\frac1K\trace\Am^2}{\LB 1+\delta_0(\sigma^2)\RB^2} +\frac{K}{n}\LB c -\sigma^4\delta_1(\sigma^2)\RB \right\}\label{eq:defthetan}\\
	\bar{\varepsilon}_n^{\Xm^n}(t) &= \Oc\LB \frac{P_1(t)}{\sqrt{K}} + \frac{tP_2(t)}{\sqrt{K}} \frac1K\trace \Am^2 \RB\label{eq:tmpvareps}
\end{align}
\hrulefill
\end{figure*}

Using Lemma~\ref{lem:integrals} in Appendix~\ref{app:Gaussian_relResults} and the definition of $\gamma_1, \delta_1$ in Proposition~\ref{prop:dettrace}, the expressions of $\mu_{n}^{\Xm^n}$ and $\theta_n^{\Xm^n}$ can be simplified as:
\begin{align}\label{eq:integralmun}
\mu_{n}^{\Xm^n} &= \sqrt{nK} C \\\label{eq:integralthetan}
\theta_n^{\Xm^n} &= \LSB\theta_{-}^2 + \zeta \frac1K\trace\Am^2\RSB^{\frac12}
\end{align}
where $\theta_{-}$ is defined in the statement of Theorem~\ref{th-main} and $\zeta=-\delta_0'(\sigma^2)(1+\delta_0(\sigma^2))^{-1}$. Note that we have used the relation $\delta_0(\sigma^2)=-\delta_1(\sigma^2)$.
Before we continue with the main proof, we will show that $\theta_{-}^2>0$ and $\zeta> 0$.

For the former, first note that the logarithm term of $\theta_{-}$ in \eqref{def-theta_-} is well defined. Indeed, for $c\geq 1$, the argument is clearly positive. For $c<1$, by Property~\ref{properties:delta0}~$(iv)$ in Appendix~\ref{app:Gaussian_relResults}, $\delta_0(\sigma^2)^2(1+\delta_0(\sigma^2))^{-2}=(c-\sigma^2\delta_0(\sigma^2))^2<c^2$, with the inequality arising from Property~\ref{properties:delta0}~$(i)$ and $(ii)$ in Appendix~\ref{app:Gaussian_relResults}; this then implies that the argument is greater than $1-c>0$. Obviously, in both cases, as the argument of the logarithm is less than one, the logarithm itself is negative. This implies that
\begin{align}
\theta_{-}^2 &= -\beta \log\left( 1 - \frac1c \frac{\delta_0(\sigma^2)^2}{(1+\delta_0(\sigma^2))^2} \right) + \left(c + \sigma^4\delta_0'(\sigma^2)\right)\\
&\overset{(a)}{>} c - \frac{\sigma^4\delta_0\LB\sigma^2\RB\LB 1+\delta_0\LB\sigma^2\RB\RB}{1-c+\sigma^2\LB1+ \delta_0\LB\sigma^2\RB\RB + \sigma^2\delta_0\LB\sigma^2\RB}\\
&\overset{(b)}{=} c - \frac{c\sigma^2\LB 1+\delta_0\LB\sigma^2\RB\RB}{\frac{c}{\delta_0\LB\sigma^2\RB} + \sigma^2\delta_0\LB\sigma^2\RB} \\
&\overset{(c)}{>} c\LB 1 - \frac{\sigma^2\LB 1+\delta_0\LB\sigma^2\RB\RB}{\sigma^2+\sigma^2\delta_0\LB\sigma^2\RB}\RB\\\label{eq:theta_-inequ}
&= 0
\end{align}
where $(a)$ follows from the definition of $\delta_0'(x)$ established in Property~\ref{properties:delta0}~$(vi)$ in Appendix~\ref{app:Gaussian_relResults}, $(b)$ follows from Property~\ref{properties:delta0}~$(iii)$ in Appendix~\ref{app:Gaussian_relResults}, and $(c)$ is due to Property~\ref{properties:delta0}~$(ii)$ in Appendix~\ref{app:Gaussian_relResults} which implies that $\frac{c}{\delta_0\LB \sigma^2\RB}>\sigma^2$.

Concerning $\zeta$, we first show that $\delta_1(\sigma^2)=-\delta_0'(\sigma^2)>0$ (where this identity follows from Property~\ref{properties:delta0}~$(vi)$ in Appendix~\ref{app:Gaussian_relResults}). Since $c^{-1}\delta_0(\sigma^2)$ is the Stieltjes transform of the Mar\u{c}enko-Pastur law $\mu_c$ taken in $-\sigma^2$ (see, e.g., \cite[Chapter 3.2]{COUbook}), we can conclude that
\begin{align}
 \delta_1(\sigma^2) &= -\delta_0'(\sigma^2) = \frac1c\int \frac1{(t+\sigma^2)^2}\mu_c(dt) > 0.
\end{align}
Since also $\delta_0(\sigma^2)>0$, it follows that $\zeta=-\beta\delta_0'(\sigma^2)(1+\delta_0(\sigma^2))>0$.

We now relate $\phi_n^{\Xm^n}(t) = \mathbb{E}\LSB \phi_n^{\Xm^n,H^n}(t)\RSB$ and $\mathbb{E}\LSB e^{\gamma_n^{\Xm^n,H^n}}\RSB$ with the help of the previously established results. Starting from \eqref{eq:equ_diff1}, one can easily show  that
\begin{align}
\left| \phi_n^{\Xm^n,\Hm^n}(t) - e^{\gamma_n^{\Xm^n,\Hm^n}} \right| \leq M t^4 n^{-2}
\end{align}
for some constant $M$ independent of $\Hm$, $t$, and $n$, from which
\begin{align}
\phi_n^{\Xm^n}(t) = \mathbb{E}\LSB \phi_n^{\Xm^n,H^n}(t)\RSB = \mathbb{E}\LSB e^{\gamma_n^{\Xm^n,H^n}}\RSB + \Oc\LB t^4n^{-2}\RB
\end{align}
or, equivalently,
\begin{align}
\mathbb{E}\LSB e^{\gamma_n^{\Xm^n,H^n}}\RSB = \phi_n^{\Xm^n}(t)  + \Oc\LB t^4n^{-2}\RB.
\end{align}
Replacing the last equation in \eqref{eq:phiprime2} leads to
\begin{align}
\frac{\partial \phi_n^{\Xm^n}(t)}{\partial t} &= \LB \iv \mu_{n}^{\Xm^n} - t\LB\theta_n^{\Xm^n}\RB^2\RB \phi_n^{\Xm^n}(t)\nonumber\\
&\quad  +  \LB \iv \mu_{n}^{\Xm^n} - t\LB\theta_n^{\Xm^n}\RB^2\RB \Oc\LB t^4n^{-2}\RB + \bar{\varepsilon}_n^{\Xm^n}(t).
\end{align}
One can verify from \eqref{eq:defmun} and \eqref{eq:defthetan} that
\begin{align}
 \mu_{n}^{\Xm^n} &= \Oc(n)\\
\LB\theta_n^{\Xm^n}\RB^2 &= \Oc\LB 1+\frac1K\trace\Am^2\RB.\label{eq:growththeta}
\end{align}
Hence
\begin{align}\label{eq:phinprimeX}
 \frac{\partial \phi_n^{\Xm^n}(t)}{\partial t} &= \LB{\bf i}\mu_{n}^{\Xm^n} -t\LB\theta_n^{\Xm^n}\RB^2\RB\phi_n^{\Xm^n}(t) + {\varepsilon}_n^{\Xm^n}(t)
\end{align}
where ${\varepsilon}_n^{\Xm^n}(t)$ satisfies
\begin{align}\label{eq:defvarepsilonprime}
	{\varepsilon}_n^{\Xm^n}(t) &= \Oc\left( \frac{P_1(t)}{\sqrt{K}} + \frac{tP_2(t)}{\sqrt{K}} \frac1K\trace \Am^2 \right).
\end{align}

\subsection*{\underline{Step 3:}}
Solving the differential equation \eqref{eq:phinprimeX}, we arrive at
\begin{align}
\phi_n^{\Xm^n}(t) &= e^{{\bf i}t\mu_n^{\Xm^n} - \frac{t^2}{2}\LB \theta_n^{\Xm^n}\RB^2}\Bigg( 1 + \nonumber\\
&\qquad  \int_{0}^t e^{-{\bf i}x\mu_n^{\Xm^n} + \frac{x^2}{2}\LB \theta_n^{\Xm^n}\RB^2} \varepsilon_n^{\Xm^n}(x)dx\Bigg)\\\label{eq:difeqphi5}
&= e^{{\bf i}t\mu_n^{\Xm^n} - \frac{t^2}{2}\LB \theta_n^{\Xm^n}\RB^2} + \tilde{\varepsilon}_n^{\Xm^n}(t)
 \end{align}
with $\tilde{\varepsilon}_n^{\Xm^n}(t) = \Oc\LB t {\varepsilon}_n^{\Xm^n}(t)\RB$.

Denote $\tilde{\phi}_n^{\Xm^n}(t) = \mathbb{E}\LSB e^{{\bf i}\frac{t}{\theta_n^{\Xm^n}}\LB \Gamma_n^{\Xm^n} - \mu_n^{\Xm^n}\RB}\RSB$. Then, from \eqref{eq:difeqphi5},
\begin{align}
\tilde{\phi}_n^{\Xm^n}(t)&= \mathbb{E}\LSB e^{{\bf i}\frac{t}{\theta_n^{\Xm^n}}\LB \Gamma_n^{\Xm^n} - \mu_n^{\Xm^n}\RB}\RSB\\
& = \phi_n^{\Xm^n}\LB \frac{t}{\theta_n^{\Xm^n}}\RB e^{-{\bf i} t\frac{\mu_n^{\Xm^n}}{\theta_n^{\Xm^n}}}\\
&= e^{-\frac{t^2}{2}} +\tilde{\varepsilon}_n^{\Xm^n}\LB \frac{t}{\theta_n^{\Xm^n}}\RB e^{-{\bf i} t\frac{\mu_n^{\Xm^n}}{\theta_n^{\Xm^n}}}\label{eq:convtmp10}.
\end{align}

To conclude, we need to control the term $\tilde{\varepsilon}_n^{\Xm^n}\LB t(\theta_n^{\Xm^n})^{-1}\RB$. This is where the precision on $\varepsilon_n^{\Xm_n}(t)$ from \eqref{eq:defvarepsilonprime} is used. Take $t\geq 0$ fixed. First, observe from \eqref{eq:theta_-inequ} that $\theta_n^{\Xm^n}\ge \theta_{-} >0$.  

We then have
\begin{align}
&\tilde{\varepsilon}_n^{\Xm^n}\LB \frac{t}{\theta_n^{\Xm^n}}\RB\nonumber\\
&= \Oc \left( \frac{t}{\theta_n^{\Xm^n}} \varepsilon_n^{\Xm^n}\LB \frac{t}{\theta_n^{\Xm^n}}\RB \right) \\
&= \Oc\left( \frac{P_1\left(t(\theta_n^{\Xm^n})^{-1}\right)}{\sqrt{K}\theta_n^{\Xm^n}} + \frac{P_2\left(t(\theta_n^{\Xm^n})^{-1}\right)}{\sqrt{K}\left(\theta_n^{\Xm^n}\right)^2} \frac1K\trace \Am^2 \right) \\
&= \Oc\left( \frac1{\sqrt{K}} \right)
\end{align}
where, in the last equality, we used $P_1\left(t(\theta_n^{\Xm^n})^{-1}\right)(\theta_n^{\Xm^n})^{-1}\leq P_1(t\theta_{-}^{-1})\theta_{-}^{-1}$, $P_2\left(t(\theta_n^{\Xm^n})^{-1}\right)\leq P_2(t\theta_{-}^{-1})$, both bounded for $t$ fixed, and
\begin{align}
	\frac{\frac1K\trace\Am^2}{\LB{\theta_n^{\Xm^n}}\RB^2} &= \frac{\frac1K\trace\Am^2}{\theta_{-}^2 + \zeta \frac1K\trace\Am^2 } \le \frac1{\zeta}<\infty.
\end{align}

We conclude that
\begin{align}\label{eq:convtmp11}
 \left|\tilde{\varepsilon}_n^{\Xm^n}\LB \frac{t}{\theta_n^{\Xm^n}}\RB e^{-{\bf i} t\frac{\mu_n^{\Xm^n}}{\theta_n^{\Xm^n}}}\right| = \Oc\LB \frac1 {\sqrt{K}}\RB.
\end{align}

Take now $\mathbb{P}_{X^n}\in\Pc\LB \Sc^n_=\RB$ for all $n$ and let  $\tilde{\phi}_n(t)\defines \mathbb{E}\LSB \tilde{\phi}_n^{X^n}(t) \RSB$. Then, from \eqref{eq:convtmp10} and \eqref{eq:convtmp11},
\begin{align}
	\tilde{\phi}_n(t) = e^{-\frac{t^2}{2}} + \Oc\left( \frac1{\sqrt{K}} \right).
\end{align}
Taking $t<0$, and using $\tilde{\phi}_n(-t)=\tilde{\phi}_n(t)^*$, the result above generalizes to $t\in\mathds{R}$.

This implies by L\'{e}vy's continuity theorem that
\begin{align}
\frac{\Gamma_n -\mu_n}{\theta_n}\Rightarrow \Nc\LB 0,1\RB 
\end{align}
where we have defined $\mu_n=\mu_n^{X^n}$ and $\theta_n=\theta_n^{X^n}$. This terminates the proof. 

\section{Additional random matrix results}\label{app:aux_RMT}
\subsection{Auxiliary results}\label{app:aux_RMTaux}

\begin{lemma}\label{lem:varbnds}
Let $G\in\mathds{C}^{M\times L}$ have i.i.d.\@ entries $G_{ij}\sim\Cc\Nc(0,1)$ and let $\Sm\in\mathds{C}^{L\times M}$ and $\Tm\in\mathds{C}^{M\times M}$. Then,
\begin{align}
 (i)\qquad\quad \var\LSB \trace \Sm  G\RSB &= \trace\Sm\Sm\htp\\
 (ii)\quad\var\LSB \trace \Tm  G G\htp\RSB &\le 2L\trace\Tm\Tm\htp.
\end{align}
\end{lemma}
\begin{IEEEproof}
The proof of part $(i)$ is obvious. Part $(ii)$ is proved by a mere application of Lemma~\ref{lem:PNI} and Lemma~\ref{lem:useful_derivatives} in Appendix~\ref{app:Gaussian_relResults}.
\end{IEEEproof}

\begin{lemma}\label{lem:XXomegader}
Let $ G \in\mathds{C}^{M\times L}$ have i.i.d.\@ entries $G_{ij}\sim\Cc\Nc(0,1)$. Let $\Tm\in\mathds{C}^{M\times M}$ be a deterministic matrix and $\omega$ be a function of $G$. Then,
\begin{align}
\mathbb{E}\LSB \trace\Tm GG\htp e^{\omega}\RSB &= L\trace\Tm\mathbb{E}\LSB e^{\omega}\RSB \nonumber\\
&\quad + \mathbb{E}\LSB\sum_{i,j}\frac{\partial \omega}{\partial G_{ij}^*}\LSB G\htp\Tm\RSB_{ji} e^{\omega}\RSB.
\end{align}
\end{lemma}
\begin{IEEEproof}
This follows immediately from Lemma~\ref{lem:IBP} in Appendix~\ref{app:Gaussian_relResults}.
\end{IEEEproof}

\begin{prop}\label{prop:vartrace}
Let $H\in\mathds{C}^{N\times K}$ have i.i.d.\@ elements $H_{ij}\sim\Cc\Nc(0,1)$ and define the functionals $Q(x) = \LB \frac1K HH\htp + x\Id_N\RB^{-1}$ and $\tilde{Q}(x) = \LB \frac1K H\htp H + x\Id_K\RB^{-1}$ for $x>0$. Further, let $\Cm,\Dm\in\mathds{C}^{N\times N}$ and $\tilde{\Cm},\tilde{\Dm}\in\mathds{C}^{K\times K}$. Then, for $u,v>0$ and any nonnegative integer $m$, the following holds:
\begin{align}
	&(i)\  \var\LSB \frac1K\trace\Cm Q(u)\Dm Q(v)^m\RSB\nonumber\\
	&\qquad\qquad \le  2\frac{\LB \sqrt{\frac{v}{u}} + m\RB^2}{u^2v^{2m+1}} \frac{\lVert\Dm\rVert^2}{K^3}\trace\Cm\Cm\htp\\
	&(ii)\ \var\LSB \frac1K\trace\tilde{\Cm}\tilde{Q}(u)\tilde{\Dm}\tilde{ Q}(v)^m\RSB\nonumber\\
	&\qquad\qquad \le   2\frac{\LB \sqrt{\frac{v}{u}} + m\RB^2}{u^2v^{2m+1}}\frac{\lVert\tilde{\Dm}\rVert^2}{K^3}\trace\tilde{\Cm}\tilde{\Cm}\htp\\
	&(iii)\ \var\LSB \frac1K\trace Q(u) Q(v)^m\frac{ H\tilde{\Cm} H\htp}{K}\RSB\nonumber\\
	&\qquad\qquad \le 2\frac{\LB \sqrt{\frac{v}{u}} + 2m\RB^2}{v^{2m+1}}\frac1{K^3}\trace\tilde{\Cm}\tilde{\Cm}\htp\\
	&(iv)\ \var\LSB \frac1K\trace Q(u) Q(v)^m\frac{ H\tilde{\Cm} H\htp}{K}\RSB \nonumber\\
	&\qquad\qquad\le 2\frac{\LB 2\sqrt{\frac{v}{u}} + 2m-1\RB^2}{u^2v^{2m-1}}\frac1{K^3}\trace\tilde{\Cm}\tilde{\Cm}\htp,\quad m\ge 1.
\end{align}

Moreover, for $\Cm$ and $\tilde{\Cm}$ Hermitian,
\begin{align}
	(v)\ \var\LSB \frac1K\trace\Cm Q(u)\Cm Q(v)^m\RSB &\le 2 \frac{\LB \sqrt{\frac{v}{u}} + m\RB^2}{u^2v^{2m+1}}\frac1{K^3}\trace\Cm^4\\
	(vi)\ \var\LSB \frac1K\trace\tilde{\Cm}\tilde{ Q}(u)\tilde{\Cm}\tilde{ Q}(v)^m\RSB &\le  2 \frac{\LB \sqrt{\frac{v}{u}} + m\RB^2}{u^2v^{2m+1}}\frac1{K^3}\trace\tilde{\Cm}^4.
\end{align}
\end{prop}
\begin{IEEEproof}
The results follow from the successive applications of Lemma~\ref{lem:PNI} and Lemma~\ref{lem:matrixineq} in Appendix~\ref{app:Gaussian_relResults}.
\end{IEEEproof}

\begin{prop}\label{prop:dettrace}
Let $\{H^n\}_{n=1}^\infty$, where $H^n\in\mathds{C}^{N\times K}$ has i.i.d.\@ elements $H^n_{ij}\sim\Cc\Nc(0,1)$, and define $Q^n(x) = \LB \frac1K H^n \LB H^n\RB \htp + x\Id_N\RB^{-1}$ for $x>0$. Let $\{\Cm^n\}_{n=1}^\infty$, where $\Cm^n\in\mathds{C}^{N\times N}$. Then, for $u\ge\sigma^2>0$ and any nonnegative integer $m$, the following holds as $n\xrightarrow[]{(\beta,c)}\infty$:
\begin{align}
&(i)\ \mathbb{E}\LSB\frac1K\trace Q^n(u) Q^n(\sigma^2)^m\frac{H^n\Cm^n \LB H^n\RB\htp}{K}\RSB \nonumber\\
&\quad=\gamma_m\LB u\RB\frac1K\trace\Cm^n + \Oc\LB\sqrt{\frac1{u^2K^5}\trace\Cm^n\LB\Cm^n\RB\htp}\RB\\
&(ii)\ \mathbb{E}\LSB\frac1K\trace Q^n(u)Q^n(\sigma^2)^m\RSB = \delta_m\LB u\RB + \Oc\LB\frac1{un^2}\RB
\end{align}
where, for $m\ge 1$,
\begin{align}
 \gamma_m(u) &= \delta_{m-1}(u) - \sigma^2\delta_m(u) \\
\delta_m(u) &= \frac{\delta_{m-1}(u)\LSB 1+\delta_0\LB\sigma^2\RB\RSB}{1-c + \sigma^2 \LSB 1 + \delta_0\LB\sigma^2\RB \RSB + u\delta_0(u)}\nonumber\\
&\quad+\frac{\sum_{k=1}^{m-1}\LSB\delta_{k-1}(u) - \sigma^2\delta_k(u)\RSB\delta_{m-k}\LB\sigma^2\RB}{1-c + \sigma^2 \LSB 1 + \delta_0\LB\sigma^2\RB \RSB + u\delta_0(u)}
\end{align}
and
\begin{align}
\gamma_0\LB u\RB &= c - u\delta_0(u)
\end{align}
with $\delta_0(u)$ as defined in Theorem~\ref{thm:detequ_stieltjes} in Appendix~\ref{app:Gaussian_relResults}.
\end{prop}
\begin{IEEEproof}
In order to simplify the notations, we drop the dependence of $n$, e.g., we write $H$ instead of $H^n$. We begin by standard Gaussian calculus based on the integration by parts formula (Lemma~\ref{lem:IBP} in Appendix~\ref{app:Gaussian_relResults}):
\begin{align}
&\mathbb{E}\LSB\frac1K\trace Q(u) Q(\sigma^2)^m\frac{ H\Cm H\htp}{K}\RSB\nonumber\\
&= \frac1{K^2}\sum_{i,j,k,r,s} \mathbb{E}\LSB H_{ij}C_{jk}H^*_{rk}Q(u)_{rs}\LSB Q(\sigma^2)^m\RSB_{si}\RSB\\
&= \frac1{K^2}\sum_{i,j,k,r,s}C_{jk} \mathbb{E}\LSB\frac{\partial \LB H^*_{rk}Q(u)_{rs}\LSB Q(\sigma^2)^m\RSB_{si}\RB }{\partial H_{ij}^*}\RSB\\
&= \frac1{K^2}\sum_{i,j,k,r,s}C_{jk} \mathbb{E}\Bigg[ \delta_{ir}\delta_{jk}Q(u)_{ls}\LSB Q(\sigma^2)^m\RSB_{si}\nonumber\\
&\qquad-\frac1K H_{rk}^*\LSB Q(u) H\RSB_{rj}Q(u)_{is}\LSB Q(\sigma^2)^m\RSB_{si} \nonumber\\
&\qquad+ H^*_{rk}Q(u)_{rs}\frac{\partial \LSB Q(\sigma^2)^m\RSB_{si}}{\partial H_{ij}^*}\Bigg]\\
&= \frac1K\trace\Cm\mathbb{E}\LSB \frac1K\trace Q(u) Q(\sigma^2)^m\RSB\nonumber\\
&\quad - \mathbb{E}\LSB \frac1K\trace Q(u)\frac{ H\Cm H\htp}{K}\frac1K\trace Q(u) Q(\sigma^2)^m\RSB\nonumber\\\label{eq:Qtcalcl}
&\quad+ \frac1{K^2}\sum_{i,j,s}\mathbb{E}\LSB \LSB\Cm H\htp Q(u)\RSB_{js} \frac{\partial \LSB Q(\sigma^2)^m\RSB_{si}}{\partial H_{ij}^*} \RSB.
\end{align}

To continue, we will develop the term  $\frac{\partial \LSB Q(\sigma^2)^m\RSB_{si}}{\partial H_{ij}^*}$ as follows:
\begin{align}
&\frac{\partial \LSB Q(\sigma^2)^m\RSB_{si}}{\partial H_{ij}^*}\nonumber\\
&= \sum_{k=1}^m\sum_{p,q} \LSB  Q(\sigma^2)^{k-1}\RSB_{sp}\frac{\partial Q(\sigma^2)_{pq}}{\partial H_{ij}^*}\LSB Q(\sigma^2)^{m-k}\RSB_{qi}\\
&= -\frac1K \sum_{k=1}^m\sum_{p,q} \LSB  Q(\sigma^2)^{k-1}\RSB_{sp} \LSB  Q(\sigma^2) H\RSB_{pj}\nonumber\\
&\qquad\qquad\qquad\qquad\times Q(\sigma^2)_{iq} \LSB Q(\sigma^2)^{m-k}\RSB_{qi}\\\label{eq:Qtderiv}
&= -\frac1K\sum_{k=1}^m\LSB Q(\sigma^2)^k H\RSB_{sj}\LSB Q(\sigma^2)^{m-k+1}\RSB_{ii}.
\end{align}

Replacing \eqref{eq:Qtderiv} in \eqref{eq:Qtcalcl}, we arrive at
\begin{align}
 &\mathbb{E}\LSB\frac1K\trace Q(u) Q(\sigma^2)^m\frac{ H\Cm H\htp}{K}\RSB \nonumber\\
 &= \frac1K\trace\Cm\mathbb{E}\LSB \frac1K\trace Q(u) Q(\sigma^2)^m\RSB\nonumber\\
 &\quad - \mathbb{E}\LSB \frac1K\trace Q(u)\frac{ H\Cm H\htp}{K}\frac1K\trace Q(u) Q(\sigma^2)^m\RSB \nonumber\\
 &\quad -\sum_{k=1}^m\mathbb{E}\LSB\frac1K\trace Q(u) Q(\sigma^2)^k\frac{ H\Cm H\htp}{K}\frac1K\trace Q(\sigma^2)^{m-k+1}\RSB.
\end{align}

By Proposition~\ref{prop:vartrace} in Appendix~\ref{app:aux_RMTaux} and Remark~\ref{rem:cauchyschwarz} in Appendix~\ref{app:Gaussian_relResults}, we have
\begin{align}
&\mathbb{E}\LSB \frac1K\trace Q(u)\frac{ H\Cm H\htp}{K}\frac1K\trace Q(u) Q(\sigma^2)^m\RSB\nonumber\\
&= \mathbb{E}
\LSB\frac1K\trace Q(u)\frac{ H\Cm H\htp}{K}\RSB\mathbb{E}\LSB\frac1K\trace Q(u) Q(\sigma^2)^m\RSB\nonumber\\
&\quad + \Oc\LB\sqrt{\frac1{u^3K^5}\trace\Cm\Cm\htp}\RB\\
&\mathbb{E}\LSB\frac1K\trace Q(u) Q^k\frac{ H\Cm H\htp}{K}\frac1K\trace Q^{m-k+1}\RSB\nonumber\\
&= \mathbb{E}\LSB\frac1K\trace Q(u) Q^k\frac{ H\Cm H\htp}{K}\RSB \mathbb{E}\LSB\frac1K\trace Q^{m-k+1}\RSB\nonumber\\ &\quad+\Oc\LB\sqrt{\frac1{u^2K^5}\trace\Cm\Cm\htp}\RB
\end{align}
and, thus,
\begin{align}
&\mathbb{E}\LSB\frac1K\trace Q(u) Q(\sigma^2)^m\frac{ H\Cm H\htp}{K}\RSB \nonumber\\
 &= \frac1K\trace\Cm\mathbb{E}\LSB \frac1K\trace Q(u) Q(\sigma^2)^m\RSB\nonumber\\
 &\quad -  \mathbb{E}\LSB\frac1K\trace Q(u)\frac{ H\Cm H\htp}{K}\RSB\mathbb{E}\LSB\frac1K\trace Q(u) Q(\sigma^2)^m\RSB\nonumber\\
&\quad - \sum_{k=1}^m\mathbb{E}\LSB\frac1K\trace Q(u) Q(\sigma^2)^k\frac{ H\Cm H\htp}{K}\RSB \mathbb{E}\LSB\frac1K\trace Q(\sigma^2)^{m-k+1}\RSB\nonumber\\
\label{eq:traceapprox}
&\quad +\Oc\LB\sqrt{\frac1{u^2K^5}\trace\Cm\Cm\htp}\RB.
\end{align}

Define the following quantities
\begin{align}
\gamma^{\circ}_m\LB u, \Cm\RB &= \mathbb{E}\LSB\frac1K\trace Q(u) Q(\sigma^2)^m\frac{ H\Cm H\htp}{K}\RSB\ ,\ m=0,1,\dots\\
\delta_m^{\circ}\LB u\RB &= \mathbb{E}\LSB\frac1K\trace Q(u) Q(\sigma^2)^m\RSB\ ,\ m=-1,0,1,\dots
\end{align}
which satisfy the relations
\begin{align}
\delta_{-1}^\circ(\sigma^2) &= c\\
\gamma^{\circ}_m\LB u, \Id_K\RB &= \mathbb{E}\LSB \frac1K\trace Q(u) Q(\sigma^2)^m\frac{ H H\htp}{K}\RSB\\
&= \mathbb{E}\LSB \frac1K\trace Q(\sigma^2)^m\RSB -u\mathbb{E}\LSB\frac1K\trace Q(u) Q(\sigma^2)^m\RSB\\\label{eq:gamma_relation}
&= \delta^{\circ}_{m-1}(\sigma^2) - u\delta_m^{\circ}(u)\ ,\quad \forall m.
\end{align}
For $m\ge 1$, we also have from the relations in \eqref{eq:Qrelation}
\begin{align}\label{eq:gamma_relation2}
\gamma^{\circ}_m\LB u, \Id_K\RB &= \delta^{\circ}_{m-1}(u) - \sigma^2\delta^{\circ}_{m}(u).
\end{align}

Using these definitions, we can express \eqref{eq:traceapprox} as
\begin{align}
&\gamma^{\circ}_m\LB u, \Cm\RB = \frac1K\trace\Cm\delta_m^{\circ}(u) - \gamma_0^{\circ}\LB u,\Cm\RB\delta_m^{\circ}(u)\nonumber\\
\label{eq:traceapprox_re}
&\quad - \sum_{k=1}^m \gamma_k^{\circ}\LB u,\Cm\RB\delta^{\circ}_{m-k}(\sigma^2) +\Oc\LB\sqrt{\frac1{u^2K^5}\trace\Cm\Cm\htp}\RB.
\end{align}
Evaluating the last equation for $m=0$ and collecting the terms in $\gamma_0^\circ\LB u,\Cm\RB$ on one side, leads to
\begin{align}
\gamma^{\circ}_0\LB u, \Cm\RB =  \frac{\delta_0^{\circ}(u)}{1+\delta_0^{\circ}(u)}\frac1K\trace\Cm + \Oc\LB\sqrt{\frac1{u^2K^5}\trace\Cm\Cm\htp}\RB.
\end{align}
By Theorem~\ref{thm:detequ_stieltjes} in Appendix~\ref{app:Gaussian_relResults},
\begin{align}\label{eq:approx1}
\delta_0^{\circ}(u) = \delta_0(u) + \Oc\LB \frac1{u^4K^2}\RB.
\end{align}
Thus, we can define
\begin{align}
\gamma_0\LB u\RB \triangleq \frac{\delta_0(u)}{1+\delta_0(u)}
\end{align}
such that
\begin{align}
\gamma^{\circ}_0\LB u, \Cm\RB = \gamma_0\LB u\RB\frac1K\trace\Cm +  \Oc\LB\sqrt{\frac1{u^2K^5}\trace\Cm\Cm\htp}\RB
\end{align}
where we use the fact that $\left|\frac1K\trace\Cm\right|\leq \sqrt{\frac1{K^5}\trace \Cm\Cm\htp}$ and $u^{-4}\leq u^{-1}\sigma^6$ (since $u\geq \sigma^2$) to discard the term $\Oc(u^{-4}K^{-3}\trace \Cm)$.

For $m\ge 1$, we can gather the terms involving $\gamma^{\circ}_m\LB u, \Cm\RB$ in \eqref{eq:traceapprox_re} on one side, replace $\gamma_0^{\circ}\LB u, \Cm\RB$ by $\gamma_0\LB u\RB\frac1K\trace\Cm$ and $\delta_0^{\circ}(u)$ by $\delta_0(u)$, to obtain, iteratively on $m$,
\begin{align}
\gamma^{\circ}_m\LB u, \Cm\RB &= \frac{\frac1K\trace\Cm\delta_m^{\circ}(u) - \gamma_0\LB u\RB\frac1K\trace\Cm\delta_m^{\circ}(u)}{1+\delta_0\LB\sigma^2\RB}\nonumber\\
&\quad -\frac{\sum_{k=1}^{m-1} \gamma_k^{\circ}\LB u,\Cm\RB\delta^{\circ}_{m-k}(\sigma^2)}{1+\delta_0\LB\sigma^2\RB} \nonumber\\
\label{eq:gammrec}
&\quad +\Oc\LB\sqrt{\frac1{u^2K^5}\trace\Cm\Cm\htp}\RB.
\end{align}
From the last equation, we can obtain a recursive expression of $\delta_m(u)^{\circ}$ by letting $\Cm=\Id_K$ and using the relations \eqref{eq:gamma_relation} and \eqref{eq:gamma_relation2}:
\begin{align}
\delta_m^{\circ}(u) &= \frac{\delta_{m-1}^{\circ}(u)\LSB 1+\delta_0\LB\sigma^2\RB\RSB}{1-c + \sigma^2\LSB 1+\delta_0\LB\sigma^2\RB\RSB + u\delta_0(u)}\nonumber\\
&\quad +\frac{\sum_{k=1}^{m-1}\LSB \delta^{\circ}_{k-1}(u)- \sigma^2\delta^{\circ}_{k}(u)\RSB\delta^{\circ}_{m-k}(\sigma^2)}{1-c + \sigma^2\LSB 1+\delta_0\LB\sigma^2\RB\RSB + u\delta_0(u)}\nonumber\\
&\quad  + \Oc\LB\frac1{uK^2}\RB.
\end{align}
Note that the denominator of the RHS of the last equation is strictly positive (see Property~\ref{properties:delta0}~$(i)-(iii)$ in Appendix~\ref{app:Gaussian_relResults}).
For $m=1$, we obtain with the help of \eqref{eq:approx1}
\begin{align}
\delta_1^\circ(u) &= \frac{\delta_0^\circ(u)\LSB 1+\delta_0\LB\sigma^2\RB\RSB}{1-c + \sigma^2\LSB 1+\delta_0\LB\sigma^2\RB\RSB + u\delta_0(u)} + \Oc\LB\frac1{uK^2}\RB\\
&= \frac{\delta_0\LSB 1+\delta_0(u)\LB\sigma^2\RB\RSB}{1-c + \sigma^2\LSB 1+\delta_0\LB\sigma^2\RB\RSB + u\delta_0(u)} + \Oc\LB\frac1{uK^2}\RB.
\end{align}

Due to the recursive definition of $\delta_m^{\circ}(u)$, we can now conclude that
\begin{align}\label{eq:deltaorder}
\delta_m^{\circ}(u) = \delta_m(u) +\Oc\LB\frac1{uK^2}\RB
\end{align}
where
\begin{align}
\delta_m(u) &= \frac{\delta_{m-1}(u)\LSB 1+\delta_0\LB\sigma^2\RB\RSB}{1-c + \sigma^2\LSB 1+\delta_0\LB\sigma^2\RB\RSB + u\delta_0(u)}\nonumber\\
\label{eq:deltatrel1}
&\quad +\frac{ \sum_{k=1}^{m-1}\LSB \delta_{k-1}(u)- \sigma^2\delta_{k}(u)\RSB\delta_{m-k}\LB\sigma^2\RB}{1-c + \sigma^2\LSB 1+\delta_0\LB\sigma^2\RB\RSB + u\delta_0(u)}
,\ m\ge 1.
\end{align}

Using \eqref{eq:deltaorder} in \eqref{eq:gammrec}, we have so far proved that, for $m\ge1$,
\begin{align}
\gamma^{\circ}_m\LB u, \Cm\RB &= \frac{\frac1K\trace\Cm\delta_m(u) - \gamma_0\LB u\RB\frac1K\trace\Cm\delta_m(u)}{1+\delta_0\LB\sigma^2\RB}\nonumber\\
&\qquad -\frac{\sum_{k=1}^{m-1} \gamma_k^{\circ}\LB u,\Cm\RB\delta_{m-k}\LB\sigma^2\RB}{1+\delta_0\LB\sigma^2\RB} \nonumber\\
&\qquad  +\Oc\LB\sqrt{\frac1{u^2K^5}\trace\Cm\Cm\htp}\RB
\end{align}
where we have relied on the fact that $\gamma^\circ_k\LB u, \Cm\RB  \frac1{uK^2}  \le  \frac1{u^2K^3}\trace\Cm=\Oc\LB\sqrt{\frac1{u^2K^5}\trace\Cm\Cm\htp}\RB$.
In particular, for $m=1$, we obtain
\begin{align}
\gamma^{\circ}_1\LB u, \Cm\RB &= \frac{\delta_1(u) - \gamma_0\LB u\RB\delta_1(u) }{1+\delta_0\LB\sigma^2\RB}\frac1K\trace\Cm \nonumber\\
&\qquad +\Oc\LB\sqrt{\frac1{u^2K^5}\trace\Cm\Cm\htp}\RB.
\end{align}
Iterating the recursion $m-1$ times, we have proved that
\begin{align}
\gamma^{\circ}_m\LB u, \Cm\RB = \gamma_m\LB u\RB \frac1K\trace\Cm +\Oc\LB\sqrt{\frac1{u^2K^5}\trace\Cm\Cm\htp}\RB
\end{align}
where, for $m\ge 1$,
\begin{align}
 \gamma_t\LB u\RB = \frac{\delta_m(u)\LB 1 - \gamma_0\LB u\RB\RB - \sum_{k=1}^{m-1} \gamma_k\LB u\RB\delta_{m-k}\LB\sigma^2\RB}{1+\delta_0\LB\sigma^2\RB}.
\end{align}
Using now the relation $\gamma_0(u) = c - u\delta_0(u)$ (see Property~\ref{properties:delta0}~$(iv)$ in Appendix~\ref{app:Gaussian_relResults}), we write the last equation as
\begin{align}
\gamma_m\LB u\RB\LB 1+\delta_0\LB\sigma^2\RB\RB &= \delta_m(u)\LB 1 - c +u\delta_0(u)\RB \nonumber\\
&\qquad - \sum_{k=1}^{m-1} \gamma_k\LB u\RB\delta_{m-k}\LB\sigma^2\RB.
\end{align}
Adding $\delta_m(u)\sigma^2\LSB 1+\delta_0\LB\sigma^2\RB\RSB$ to both sides, we can express $\delta_m(u)$ as
\begin{align}
\delta_m\LB u\RB &= \frac{\LSB\gamma_m(u) + \sigma^2\delta_m(u)\RSB\LSB 1+\delta_0\LB\sigma^2\RB\RSB}{1 - c + \sigma^2\LSB 1+\delta_0\LB\sigma^2\RB\RSB + u\delta_0(u)}\nonumber\\
\label{eq:deltatrel2}
&\qquad +\frac{\sum_{k=1}^{m-1} \gamma_k\LB u\RB\delta_{m-k}\LB\sigma^2\RB}{1 - c + \sigma^2\LSB 1+\delta_0\LB\sigma^2\RB\RSB + u\delta_0(u)}.
\end{align}
Equating \eqref{eq:deltatrel2} and \eqref{eq:deltatrel1}, we can see that $\gamma_{m}(u)$ must satisfy the following relation
\begin{align}
\gamma_m(u) = \delta_{m-1}(u) - \sigma^2\delta_m(u)\ ,\qquad m\ge 1.
\end{align}
This terminates the proof.
\end{IEEEproof}

\subsection{Proof of Proposition~\ref{prop:tracexp} in Appendix~\ref{proof:CLT}}\label{proof:prop:tracexp}
We want to derive asymptotically exact approximations of $\mathbb{E}\LSB \trace Q(u)\frac{ H H\htp}{K} e^{\gamma_n^{\Xm^n,H^n}}\RSB$ (part $(i)$) and $\mathbb{E}\LSB \trace Q(\sigma^2)\frac{ H\Am H\htp}{K} e^{\gamma_n^{\Xm^n,H^n}}\RSB$ (part $(ii)$). 

In the proofs below, we will often use the notation $P(t)$ or $P_i(t)$ to refer to some non-zero polynomials in $t$ with nonnegative coefficients. These polynomials may take different values from one equation to the next.

\subsubsection*{Proof of part $(i)$}
By the product rule of differentiation, Lemma~\ref{lem:IBP} and Lemma~\ref{lem:useful_derivatives} in Appendix~\ref{app:Gaussian_relResults},
we obtain the chain of equations \eqref{eq:trQHHchiexp_part1}--\eqref{eq:trQHHchiexp} on the top of the next page.
\begin{figure*}
\begin{align}\nonumber
& \mathbb{E}\LSB \trace Q(u)\frac{ H H\htp}{K} e^{\gamma_n^{\Xm^n,H^n}} \RSB\\
\label{eq:trQHHchiexp_part1}
&= \mathbb{E}\LSB \frac1K\sum_{i,j,k} \frac{\partial\LB H^*_{kj} Q(u)_{ki} e{\gamma_n^{\Xm^n,H^n}}\RB}{\partial H_{ij}^*} \RSB\\
&= \mathbb{E}\LSB \frac1K\sum_{i,j,k} \left(\delta_{ik} Q(u)_{ki} e^{\gamma_n^{\Xm^n,H^n}}   -\frac{H_{kj}^*\LSB Q(u) H\RSB_{kj}Q(u)_{ii} }{K} e^{\gamma_n^{\Xm^n,H^n}} + H^*_{kj}Q(u)_{ki}\frac{\partial \gamma_n^{\Xm^n,H^n}}{\partial H_{ij}^*}e^{\gamma_n^{\Xm^n,H^n}}\right)\RSB\\
&= \mathbb{E}\LSB\LB \trace Q(u) - \frac1{K}\trace Q(u)\frac{ H H\htp}{K}\trace Q(u) + \frac1K\sum_{i,j} \frac{\partial {\gamma_n^{\Xm^n,H^n}}}{\partial H_{ij}^*} \LSB  H\htp  Q(u)\RSB_{ji}\RB e^{\gamma_n^{\Xm^n,H^n}} \RSB\\\label{eq:trQHHchiexp}
&= \mathbb{E}\LSB\LB \frac{N}{u} - \frac1u\trace Q(u)\frac{ H H\htp}{K} - \frac1{K}\trace Q(u)\frac{ H H\htp}{K}\trace Q(u) + \frac1K\sum_{i,j} \frac{\partial {\gamma_n^{\Xm^n,H^n}}}{\partial H_{ij}^*} \LSB  H\htp  Q(u)\RSB_{ji}\RB e^{\gamma_n^{\Xm^n,H^n}} \RSB
\end{align}
\hrulefill
\end{figure*}

Gathering the terms involving $\trace Q(u)\frac{ H H\htp}{K}$ on the LHS yields 
\begin{align}
&\mathbb{E}\LSB \trace Q(u)\frac{ H H\htp}{K}\LB1+\frac1{u} + \frac1K\trace Q(u)\RB e^{{\gamma_n^{\Xm^n,H^n}}}\RSB\nonumber\\
& = \frac{N}{u}\mathbb{E}\LSB e^{{\gamma_n^{\Xm^n,H^n}}}\RSB \nonumber\\
\label{eq:almostfinal}
&\quad + \frac1K\sum_{i,j}\mathbb{E}\LSB \frac{\partial {\gamma_n^{\Xm^n,H^n}}}{\partial H_{ij}^*} \LSB H\htp Q(u)\RSB_{ji}e^{{\gamma_n^{\Xm^n,H^n}}}\RSB.
\end{align}

Recall that $\gamma_n^{\Xm^n,H^n} = \iv t\mu_{n}^{\Xm^n,H^n}-\frac{t^2}2\LB\theta_{n}^{\Xm^n,H^n}\RB^2 + \iv \frac{t^3}{3}\kappa_{n}^{\Xm^n,H^n}$ \eqref{eq:defgamma}. From the standard derivation rules as provided in Lemma~\ref{lem:useful_derivatives} and Corollary~\ref{cor:derivatives} in Appendix~\ref{app:Gaussian_relResults}, denoting $Q=Q(\sigma^2)$ for brevity,
\begin{align}
	\frac{\partial \mu_{n}^{\Xm^n,H^n}}{\partial H_{ij}^*} &= \sqrt{\frac{n}{{K^3}}}\left[ Q H\right]_{ij} - \frac{n+1}{\sqrt{nK^3}}\LSB QHA\RSB_{ij} \nonumber\\
	&\qquad + \frac{n+1}{\sqrt{nK^5}}\left[  Q H\Am H\htp Q H \right]_{ij}\\
	&= -\frac{1}{\sqrt{nK^3}}\left[ Q H\right]_{ij} + \frac{1}{\sqrt{nK^3}}\LSB QH\Xm\Xm\htp\RSB_{ij}\nonumber\\
	\label{eq:difgam1}
	& \qquad + \frac{n+1}{\sqrt{nK^5}}\left[  Q H\Am H\htp Q H \right]_{ij}.
\end{align}
Similarly,
\begin{align}
	&\frac{\partial \LB\theta_{n}^{\Xm^n,H^n}\RB^2}{\partial H_{ij}^*}\nonumber\\
	&=-\frac{2(n+1)}{nK}\left[ \left( Q\frac1K H H\htp \right)^2 Q\frac1K H \right]_{ij} \nonumber\\
	&\quad + \frac{2(n+1)}{nK}\left[  Q\frac1K H H\htp Q\frac1K H \right]_{ij}\nonumber\\
	&\quad - \frac{2\sigma^2}{nK} \left[ Q\frac1K H\Xm\Xm\htp H\htp Q^2\frac1K H \right]_{ij} \nonumber \\
	&\quad + \frac{2\sigma^2}{nK}\left[  Q^2\frac1K H\Xm\Xm\htp\right]_{ij} \nonumber\\
	&\quad -\frac{2\sigma^2}{nK}\left[ Q^2\frac1K H\Xm\Xm\htp H\htp Q\frac1K H \right]_{ij} \\
	&= \frac{2\sigma^2(n+1)}{nK}\left[  Q\frac1K H H\htp Q^2\frac1K H \right]_{ij} \nonumber\\
	&\quad + \frac{2\sigma^4}{nK} \left[ Q^2\frac1K H\Xm\Xm\htp\tilde{ Q} \right]_{ij} \nonumber\\
	&\quad - \frac{2\sigma^2}K\left[  Q\frac1{Kn} H\Xm\Xm\htp H\htp Q^2\frac1K H\right]_{ij}
\end{align}
where, in the last equality, we used $\Id_N-\frac1KQHH\htp=\sigma^2Q$, $\Id_K-\frac1K H\htp Q H = \sigma^2\tilde{Q}$, and $QH=H\tilde{Q}$. Following the same derivation, we also have 
\begin{align}
	&\frac{\partial \kappa_{n}^{\Xm^n,H^n}}{\partial H_{ij}^*} = \nonumber\\
	&\frac{3\sigma^2(n+1)}{\sqrt{n^3K^3}}\left[ \left( Q\frac1K H H\htp \right)^2 Q^2\frac1K H\right]_{ij}\nonumber\\
	& + \frac{3\sigma^4}{\sqrt{n^3K^3}}\left[ Q\frac1K H\Xm\Xm\htp H\htp Q^3\frac1K H \right]_{ij} \nonumber \\
	& + \frac{3\sigma^4}{\sqrt{n^3K^3}}\left[ \frac1K HH\htp Q^3 \frac1K H \Xm\Xm\htp \tilde{Q} \right]_{ij}\nonumber\\
	& - \frac{3\sigma^2}{\sqrt{n^3K^3}} \left[ \frac1K HH\htp Q^2 \frac1K H \Xm\Xm\htp H\htp Q^2\frac1K H \right]_{ij}\label{eq:difgam3}.
\end{align}

Using these results, the second term on the RHS of \eqref{eq:almostfinal} can be developed as follows:
\begin{align}
&\frac1K\sum_{i,j}\mathbb{E}\LSB \frac{\partial {\gamma_n^{\Xm^n,H^n}}}{\partial H_{ij}^*} \LSB H\htp Q(u)\RSB_{ji}e^{{\gamma_n^{\Xm^n,H^n}}}\RSB\nonumber\\
&\overset{(a)}{=} {\bf i}t\frac{n+1}{\sqrt{nK}}\Bigg(\mathbb{E}\LSB \frac1K\trace Q\frac{ H\Xm\Xm\htp H\htp}{(n+1)K} Q(u) e^{{\gamma_n^{\Xm^n,H^n}}}\RSB \nonumber\\
&\qquad + \mathbb{E}\LSB \frac1K\trace Q \frac{ H\Am H\htp}{K} Q\frac{ H H\htp}{K} Q(u) e^{{\gamma_n^{\Xm^n,H^n}}}\RSB \Bigg) \nonumber\\
&\quad + \Oc\LB\frac{P(t)}{uK}\RB\\
&\overset{(b)}{=} {\bf i}t\frac{n+1}{\sqrt{nK}}\Bigg(\mathbb{E}\LSB \frac1K\trace Q\frac{ H H\htp}{K} Q(u) e^{{\gamma_n^{\Xm^n,H^n}}}\RSB  \nonumber\\
\label{eq:calc1}
&\quad- \mathbb{E}\LSB \frac{\sigma^2}K\trace Q \frac{ H\Am H\htp}{K} Q Q(u) e^{{\gamma_n^{\Xm^n,H^n}}}\RSB \Bigg) + \Oc\LB\frac{P(t)}{uK}\RB
\end{align}
for some polynomial $P(t)$, where $(a)$ follows from the derivative of $\gamma_n^{\Xm^n,H^n}$ as developed in \eqref{eq:difgam1}--\eqref{eq:difgam3} and the observations that all terms resulting from $\LB\theta_{n}^{\Xm^n,H^n}\RB^2$ and $\kappa_{n}^{\Xm^n,H^n}$ are $\Oc((uK)^{-1})$ and $\Oc( u^{-1}K^{-2})$, respectively, and $(b)$ follows from $ Q\frac{ H H\htp}{K} = \Id_N -\sigma^2 Q$ (see \eqref{eq:Qrelation}) and the definition of $\Am=\Id_K-\frac1{n+1}\Xm\Xm\htp$.

Based on Proposition~\ref{prop:vartrace} in Appendix~\ref{app:aux_RMTaux} and Lemma~\ref{lem:cauchyschwarz} in Appendix~\ref{app:Gaussian_relResults}, we find the following estimations:
\begin{align}
&\mathbb{E}\LSB \frac1K\trace Q\frac{ H H\htp}{K} Q(u) e^{{\gamma_n^{\Xm^n,H^n}}}\RSB \nonumber\\
\label{eq:calc2}
&= \mathbb{E}\LSB \frac1K\trace Q(u) Q\frac{ H H\htp}{K} \RSB \mathbb{E}\LSB  e^{{\gamma_n^{\Xm^n,H^n}}}\RSB + \Oc\LB\frac1{uK}\RB\\
&\mathbb{E}\LSB \frac{\sigma^2}K\trace Q \frac{ H\Am H\htp}{K} Q Q(u) e^{{\gamma_n^{\Xm^n,H^n}}}\RSB \nonumber\\
&= \mathbb{E}\LSB \frac{\sigma^2}K\trace Q(u) Q^2 \frac{ H\Am H\htp}{K} \RSB\mathbb{E}\LSB  e^{{\gamma_n^{\Xm^n,H^n}}}\RSB \nonumber\\
\label{eq:calc3}
&\quad+ \Oc\LB \sqrt{\frac1{u^2K^3}\trace\Am^2}\RB.
\end{align}

By Proposition~\ref{prop:dettrace} in Appendix~\ref{app:aux_RMTaux}, 
\begin{align}\label{eq:calc4}
&\mathbb{E}\LSB \frac1K\trace Q(u) Q\frac{ H H\htp}{K} \RSB = \delta_0(u) - \sigma^2\delta_1(u) + \Oc\LB\frac1{uK^2}\RB\\
&\mathbb{E}\LSB \frac{\sigma^2}K\trace Q(u) Q^2 \frac{ H\Am H\htp}{K} \RSB \nonumber\\
&= \sigma^2\gamma_2\LB u\RB\frac1K\trace\Am + \Oc\LB \sqrt{\frac1{u^2K^5}\trace\Am^2}\RB \\
\label{eq:calc5}
& = \Oc\LB \sqrt{\frac1{u^2K^5}\trace\Am^2}\RB.
\end{align}
Combining \eqref{eq:almostfinal}, \eqref{eq:calc1}, \eqref{eq:calc2}, \eqref{eq:calc3}, \eqref{eq:calc4}, and \eqref{eq:calc5}, we obtain
\begin{align}
&\mathbb{E}\LSB \trace Q(u)\frac{ H H\htp}{K}\LB1+\frac1{u} + \frac1K\trace Q(u)\RB e^{{\gamma_n^{\Xm^n,H^n}}}\RSB  \nonumber\\
&= \frac{N}{u}\mathbb{E}\LSB e^{{\gamma_n^{\Xm^n,H^n}}}\RSB + \Oc\left(\frac1{u\sqrt{K}} P(t) \right)  \nonumber \\
&\quad+\iv t\frac{n+1}{\sqrt{nK}}\LB \delta_0(u) - \sigma^2\delta_1(u) \RB \EE\left[ e^{\gamma_n^{\Xm^n,H^n}} \right] \\\label{eq:calc6}
&=  \frac{N}{u}\mathbb{E}\LSB e^{{\gamma_n^{\Xm^n,H^n}}}\RSB + \iv t\sqrt{\frac{n}{K}}\LB \delta_0(u) - \sigma^2\delta_1(u) 
\RB \EE\left[ e^{\gamma_n^{\Xm^n,H^n}} \right] \nonumber \\ 
&\quad+ \Oc\left(\frac1{u\sqrt{K}} P(t) \right)
\end{align}
for some other polynomial $P(t)$, where we used in particular $\sqrt{K^{-3}\trace \Am^2}\leq 1/\sqrt{K}$ and $\frac1{\sqrt{nK}}\LB \delta_0(u) - \sigma^2\delta_1(u) \RB = \Oc((uK)^{-1})$ by Property~\ref{properties:delta0} in Appendix~\ref{app:Gaussian_relResults}.

Next, we consider the LHS of \eqref{eq:almostfinal}. Let us first define the following quantities:
\begin{align}
\Psi &= \frac1K\trace Q(u), \qquad \Phi = \trace Q(u)\frac{ H H\htp}{K}.
\end{align}
Using these definitions, we can express the LHS of \eqref{eq:almostfinal} as 
\begin{align}
&\mathbb{E}\LSB \trace Q(u)\frac{ H H\htp}{K}\LB1+\frac1{u} + \frac1K\trace Q(u)\RB e^{{\gamma_n^{\Xm^n,H^n}}}\RSB \nonumber\\
& = \LB 1+\frac1u\RB  \mathbb{E}\LSB \Phi e^{{\gamma_n^{\Xm^n,H^n}}} \RSB + \mathbb{E}\LSB \Phi \Psi e^{{\gamma_n^{\Xm^n,H^n}}}\RSB.\label{eq:tmpref-0001}
\end{align}

We can now develop the second term on the RHS of the last equation as follows:
\begin{align}
&\mathbb{E}\LSB \Phi \Psi e^{{\gamma_n^{\Xm^n,H^n}}}\RSB\nonumber\\\label{eq:inequal-21}
&= \mathbb{E}\LSB \Psi\RSB\mathbb{E}\LSB \Phi e^{{\gamma_n^{\Xm^n,H^n}}}\RSB + \mathbb{E}\LSB \Phi \LB\Psi - \mathbb{E}\LSB\Psi\RSB \RB e^{{\gamma_n^{\Xm^n,H^n}}}\RSB\\
&\overset{(a)}{=} \mathbb{E}\LSB \Psi\RSB\mathbb{E}\LSB \Phi e^{{\gamma_n^{\Xm^n,H^n}}}\RSB - \mathbb{E}\LSB \Phi\RSB \mathbb{E}\LSB\Psi\RSB \mathbb{E}\LSB e^{{\gamma_n^{\Xm^n,H^n}}}\RSB\nonumber\\
&\quad + \mathbb{E}\LSB \Phi\RSB \mathbb{E}\LSB\LB \frac1u\frac{N}{K} - \frac1u \frac1K\trace Q(u)\frac{ H H\htp}{K} \RB e^{\gamma_n^{\Xm^n,H^n}}\RSB \nonumber\\
&\quad + \Oc\LB\frac1{u^2K}\RB\\
&= \mathbb{E}\LSB \Psi\RSB\mathbb{E}\LSB \Phi e^{{\gamma_n^{\Xm^n,H^n}}}\RSB - \mathbb{E}\LSB \Phi\RSB \mathbb{E}\LSB\Psi\RSB \mathbb{E}\LSB e^{{\gamma_n^{\Xm^n,H^n}}}\RSB \nonumber\\
&\quad + \frac1u\frac NK\mathbb{E}\LSB \Phi\RSB\mathbb{E}\LSB e^{{\gamma_n^{\Xm^n,H^n}}}\RSB - \frac1u \mathbb{E}\LSB \frac1K\Phi\RSB \mathbb{E}\LSB \Phi e^{{\gamma_n^{\Xm^n,H^n}}}\RSB\nonumber\\
&\quad + \Oc\LB\frac1{u^2K}\RB\\
&= \mathbb{E}\LSB \Phi e^{{\gamma_n^{\Xm^n,H^n}}}\RSB \LB \mathbb{E}\LSB\Psi\RSB -\frac1u \mathbb{E}\LSB \frac1K\Phi\RSB \RB \nonumber\\
&\quad - \mathbb{E}\LSB \Phi\RSB \mathbb{E}\LSB\Psi\RSB \mathbb{E}\LSB e^{{\gamma_n^{\Xm^n,H^n}}}\RSB + \frac1u\frac NK\mathbb{E}\LSB \Phi\RSB\mathbb{E}\LSB e^{{\gamma_n^{\Xm^n,H^n}}}\RSB \nonumber\\
&\quad + \Oc\LB\frac1{u^2K}\RB\\
&\overset{(b)}{=}  \mathbb{E}\LSB \Phi e^{{\gamma_n^{\Xm^n,H^n}}}\RSB\LB \delta_0(u) - \frac1u \gamma_0(u) \RB \nonumber\\
&\quad - K\gamma_0\LB u\RB\delta_0(u) \mathbb{E}\LSB e^{{\gamma_n^{\Xm^n,H^n}}}\RSB + \frac1u c K\gamma_0\LB u\RB \mathbb{E}\LSB e^{\gamma_n^{\Xm^n,H^n}}\RSB\nonumber\\
&\quad + \Oc\LB\frac1{uK}\RB\\
&\overset{(c)}{=} \LB 2\delta_0(u) -\frac{c}{u} \RB \mathbb{E}\LSB \Phi e^{{\gamma_n^{\Xm^n,H^n}}}\RSB \nonumber\\
&\quad + N\LB\frac{c}{u}-2\delta_0(u) + \frac{u}{c}\delta_0(u)^2 \RB\mathbb{E}\LSB e^{\gamma_n^{\Xm^n,H^n}}\RSB\nonumber\\
\label{eq:inequal-2}
&\quad + \Oc\LB\frac1{uK}\RB
\end{align}
where $(a)$ follows from Remark~\ref{rem:cauchyschwarz} and Proposition~\ref{prop:vartrace} in Appendix~\ref{app:aux_RMTaux}, and $\Psi$ is expanded using \eqref{eq:Qrelation}, $(b)$ follows by Proposition~\ref{prop:dettrace} in Appendix~\ref{app:aux_RMTaux} and the fact that $\left|e^{\gamma_n^{\Xm^n,H^n}} \right|\le 1$, and in $(c)$ we used $\gamma_0(u) = c-u\delta_0(u)$ (see Proposition~\ref{prop:dettrace}). Thus, \eqref{eq:tmpref-0001} can be expressed as
\begin{align}
&\mathbb{E}\LSB \Phi\LB 1+\frac1u + \Psi\RB e^{{\gamma_n^{\Xm^n,H^n}}}\RSB \nonumber\\
&= \LB 1+ \frac{1-c}{u} + 2\delta_0(u)\RB\mathbb{E}\LSB \Phi e^{{\gamma_n^{\Xm^n,H^n}}}\RSB\nonumber\\
&\quad + N\LB\frac cu - 2\delta_0(u) + \frac{u}{c}\delta_0(u)^2 \RB\mathbb{E}\LSB e^{{\gamma_n^{\Xm^n,H^n}}}\RSB\nonumber\\
\label{eq:psie}
&\quad + \Oc\LB\frac1{uK}\RB.
\end{align}

Equating the RHS of \eqref{eq:psie} and the RHS of \eqref{eq:calc6} and solving for $\mathbb{E}\LSB \Phi e^{{\gamma_n^{\Xm^n,H^n}}}\RSB$ leads to
\begin{align}
&\mathbb{E}\LSB \Phi e^{{\gamma_n^{\Xm^n,H^n}}}\RSB\nonumber\\
& = N\frac{\LB 1-c + 2u\delta_0(u) - \frac{u^2}{c}\delta_0(u)^2\RB}{1-c +u\LB 1 +2\delta_0(u)\RB}\mathbb{E}\LSB e^{{\gamma_n^{\Xm^n,H^n}}}\RSB\nonumber\\
&\quad +{\bf i}t\sqrt{\frac nK} u\frac{\delta_0(u) - \sigma^2\delta_1(u) 
}{1 - c + u\LB 1 + 2\delta_0(u)\RB}\mathbb{E}\LSB e^{{\gamma_n^{\Xm^n,H^n}}}\RSB \nonumber\\
\label{eq:case1}
&\quad + \Oc\left( \frac1{u\sqrt{K}}P(t) \right)
\end{align}
for some polynomial $P(t)$.

This concludes the proof of part $(i)$.

\subsubsection*{Proof of part $(ii)$}
We begin as in the proof of part $(i)$. From the derivative of $\gamma_n^{\Xm^n,H^n}$ in \eqref{eq:difgam1}--\eqref{eq:difgam3} and standard Gaussian calculus, we have
\begin{align}\nonumber
&\mathbb{E}\LSB \trace Q\frac{ H\Am H\htp}{K}e^{{\gamma_n^{\Xm^n,H^n}}}\RSB\nonumber\\
  & = \frac1K \sum_{i,j,k,l}\mathbb{E}\LSB H_{ij}A_{jk}H^*_{lk}Q_{li}e^{{\gamma_n^{\Xm^n,H^n}}}\RSB\\
& = \frac1K \sum_{i,j,k,l}A_{jk}\mathbb{E}\LSB\frac{\partial \LB H^*_{lk}Q_{li}e^{{\gamma_n^{\Xm^n,H^n}}}\RB}{\partial H_{ij}^*}\RSB\\
&= \frac1K\trace\Am \mathbb{E}\LSB \trace Q e^{{\gamma_n^{\Xm^n,H^n}}}\RSB - \mathbb{E}\LSB\frac1K\trace Q\trace Q\frac{ H\Am H\htp}{K}e^{{\gamma_n^{\Xm^n,H^n}}}\RSB\nonumber\\
&\quad + \frac1K\sum_{i,j}\mathbb{E}\LSB \LSB\Am  H\htp Q\RSB_{ji}\frac{\partial {\gamma_n^{\Xm^n,H^n}}}{\partial H_{ij}^*}e^{{\gamma_n^{\Xm^n,H^n}}}\RSB\\
&=  - \mathbb{E}\LSB\frac1K\trace Q\trace Q\frac{ H\Am H\htp}{K}e^{{\gamma_n^{\Xm^n,H^n}}}\RSB\nonumber\\
&\quad +{\bf i}t \sqrt{\frac nK} \mathbb{E}\Bigg[ \Bigg(\frac1K\trace  Q^2 \frac{ H\Xm\Xm\htp\Am H\htp}{K(n+1)} \nonumber\\
&\qquad + \frac1K\trace Q\LB\frac{Q H\Am H\htp}{K}\RB^2  \Bigg) e^{{\gamma_n^{\Xm^n,H^n}}}\Bigg] \nonumber\\
&\quad + \Oc\left( \frac{P_1(t)}{K}\LB 1+\frac1K\trace\Am^2\RB \right)
\end{align}
for some polynomial $P_1(t)$, where the last line follows from the observation that $\trace\Am =0$ and that the terms in the derivative of ${\gamma_n^{\Xm^n,H^n}}$ resulting from $\LB\theta_{n}^{\Xm^n,H^n}\RB^2$ and $\kappa_{n}^{\Xm^n,H^n}$ are of order $\Oc({\frac {t^2}{K}(1+\frac1K\trace\Am^2}))$ and $\Oc({\frac {t^3}{K^2}(1+\frac1K\trace\Am^2}))$, respectively.

Rearranging the terms, one arrives at
\begin{align}
 &\mathbb{E}\LSB \trace Q\frac{ H\Am H\htp}{K}\LB 1+\frac1K\trace Q\RB e^{{\gamma_n^{\Xm^n,H^n}}}\RSB\nonumber\\
 &= {\bf i}t \sqrt{\frac nK} \mathbb{E}\Bigg[ \Bigg(\frac1K\trace  Q^2 \frac{ H\Xm\Xm\htp\Am H\htp}{K(n+1)} \nonumber\\
 &\qquad + \frac1K\trace Q\LB Q\frac{ H\Am H\htp}{K}\RB^2  \Bigg) e^{{\gamma_n^{\Xm^n,H^n}}}\Bigg]\nonumber\\
  \label{eq:calculus1}
 &\quad + \Oc\left( \frac{P_1(t)}{K}\LB 1+\frac1K\trace\Am^2\RB \right).
\end{align}
Using the identity $\Am-\Am^2 = \frac{\Xm\Xm\htp}{n+1}\Am$, we obtain
\begin{align}
 &{\bf i}t \sqrt{\frac nK} \mathbb{E}\Bigg[ \Bigg(\frac1K\trace  Q^2 \frac{ H\Xm\Xm\htp\Am H\htp}{K(n+1)} \nonumber\\
 &\qquad + \frac1K\trace Q\LB Q\frac{ H\Am H\htp}{K}\RB^2  \Bigg) e^{{\gamma_n^{\Xm^n,H^n}}}\Bigg]\nonumber\\
&= {\bf i}t \sqrt{\frac nK} \mathbb{E}\Bigg[ \Bigg(\frac1K\trace  Q^2 \frac{ H\Am H\htp}{K} - \frac1K\trace  Q^2 \frac{ H\Am^2 H\htp}{K} \nonumber\\
\label{eq:calculus2}
&\qquad+ \frac1K\trace Q\LB Q\frac{ H\Am H\htp}{K}\RB^2  \Bigg) e^{{\gamma_n^{\Xm^n,H^n}}}\Bigg].
\end{align}
Note now that
\begin{align}
&\var\LSB\frac1K\trace Q\LB Q\frac{ H\Am H\htp}{K}\RB^2\RSB\nonumber\\
&= \var\LSB\frac1K\trace\tilde{ Q}^2\frac{ H\htp H}{K}\Am\tilde{ Q}\frac{ H\htp H}{K}\Am\RSB\\
&= \var\LSB \frac1K\trace\tilde{ Q}\LB \Id_K-\sigma^2\tilde{ Q}\RB\Am\LB \Id_K-\sigma^2\tilde{ Q}\RB\Am \RSB\\
&\le \Bigg( \sqrt{\var\LSB\frac1K\trace\tilde{ Q}\Am^2\RSB}+\sqrt{\var\LSB\frac{\sigma^2}K\trace\tilde{ Q}^2\Am^2\RSB} \nonumber\\
&\quad+\sqrt{\var\LSB\frac{\sigma^2}{K}\trace\LB\tilde{ Q}\Am\RB^2\RSB}+\sqrt{\var\LSB\frac{\sigma^4}{K}\trace\tilde{ Q}^2\Am\tilde{ Q}\Am\RSB}\Bigg)^2\\\label{eq:approx_1}
&= \Oc\LB\frac1{K^3}\trace\Am^4\RB = \Oc \left( \frac1{K} \left( \frac1K \trace \Am^2 \right)^2 \right)
\end{align}
where the inequality follows from Remark~\ref{rem:cauchyschwarz} in Appendix~\ref{app:Gaussian_relResults} and the last line follows from a direct application of Proposition~\ref{prop:vartrace} in Appendix~\ref{app:aux_RMTaux} to each of the individual terms, along with $\trace \Am^4\leq (\trace \Am^2)^2$.
By Proposition~\ref{prop:vartrace},
\begin{align}
\var\LSB\frac1K\trace  Q^2 \frac{ H\Am H\htp}{K}\RSB &= \Oc\LB\frac1{K^3}\trace\Am^2\RB = \Oc \left( \frac1K \right) \\
\var\LSB\frac1K\trace  Q^2 \frac{ H\Am^2 H\htp}{K}\RSB &= \Oc\LB\frac1{K^3}\trace\Am^4\RB\\
& = \Oc \left( \frac1{K} \left( \frac1K \trace \Am^2 \right)^2 \right).
\end{align}
Thus, by Lemma~\ref{lem:cauchyschwarz} in Appendix~\ref{app:Gaussian_relResults}, the RHS of \eqref{eq:calculus2} can be written as in \eqref{eq:calculus3} on the top of the next page.
\begin{figure*}
\begin{align}\label{eq:calculus3}
&{\bf i}t \sqrt{\frac nK} \mathbb{E}\LSB \LB\frac1K\trace  Q^2 \frac{ H\Am H\htp}{K} - \frac1K\trace  Q^2 \frac{ H\Am^2 H\htp}{K}+ \frac1K\trace Q\LB Q\frac{ H\Am H\htp}{K}\RB^2  \RB e^{{\gamma_n^{\Xm^n,H^n}}}\RSB\nonumber\\
&= {\bf i}t \sqrt{\frac nK} \mathbb{E}\LSB \LB\frac1K\trace  Q^2 \frac{ H\Am H\htp}{K} - \frac1K\trace  Q^2 \frac{ H\Am^2 H\htp}{K}+ \frac1K\trace Q\LB Q\frac{ H\Am H\htp}{K}\RB^2  \RB \RSB\mathbb{E}\LSB e^{{\gamma_n^{\Xm^n,H^n}}}\RSB + \Oc\LB \frac{t}{\sqrt{K}}\LB 1 + \frac1K\trace \Am^2\RB\RB
\end{align}
\hrulefill
\end{figure*}
By Proposition~\ref{prop:dettrace} in Appendix~\ref{app:aux_RMTaux}, we can approximate the first two terms in \eqref{eq:calculus3} by
\begin{align}
&\mathbb{E}\LSB\frac1K\trace  Q^2 \frac{ H\Am H\htp}{K}\RSB\nonumber\\
&= \gamma_1\LB\sigma^2\RB\frac1K\trace\Am + \Oc\LB \sqrt{\frac1{K^5}\trace\Am^2}\RB \\ 
\label{eq:calculus5}
&= \Oc\LB \frac1{\sqrt{K^3}} \RB \\
&\mathbb{E}\LSB\frac1K\trace  Q^2 \frac{ H\Am^2 H\htp}{K}\RSB\nonumber\\
&= \gamma_1\LB\sigma^2\RB\frac1K\trace\Am^2  + \Oc\LB \sqrt{\frac1{K^5}\trace\Am^4}\RB \\ \label{eq:calculus6}
&= \gamma_1\LB\sigma^2\RB\frac1K\trace\Am^2  + \Oc\LB \frac1{\sqrt{K^3}} \frac1K\trace \Am^2 \RB.
\end{align}
It remains to find an approximation of the term $\mathbb{E}\LSB\frac1K\trace Q\LB Q\frac{ H\Am H\htp}{K}\RB^2\RSB$. By Lemma~\ref{lem:IBP} in Appendix~\ref{app:Gaussian_relResults},
\begin{align}
&\mathbb{E}\LSB\frac1K\trace Q\LB Q\frac{ H\Am H\htp}{K}\RB^2\RSB\nonumber\\
&= \mathbb{E}\LSB\frac1{K^3}\trace H\Am H\htp Q^2 H\Am H\htp Q\RSB\\
&= \frac1{K^3}\sum_{i,j}\mathbb{E}\LSB H_{ij} \LSB\Am H\htp Q^2 H\Am H\htp Q\RSB_{ji}\RSB\\\label{eq:dbltrace1}
&= \frac1{K^3}\sum_{i,j}\mathbb{E}\LSB \frac{\partial \LSB\Am H\htp Q^2 H\Am H\htp Q\RSB_{ji}}{\partial H_{ij}^*}\RSB.
\end{align}

The derivative further develops as
\begin{align}
 &\frac{\partial \LSB\Am H\htp Q^2 H\Am H\htp Q\RSB_{ji}}{\partial H_{ij}^*}=\nonumber\\
 & A_{jj}\LSB Q^2 H\Am H\htp Q\RSB_{ii} - \frac1K \LSB\Am H\htp Q H\RSB_{jj}\LSB Q^2 H\Am H\htp Q\RSB_{ii} \nonumber\\
 & - \frac1K\LSB\Am H\htp Q^2 H\RSB_{jj}\LSB Q H\Am H\htp Q\RSB_{ii} + \LSB\Am H\htp Q^2 H\Am\RSB_{jj} Q_{ii}\nonumber\\
 \label{eq:dbltr1der}
 & -\frac1K\LSB\Am H\htp Q^2 H\Am H\htp Q H\RSB_{jj}Q_{ii}.
\end{align}

Replacing \eqref{eq:dbltr1der} in \eqref{eq:dbltrace1} and rearranging the resulting terms, we arrive at
\begin{align}
&\mathbb{E}\LSB\frac1K\trace Q\LB Q\frac{ H\Am H\htp}{K}\RB^2\LB1+\frac1K\trace Q\RB\RSB \nonumber\\
 &= \mathbb{E}\LSB \frac1K\trace Q^3 \frac{ H\Am H\htp}{K}\LB \frac1K\trace\Am - \frac1K\trace Q\frac{ H\Am H\htp}{K}\RB\RSB\nonumber\\
 &\quad - \mathbb{E}\LSB\LB \frac1K\trace Q^2\frac{ H\Am H\htp}{K}\RB^2\RSB\nonumber\\
 &\quad + \mathbb{E}\LSB \frac1K\trace Q \frac1K\trace Q^2 \frac{ H\Am^2 H\htp}{K}\RSB.
\end{align}

Applying Proposition~\ref{prop:dettrace} in Appendix~\ref{app:aux_RMTaux} together with Proposition~\ref{prop:vartrace} in Appendix~\ref{app:aux_RMTaux} and Lemma~\ref{lem:cauchyschwarz} in Appendix~\ref{app:Gaussian_relResults} to the individual terms leads to
\begin{align}
&\mathbb{E}\LSB\frac1K\trace Q\LB Q\frac{ H\Am H\htp}{K}\RB^2\LB1+\frac1K\trace Q\RB\RSB \nonumber\\
\label{eq:approx_11}
&=  \delta_0\LB\sigma^2\RB \gamma_1\LB \sigma^2\RB\frac1K\trace\Am^2 + \Oc \left( \frac1{\sqrt{K^5}}\trace \Am^2 \right). 
\end{align}

Similarly, by Lemma~\ref{lem:cauchyschwarz}, Proposition~\ref{prop:vartrace}, the variance bound in \eqref{eq:approx_1}, and Proposition~\ref{prop:dettrace},
\begin{align}
&\mathbb{E}\LSB\frac1K\trace Q\LB Q\frac{ H\Am H\htp}{K}\RB^2\LB1+\frac1K\trace Q\RB\RSB \nonumber\\
&= \mathbb{E}\LSB\frac1K\trace Q\LB Q\frac{ H\Am H\htp}{K}\RB^2\RSB \mathbb{E}\LSB 1+\frac1K\trace Q\RSB\nonumber\\
&\quad + \Oc\LB\sqrt{\frac1{K^5}\trace\Am^4}\RB\\
&= \mathbb{E}\LSB\frac1K\trace Q\LB Q\frac{ H\Am H\htp}{K}\RB^2\RSB  \LB 1 + \delta_0(\sigma^2) \RB\nonumber\\
\label{eq:approx_2}
&\quad + \Oc\LB \frac1{\sqrt{K^5}}\trace\Am^2 \RB.
\end{align}

Equating the RHSs of \eqref{eq:approx_11} and \eqref{eq:approx_2} and solving for $\mathbb{E}\LSB\frac1K\trace Q\LB Q\frac{ H\Am H\htp}{K}\RB^2\RSB$ yields 
\begin{align}
\mathbb{E}\LSB\frac1K\trace Q\LB Q\frac{ H\Am H\htp}{K}\RB^2\RSB
 &= \frac{\delta_0\LB\sigma^2\RB \gamma_1\LB \sigma^2\RB\frac1K\trace\Am^2}{1+\delta_0(\sigma^2)}\nonumber\\
 \label{eq:approx_3}
 &\quad   +  \Oc\LB \frac1{\sqrt{K^5}} \trace\Am^2\RB.
\end{align}

Similar to the proof of Part $(i)$, let us define
\begin{align}
\Psi &= \frac1K\trace Q\\
\Phi &= \trace Q\frac{ H\Am H\htp}{K}.
\end{align}
Putting the results from \eqref{eq:calculus1}, \eqref{eq:calculus2}, \eqref{eq:calculus3}, \eqref{eq:calculus5}, \eqref{eq:calculus6}, and \eqref{eq:approx_3} together, we conclude that
\begin{align}
\mathbb{E}\LSB \Phi\LB 1+\Psi\RB e^{{\gamma_n^{\Xm^n,H^n}}}\RSB &= -{\bf i}t \sqrt{\frac nK} \frac{\gamma_1(\sigma^2)}{1+\delta_0(\sigma^2)}\frac1K\trace\Am^2\nonumber\\
\label{eq:exptrace_1}
&\quad +\Oc\left( \frac{P_1(t)}{\sqrt{K}} + \frac{tP_2(t)}{\sqrt{K}} \frac1K\trace \Am^2 \right)
\end{align}
for two polynomials $P_1(t)$ and $P_2(t)$, where the term $t$ in front of $P_2(t)$ arises from the pre-multiplication by at least $\iv t$ of the various estimators involved.

We now need to find an alternative representation of the term $\mathbb{E}\LSB \Phi \Psi e^{{\gamma_n^{\Xm^n,H^n}}}\RSB$ in the LHS of the last equation. Following the same arguments as in \eqref{eq:inequal-21}--\eqref{eq:inequal-2}, and using $\sqrt{K^{-3}\trace \Am^2}\leq 1/\sqrt{K}$, we can write
\begin{align}
&\mathbb{E}\LSB \Phi \Psi e^{{\gamma_n^{\Xm^n,H^n}}}\RSB\nonumber\\
 &= \mathbb{E}\LSB \Psi\RSB\mathbb{E}\LSB \Phi e^{{\gamma_n^{\Xm^n,H^n}}}\RSB + \mathbb{E}\LSB \Phi \LB\Psi - \mathbb{E}\LSB\Psi\RSB \RB e^{{\gamma_n^{\Xm^n,H^n}}}\RSB\\
&=\mathbb{E}\LSB \Psi\RSB\mathbb{E}\LSB \Phi e^{{\gamma_n^{\Xm^n,H^n}}}\RSB + \mathbb{E}\LSB \Phi\RSB\mathbb{E}\LSB \LB \Psi-\mathbb{E}\LSB \Psi\RSB\RB e^{{\gamma_n^{\Xm^n,H^n}}}\RSB \nonumber\\
&\quad + \Oc\LB\frac1{\sqrt{K}}\RB\\
&= \mathbb{E}\LSB \Psi\RSB\mathbb{E}\LSB \Phi e^{{\gamma_n^{\Xm^n,H^n}}}\RSB - \mathbb{E}\LSB \Phi\RSB \mathbb{E}\LSB\Psi\RSB \mathbb{E}\LSB e^{{\gamma_n^{\Xm^n,H^n}}}\RSB \nonumber\\
&\quad+ \mathbb{E}\LSB \Phi\RSB \mathbb{E}\LSB\Psi e^{\gamma_n^{\Xm^n,H^n}}\RSB + \Oc\LB\frac1{\sqrt{K}}\RB\\
&= \mathbb{E}\LSB \Psi\RSB\mathbb{E}\LSB \Phi e^{{\gamma_n^{\Xm^n,H^n}}}\RSB - \mathbb{E}\LSB \Phi\RSB \mathbb{E}\LSB\Psi\RSB \mathbb{E}\LSB e^{{\gamma_n^{\Xm^n,H^n}}}\RSB\nonumber\\
&\quad + \mathbb{E}\LSB \Phi\RSB \mathbb{E}\LSB\LB \frac{1}{\sigma^2}\frac{N}{K} - \frac1{\sigma^2}\frac1K\trace Q\frac{ H H\htp}{K}\RB e^{\gamma_n^{\Xm^n,H^n}}\RSB\nonumber\\
&\quad + \Oc\LB\frac1{\sqrt{K}}\RB\\
&= \mathbb{E}\LSB \Psi\RSB\mathbb{E}\LSB \Phi e^{{\gamma_n^{\Xm^n,H^n}}}\RSB \nonumber\\
&\quad + \LB\frac{N}{\sigma^2} -K \mathbb{E}\LSB\Psi\RSB\RB \mathbb{E}\LSB \frac{\Phi}{K}\RSB\mathbb{E}\LSB e^{{\gamma_n^{\Xm^n,H^n}}}\RSB\nonumber\\
&\quad - \frac1{\sigma^2} \mathbb{E}\LSB \frac{\Phi}{K}\RSB \mathbb{E}\LSB \trace Q\frac{ H H\htp}{K}e^{{\gamma_n^{\Xm^n,H^n}}}\RSB + \Oc\LB\frac1{\sqrt{K}}\RB\\
&= \delta_0(\sigma^2)\mathbb{E}\LSB \Phi e^{{\gamma_n^{\Xm^n,H^n}}}\RSB + \Oc\LB\frac1{\sqrt{K}}\RB.
\end{align}

From the last result and \eqref{eq:exptrace_1}, we have
\begin{align}
&\mathbb{E}\LSB \Phi\LB 1+\Psi\RB e^{{\gamma_n^{\Xm^n,H^n}}}\RSB\nonumber\\\label{eq:solve-1}
&= (1+\delta_0(\sigma^2))\mathbb{E}\LSB \Phi e^{{\gamma_n^{\Xm^n,H^n}}}\RSB + \Oc\LB\frac1{\sqrt{K}}\RB\\
&= -{\bf i}t \sqrt{\frac nK} \frac{\gamma_1\LB\sigma^2\RB}{1+\delta_0(\sigma^2)}\frac1K\trace\Am^2 \mathbb{E}\LSB  e^{{\gamma_n^{\Xm^n,H^n}}}\RSB \nonumber\\
&\quad +\Oc\LB \frac{P_1(t)}{\sqrt{K}} + \frac{tP_2(t)}{\sqrt{K}}\frac1K\trace \Am^2 \RB\label{eq:solve-2}
\end{align}
for some polynomials $P_1(t)$ and $P_2(t)$.

Solving \eqref{eq:solve-1} and \eqref{eq:solve-2} for $\mathbb{E}\LSB \Phi e^{{\gamma_n^{\Xm^n,H^n}}}\RSB$ yields
\begin{align}
\mathbb{E}\LSB \Phi e^{{\gamma_n^{\Xm^n,H^n}}}\RSB &= -{\bf i}t \sqrt{\frac{n}{K}}\frac{\gamma_1\LB\sigma^2\RB\frac1K\trace\Am^2}{\LB 1+\delta_0(\sigma^2)\RB^2}\mathbb{E}\LSB e^{{\gamma_n^{\Xm^n,H^n}}}\RSB \nonumber\\
&\quad + \Oc\LB \frac{P_1(t)}{\sqrt{K}} + \frac{tP_2(t)}{\sqrt{K}}\frac1K\trace \Am^2 \RB.
\end{align}
This concludes the proof of part $(ii)$.

\section*{Acknowledgments}
We would like to thank the anonymous reviewers for their valuable comments which helped to improve and significantly shorten some of the proofs. In particular, we are indebted to one of the reviewers for providing the proof of Lemma~\ref{lemma:yuri}. We would like to thank Dr.\@ Laurent Schmalen for the generation of the simulation results for Figure~\ref{fig:Pe_n_simu} and Prof.\@ Stephan ten Brink for various discussions around the topic of iterative coding systems. We are grateful to Prof.\@ M\'{e}rouane Debbah for discussions at the early stage of this work. 

\bibliographystyle{IEEEtran}
\bibliography{IEEEabrv,bibliography}

\begin{IEEEbiographynophoto}{Jakob Hoydis}(S'08--M'12) received the diploma degree (Dipl.-Ing.) in electrical engineering and information technology from RWTH Aachen University, Germany, and the Ph.D. degree from Sup\'{e}lec, Gif-sur-Yvette, France, in 2008 and 2012, respectively. From 2012-2013, he worked for Bell Laboratories, Alcatel-Lucent, Stuttgart, Germany, on next-generation mobile communication systems. From 2014-2015, he was co-founder and technical director of Spraed, Orsay, France. In September 2015, he joined Bell Labs France in the ''Software-Defined Wireless Networks"-Department. He is recipient of the 2012 Publication Prize of the Supélec Foundation, the 2013 VDE ITG Förderpreis and the 2015 Leonard G. Abraham Prize of the IEEE Communications Society. He received the WCNC'2014 best paper award and has been nominated as an Exemplary Reviewer 2012 for the IEEE Communication letters. His research interests are in the areas of cloud computing, SDR, large random matrix theory, information theory, signal processing and their applications to wireless communications.
\end{IEEEbiographynophoto}
\vspace{-10pt}

\begin{IEEEbiographynophoto}{Romain Couillet}(S'07--M'11) received his MSc in Mobile Communications at the Eurecom Institute and his MSc in Communication Systems in Telecom ParisTech, France in 2007. From 2007 to 2010, he worked with ST-Ericsson as an Algorithm Development Engineer on the Long Term Evolution Advanced project, where he prepared his PhD with Supelec, France, which he graduated in November 2010. He is currently an assistant professor in the Telecommunication department of CentraleSupélec, France. His research topics are in random matrix theory applied to wireless communications, signal processing, and statistics. In 2015, he received the HDR title from University ParisSud. He is the recipient of the 2013 CNRS Bronze Medal in the section "science of information and its interactions", of the 2013 IEEE ComSoc Outstanding Young Researcher Award (EMEA Region), of the 2011 EEA/GdR ISIS/GRETSI best PhD thesis award, and of the Valuetools 2008 best student paper award.
\end{IEEEbiographynophoto}
\vspace{-10pt}
\begin{IEEEbiographynophoto}{Pablo Piantanida}(S'04--M'08) received both B.Sc. in Electrical Engineering  (with honors)  and B.Sc. in Mathematics, and M.Sc degrees from the University of Buenos Aires (Argentina) in 2003, and the Ph.D. from Université Paris-Sud (Orsay, France) in 2007. Since October 2007 he has joined the Laboratoire des Signaux et Systèmes (L2S), CentraleSupélec-CNRS, as an Assistant Professor in Network Information Theory. His research interests include multi-terminal information theory, Shannon theory, machine learning,  cooperative communications, physical-layer security and distributed source coding.
\end{IEEEbiographynophoto}
\vfill
\end{document}

%% file: macros.tex
\newcommand{\LB}{\left(}
\newcommand{\RB}{\right)}
\newcommand{\LSB}{\left[}
\newcommand{\RSB}{\right]}
\newcommand{\htp}{^{\sf H}}
\newcommand{\tp}{^{\sf T}}

\newcommand{\mb}{\mathbf}

\newfont{\bbb}{msbm10 scaled 500}

\newfont{\bb}{msbm10 scaled 1100}
\newcommand{\CC}{\mbox{\bb C}}

\newcommand{\EE}{\mbox{\bb E}}

\newcommand{\iv}{{\bf i}}

\newcommand{\mv}{{\bf m}}

\newcommand{\wv}{{\bf w}}

\newcommand{\xv}{{\bf x}}
\newcommand{\yv}{{\bf y}}

\newcommand{\zerov}{{\bf 0}}

\newcommand{\Am}{{\bf A}}
\newcommand{\Bm}{{\bf B}}
\newcommand{\Cm}{{\bf C}}
\newcommand{\Dm}{{\bf D}}

\newcommand{\Hm}{{\bf H}}
\newcommand{\Id}{{\bf I}}

\newcommand{\Qm}{{\bf Q}}
\newcommand{\Rm}{{\bf R}}
\newcommand{\Sm}{{\bf S}}
\newcommand{\Tm}{{\bf T}}

\newcommand{\Wm}{{\bf W}}

\newcommand{\Xm}{{\bf X}}
\newcommand{\Ym}{{\bf Y}}

\newcommand{\Cc}{{\cal C}}

\newcommand{\Nc}{{\cal N}}
\newcommand{\Oc}{{\cal O}}
\newcommand{\Pc}{{\cal P}}

\newcommand{\Sc}{{\cal S}}

\newcommand{\supp}{{\hbox{Supp}\,}}

\renewcommand{\det}{{\hbox{det}}}
\newcommand{\trace}{{\hbox{tr}\,}}

\newcommand{\var}{\mathbb{V}{\rm ar}}

\renewcommand{\Re}{{\rm Re}}

\newcommand{\defines}{{\,\,\stackrel{\scriptscriptstyle \bigtriangleup}{=}\,\,}}

\newtheorem{theorem}{Theorem}
\newtheorem{definition}{Definition}
\newtheorem{lemma}{Lemma}
\newtheorem{cor}{Corollary}
\newtheorem{prop}{Proposition}

\newtheorem{remark}{Remark}
\newtheorem{property}{Property}